\documentclass[11pt]{article}
\usepackage[T1]{fontenc}
\usepackage{lmodern}
\usepackage[margin=1in]{geometry}
\usepackage{amsmath,amssymb,amsthm,mathtools}
\usepackage{mathrsfs}
\usepackage{microtype}
\usepackage{xcolor}
\usepackage{tikz}
\definecolor{linkblue}{RGB}{25,55,105}
\usepackage[colorlinks=true,allcolors=linkblue]{hyperref}
\usepackage[capitalise,nameinlink,noabbrev]{cleveref}
\usepackage{aliascnt}
\numberwithin{equation}{section}
\newtheorem{theorem}{Theorem}[section]
\newaliascnt{lemma}{theorem}
\newtheorem{lemma}[lemma]{Lemma}
\aliascntresetthe{lemma}
\newaliascnt{proposition}{theorem}
\newtheorem{proposition}[proposition]{Proposition}
\aliascntresetthe{proposition}
\newaliascnt{corollary}{theorem}
\newtheorem{corollary}[corollary]{Corollary}
\aliascntresetthe{corollary}
\newaliascnt{conjecture}{theorem}
\newtheorem{conjecture}[conjecture]{Conjecture}
\aliascntresetthe{conjecture}
\newaliascnt{definition}{theorem}
\newtheorem{definition}[definition]{Definition}
\aliascntresetthe{definition}
\theoremstyle{remark}
\newaliascnt{remark}{theorem}
\newtheorem{remark}[remark]{Remark}
\aliascntresetthe{remark}
\crefname{lemma}{Lemma}{Lemmas}
\crefname{proposition}{Proposition}{Propositions}
\crefname{corollary}{Corollary}{Corollaries}
\crefname{conjecture}{Conjecture}{Conjectures}
\crefname{remark}{Remark}{Remarks}
\crefname{definition}{Definition}{Definitions}
\newcommand{\R}{\mathbb R}
\newcommand{\Q}{\mathbb Q}
\newcommand{\Z}{\mathbb Z}
\newcommand{\C}{\mathbb C}
\newcommand{\E}{\mathbb E}
\newcommand{\T}{\mathbb T}
\newcommand{\shift}{\mathsf S}
\newcommand{\Acal}{\mathcal A}

\DeclareMathOperator{\arcosh}{arcosh}
\DeclareMathOperator{\arsinh}{arsinh}
\title{Degree Lower Bounds for Symmetric Boolean Functions:\\ Breaking the Prime-Gap Barrier}
\author{
  Xiaoming Sun\thanks{State Key Lab of Processors, Institute of Computing
    Technology, Chinese Academy of Sciences, Beijing, China; University of
    Chinese Academy of Sciences, Beijing, China.
  \texttt{sunxiaoming@ict.ac.cn}.}
  \and
  Chengu Wang\thanks{\texttt{wangchengu@gmail.com}.}
}
\date{}
\hypersetup{pdftitle={Degree Lower Bounds for Symmetric Boolean Functions: Breaking the Prime-Gap Barrier},pdfauthor={Xiaoming Sun, Chengu Wang}}

\newcommand{\Bt}{\mathcal B}
\newcommand{\Sg}{\mathcal S}
\newcommand{\Cf}{\mathcal C}
\newcommand{\Df}{\mathcal D}
\newcommand{\rmin}{{\rm in}}
\newcommand{\rmout}{{\rm out}}
\begin{document}
\hypersetup{pageanchor=false}
\pagenumbering{gobble}
\maketitle

\begin{abstract}
  We prove that every nonconstant symmetric Boolean function on $n$
  variables has exact real polynomial degree at least
  $n-O(n^{1/3}(\log n)^{2/3})$. This makes progress toward the conjecture of
  von zur Gathen and Roche that the degree is always at least $n-3$.
  The original prime-gap argument of von zur Gathen and Roche yields a degree deficit of $O(n^{0.525})$ using
  the theorem of Baker, Harman, and Pintz, or $O(n^{0.52})$ using an
  estimate claimed in a recent preprint. Our $\widetilde{O}(n^{1/3})$
  bound on the degree deficit improves upon the $\widetilde{O}(\sqrt{n})$ bound
  obtainable from the prime-gap approach, even assuming the Riemann hypothesis or
  Legendre's conjecture. Notably, our proof avoids prime-gap estimates entirely.

  The proof combines vanishing binomial moments with the arithmetic of
  linear recurrences. A quantitative form of the integer Hankel
  determinant method in the P\'olya--Carlson theorem forces the function's
  values on a central interval of Hamming weights to satisfy a short
  rational recurrence. A second determinant argument bounds the number
  of characteristic roots near $-1$. Through the binomial transform,
  these roots can then be cancelled by a polynomial test of degree
  below the deficit. An integer annihilator and contour estimates make
  the test isolate the first failure of the recurrence, yielding a contradiction
  with the vanishing-moment conditions. If the recurrence agrees with the function on
  a sufficiently long interval, arithmetic height bounds instead force
  a bounded period, which is ruled out by a separate polynomial test. Every
  numbered result in the paper has been formally verified in Lean~4.
\end{abstract}

\clearpage
\tableofcontents
\clearpage
\pagenumbering{arabic}
\hypersetup{pageanchor=true}

\section{Introduction}\label{sec:introduction}
Every function $f:\{0,1\}^n\to\R$ can be represented by a unique multilinear
polynomial in $\R[x_1,\ldots,x_n]$, meaning that the polynomial equals $f$ at
every point of $\{0,1\}^n$, with nothing required off the cube. The degree of
this polynomial is the
\emph{exact real degree} $\deg_\R(f)$.

The study of Boolean functions through their real polynomial
representations goes back to the book of Minsky and
Papert~\cite{MP17}, first published in 1969.
In 1992, Nisan and Szegedy~\cite{NS92,NS94} studied the degree as a
complexity measure in its own right. They proved that a function depending
on $n$ variables has degree at least $\log_2n-O(\log\log n)$, which is
tight up to the lower-order term for the address function. A general
Boolean function can therefore have degree logarithmic in its number of
relevant variables.

A Boolean function $f:\{0,1\}^n\to\{0,1\}$ is symmetric when its value depends
only on the Hamming weight. Write
\[
  f(x)=F(|x|),\qquad F:\{0,\ldots,n\}\to\{0,1\},
\]
and call $F$ the \emph{profile} of $f$.
Symmetry reduces a truth table of size $2^n$ to these $n+1$ values.
The symmetrization argument of Minsky and Papert averages a multilinear
polynomial over all permutations of the variables and produces a univariate
polynomial in the Hamming weight, of no larger degree. This is the step behind
the following reduction: for symmetric $f$, the degree $\deg_\R(f)$ equals the
degree of the unique polynomial $q_F\in\Q[X]$ of degree at most $n$
interpolating $F$ on the \emph{grid} $\{0,\ldots,n\}$.

Even a simple root count forces substantial degree: for a nonconstant profile
$F$, the polynomial $q_F(q_F-1)$ is nonzero and vanishes at all $n+1$ grid
points, so $\deg q_F\ge\lceil(n+1)/2\rceil$. Nonconstant symmetric functions
therefore lie at the opposite extreme from the logarithmic behavior above.
The journal version of Nisan and Szegedy closes with three open
problems~\cite[Section~4]{NS94}. The third asks for a lower bound on the
degree of a nonconstant symmetric function
going beyond the easy bound $n/2$. That is the problem studied here.
The challenge is to show that the
degree is much closer to $n$. This requires using the equal spacing
of the grid and the arithmetic of the two allowed values, beyond
the number of zeros alone.

Von zur Gathen and Roche~\cite{GR97} studied this problem as one of
polynomials with two values. They describe their work as motivated by
that of Nisan and Szegedy, and credit Szegedy with posing the problem.
Let $p(n)$ be the largest prime at most $n+1$. Their argument, carried out
modulo the prime $p(n)$~\cite[Theorem 2.6 and the proof of Theorem 2.7]{GR97}, gives
\[
  \deg q_F\ge p(n)-1
\]
for every nonconstant profile $F$. Since $p(n)$ is close to $n$, the natural
quantity is now the remaining distance from $n$. For a nonconstant profile define
\[
  \delta(F)=n-\deg q_F,\qquad
  \Delta(n)=\max_{F\text{ nonconstant}}\delta(F).
\]
We call $\delta(F)$ the \emph{degree deficit}, or simply the
\emph{deficit}, of $F$ and of $f$.
Thus $\Delta(n)\le n+1-p(n)$, and in particular $\Delta(n)=0$ when
$n+1$ is prime. Mozzochi's prime-gap theorem~\cite{Mozzochi86} gave them
$\Delta(n)=O(n^{0.548})$; Baker, Harman, and Pintz's
theorem~\cite[Theorem 1]{BHP01} improves this to $O(n^{0.525})$.
Li's preprint~\cite[Theorem 1]{Li25} claims a prime-gap estimate that would
give $O(n^{0.52})$.

Von zur Gathen and Roche also conjectured a constant bound, and more
specifically its sharp possible value~\cite[pp.~346, 349, 360]{GR97}:
\begin{conjecture}[von zur Gathen--Roche]\label{conj:deficit-at-most-three}
  For every $n\ge1$, $\Delta(n)\le3$.
\end{conjecture}
Their constructions attain deficit three in two families
\cite[Theorems 4.2 and 4.3]{GR97}, described in \Cref{sec:open-problems}.
Thus no smaller universal constant is possible. Their computer
search determined every profile of positive deficit for $n\le128$, and
the maximal deficit in this range is three.

The reduction to a prime explains both the strength and the limitation of
using primes: it gives full degree at prime-adjacent sizes, while
its guarantee at a general size is measured by the distance to the
preceding prime. A uniform improvement by this route would require
correspondingly stronger control of prime gaps. Under the Riemann
hypothesis, Cram\'er~\cite{Cramer36} proved that the prime gap near $n$ is
$O(\sqrt n\log n)$, which would give
$\Delta(n)=O(\sqrt n\log n)$. Even Legendre's conjecture, that there is a
prime between $m^2$ and $(m+1)^2$ for every integer $m\ge1$, would give only
$\Delta(n)=O(\sqrt n)$. Cram\'er's probabilistic model suggests
gaps of maximal order $(\log n)^2$, and prime gaps are unbounded in any case,
so this route cannot by itself prove \Cref{conj:deficit-at-most-three}, or any
constant bound.

Cohen, Shpilka, and Tal~\cite{CST17} study a more general problem: the
degree of polynomials mapping $\{0,\ldots,n\}$ into $\{0,\ldots,m\}$, for
an integer $m\ge1$. Symmetric Boolean functions are the case $m=1$. In that
case, they prove a square-root degree-deficit
bound at sizes $n=p^2-1$ with $p$ prime, and ask for a uniform bound
$\deg q_F\ge n-\sqrt n$ as an intermediate goal toward the constant-deficit
conjecture.

In a different direction, Sun, Sun, Wang, et al.~\cite{Sun20} investigate
the degree of symmetric Boolean functions over the finite ring $\Z_m$,
rather than over $\R$ or as integer-valued polynomials. Their arguments
build on earlier work of Wilson~\cite{Wilson06} on Mahler expansions and
periodicity properties of symmetric functions modulo prime powers.

\subsection{Main result}
Our main theorem lowers the exponent in the bound on $\Delta(n)$ from
$0.525$~\cite{GR97,BHP01}, or the $0.52$ claimed in~\cite{Li25}, to
$1/3$.

\begin{theorem}\label{thm:one-third-degree-bound}
  There is an absolute constant $C>0$ such that, for every positive integer $n$,
  \[
    \Delta(n)\le C n^{1/3}(\log n)^{2/3}.
  \]
\end{theorem}

\Cref{thm:one-third-degree-bound} goes beyond the bound $O(\sqrt n\log n)$ that
Cram\'er's theorem gives under the Riemann hypothesis and the bound
$O(\sqrt n)$ that Legendre's conjecture would give, and it gives a
smaller deficit than the uniform goal $\sqrt n$ of Cohen, Shpilka, and
Tal. The proof uses none of the prime-gap estimates above. Its arithmetic
inputs are elementary integrality facts (a nonzero integer has modulus at
least one, and $p$-adic valuations of denominators), the irreducibility of
cyclotomic polynomials, and Voutier's explicit form of Dobrowolski's
height bound.

Every numbered theorem, lemma, corollary, proposition and definition of
this paper has been formalized and checked in Lean~4 with Mathlib, but the
formal proofs do not always match the written ones.
The Lean code is available at
\url{https://github.com/wcgbg/sym-bool-func-deg}.

\subsection{Corollaries}

\Cref{thm:one-third-degree-bound} has consequences in a few other languages. We state
each with the background of its question.

\paragraph{Minimum Fourier degree.}
For $u:\{0,1\}^n\to\R$ and $S\subseteq[n]=\{1,\ldots,n\}$, write
\[
  \widehat u(S)=2^{-n}\sum_{x\in\{0,1\}^n}u(x)\chi_S(x),\qquad
  \chi_S(x)=(-1)^{\sum_{i\in S}x_i}.
\]
The exact real degree of $u$ is the largest size of a set $S\subseteq[n]$
with $\widehat u(S)\ne0$. A complementary question asks for the
\emph{minimum Fourier degree} of $u$, the smallest size of a nonempty such
set. Multiplication by the parity function $\chi_{[n]}$ exchanges a set with
its complement, connecting these two ends of the Fourier support.
Balancedness, meaning $\widehat u(\varnothing)=0$, ensures that this
exchange yields a nonempty set in the consequence below.

This question comes from computational learning theory. Mossel,
O'Donnell, and Servedio~\cite{MOS04} used low nonzero Fourier
coefficients to learn symmetric $k$-juntas, which are functions on $n$
variables that depend on at most $k$ of them and are symmetric in those,
from uniform examples, in time about $n^{2k/3}$. For a balanced function, they obtained such a
coefficient by multiplying by parity and applying the bound of von zur
Gathen and Roche; unbalanced functions required a different argument.
Let $g(n)$ be the largest length of a prime-free interval up to $n$.
Shpilka and Tal~\cite{ST14} bounded the minimum Fourier degree of symmetric
Boolean functions other than constants and parity functions by
$O(g(n)+\sqrt n)$. They also record the stronger balanced consequence
of the prime argument of von zur Gathen and Roche. The same deduction
from our theorem gives the following balanced bound.
\begin{corollary}\label{cor:balanced-minimum-fourier-degree}
  There is an absolute constant $C>0$ such that, for every positive integer $n$,
  every balanced symmetric
  $u:\{0,1\}^n\to\{-1,1\}$ other than the parity function $\chi_{[n]}$ and
  its negative has $\widehat u(S)\ne0$ for some $S\subseteq[n]$ with
  $1\le|S|\le Cn^{1/3}(\log n)^{2/3}$.
\end{corollary}
\begin{proof}
  The symmetric sign function $v=\chi_{[n]}u$ is nonconstant, so
  $(1+v)/2$ has degree at least $n-\Delta(n)$. Hence $v$ has a nonzero
  Fourier coefficient at a set $T\subseteq[n]$ of size at least
  $n-\Delta(n)$.
  Balancedness gives $\widehat v([n])=\widehat u(\varnothing)=0$, so
  $T\ne[n]$.
  The identity $\widehat v(T)=\widehat u([n]\setminus T)$ gives the
  claim with $S=[n]\setminus T$, since $1\le|S|\le\Delta(n)$, and
  \Cref{thm:one-third-degree-bound} bounds $\Delta(n)$.
\end{proof}
With $v$ and $T$ as in the proof, this deduction fails for unbalanced $u$: then
$\widehat v([n])=\widehat u(\varnothing)\ne0$, so $T$ may be $[n]$, whose
complement is empty. The bound of Shpilka and Tal does cover unbalanced functions.

\paragraph{Resilient functions.}
The same extremal problem has a separate history in cryptography. For an
integer $0\le t\le n$, a Boolean function on $n$ variables is
\emph{$t$-resilient} if it stays balanced after any $t$ of its inputs are
fixed; for its \emph{sign form} $u=2f-1:\{0,1\}^n\to\{-1,1\}$, this means
$\widehat u(S)=0$ for all $S\subseteq[n]$ with $|S|\le t$.
Chor, Goldreich, H{\aa}stad, Friedman, Rudich, and
Smolensky~\cite{CGHFRS85} studied
resilient functions in 1985 in connection with the bit extraction
problem. For symmetric $u$ and $v=\chi_{[n]}u$, the identity in the proof
of \Cref{cor:balanced-minimum-fourier-degree} shows that $u$ is
$t$-resilient exactly when $\deg_\R(v)\le n-t-1$. Hence
$\Delta(n)\ge t+1$ if and only if some symmetric $t$-resilient function
on $n$ variables differs from parity and its negative.

Chor et al.\ conjectured that parity and its negative are the only
$1$-resilient symmetric functions, which is the assertion
$\Delta(n)\le1$, and proved the corresponding statement for
resiliency $2n/3$. Mitchell~\cite{Mitchell90} raised the same question
independently. In 1993, Gopalakrishnan, Hoffman, and
Stinson~\cite{GHS93} disproved the conjecture with an infinite family
of counterexamples from a binomial identity, the smallest at $n=14$.
They asked for the smallest constant $t$, if any, such that parity and
its negative are the only $t$-resilient symmetric functions. In the present
notation this asks for $\sup_n\Delta(n)$, which
\Cref{conj:deficit-at-most-three} asserts is three.

In this language, \Cref{conj:deficit-at-most-three} states that no
nonaffine symmetric $3$-resilient function exists~\cite{Carlet21}.
\begin{corollary}\label{cor:resilient-functions}
  There is an absolute constant $C>0$ such that, for every positive integer
  $n$ and every integer $t$ with $Cn^{1/3}(\log n)^{2/3}\le t\le n$,
  every symmetric $t$-resilient function on $n$ variables has sign form
  $\chi_{[n]}$ or $-\chi_{[n]}$.
\end{corollary}
\begin{proof}
  Choose $C$ as in \Cref{thm:one-third-degree-bound}. Then
  $t\ge\Delta(n)$, so the equivalence above excludes every symmetric
  $t$-resilient function other than parity and its negative.
\end{proof}

\paragraph{Zeros at one.}
The moment conditions have a classical reading. By
\Cref{lem:degree-and-moments}, for a nonconstant profile $F$ and
$Z_k=2F(k)-1$, the polynomial
$\sum_{k=0}^n(-1)^k\binom nkZ_kz^k$ vanishes at $z=1$ to order exactly
$\delta(F)$. The multiplicity of a zero at one under coefficient
constraints is a well-studied problem. For a polynomial
$\sum_{j=0}^na_jz^j$ with complex coefficients, Erd\'elyi~\cite[Theorem 3.2]{Erdelyi15}, motivated by a
question of Szegedy, proved that $|a_0|=1$ and
$|a_j|\le L^{-1}\binom nj$ for $1\le j\le n$, with $2\le L\le e^n$, allow at most
$n-\lfloor c\sqrt{n\log L}\rfloor$ zeros at one, for an absolute constant
$c>0$. Our coefficients have
exactly binomial size, which is the excluded case $L=1$, and size alone
gives nothing there: $(1-z)^n$ arises from the constant profile. The
Boolean values themselves must be used.
\begin{corollary}\label{cor:zeros-at-one}
  There is an absolute constant $C>0$ such that the following holds for
  every positive integer $n$. Let
  \[
    p_\epsilon(z)=\sum_{k=0}^n\epsilon_k\binom nk z^k,
    \qquad \epsilon_0,\ldots,\epsilon_n\in\{-1,1\}.
  \]
  If $p_\epsilon(z)\ne\pm(1-z)^n$, then its order of vanishing at $z=1$ is at most
  $Cn^{1/3}(\log n)^{2/3}$.
\end{corollary}
\begin{proof}
  Put $Z_k=(-1)^k\epsilon_k$ and $F(k)=(1+Z_k)/2$. The condition
  $p_\epsilon(z)\ne\pm(1-z)^n$ makes $F$ nonconstant, and $p_\epsilon$ is the polynomial
  displayed above for this $Z$. Its order of vanishing at $z=1$ is therefore
  $\delta(F)\le\Delta(n)$. Apply \Cref{thm:one-third-degree-bound}.
\end{proof}

\subsection{Open problems}\label{sec:open-problems}
The principal open problem is to close the gap between the bound
$\Delta(n)=\widetilde{O}(n^{1/3})$ of \Cref{thm:one-third-degree-bound}
and the value $3$ predicted by \Cref{conj:deficit-at-most-three}.

Computation supports the conjecture. Our computer search verifies
$\Delta(n)\le3$ for every $n\le8192$, extending the range $n\le128$ of von zur
Gathen and Roche~\cite{GR97}. Let $(f_j)_{j\ge0}$ be the Fibonacci sequence with
$f_0=0$ and $f_1=1$. The search also shows that, for $n\le8192$, $\Delta(n)=3$
holds exactly at the sizes given by their two
constructions~\cite[Theorems 4.2 and 4.3]{GR97}: $n=4t^2-1$ for integers $t\ge2$,
and $n=f_{2i+2}f_{2i+3}+1$ for integers $i\ge2$ with $i\not\equiv1\pmod3$. We do
not know whether $\Delta(n)\ge3$ holds at any size outside these two families.
This computation is not part of the Lean formalization.
The search code and validation records are available in the
\href{https://github.com/wcgbg/sym-bool-func-deg}{same public repository}.

Closing the gap may require more of the Boolean structure than our proof
retains. For a profile $F$, put $Z_k=2F(k)-1$ for $0\le k\le n$. A large deficit
makes the alternating binomial moments of $Z$ vanish, by
\Cref{lem:degree-and-moments}, and our proof derives from them a linear
recurrence for $Z_k$ on an interval of indices around $n/2$. Beyond the
vanishing moments, the proof uses that $|Z_k|\le1$ at every index, that $Z$
is integral on that interval and at its first failure, and, when the
agreement is long, that $Z_k^2=1$ on that interval. It does not otherwise
use $Z_k^2=1$ outside that interval. A stronger argument could try to use
these constraints together.

\subsection{Organization}
\Cref{sec:proof-overview} outlines the proof informally.
\Cref{sec:preliminaries} fixes notation and introduces the binomial
transform, through which every polynomial test is estimated.
\Cref{sec:central-recurrence} finds the recurrence near the center, and
\Cref{sec:recurrence-arithmetic} collects its arithmetic: a short integer
annihilator, integer Hankel determinants, the radial drift of the roots,
and an angular count. \Cref{sec:prediction-and-filtering} predicts recurrences from short
observations and separates the roots of large drift by a rational
filter. \Cref{sec:circles-and-far-sector} proves the two analytic tools used
repeatedly: coefficient bounds from one circle and a contour bound for
the roots outside a sector. \Cref{sec:roots-near-minus-one} counts the roots near
$-1$, \Cref{sec:endpoint-test} constructs the endpoint test, and
\Cref{sec:proof-of-main-theorem} completes the proof.

\section{Overview of the proof}\label{sec:proof-overview}
This section explains the proof informally; nothing later depends on
it. Let $F$ be a nonconstant profile with deficit at least $r$, and let
$Z_k=2F(k)-1\in\{-1,1\}$, $0\le k\le n$, be its \emph{sign word}. Write
$\ell=\log(n+3)$. We derive a contradiction once
$r$ is a large multiple of $n^{1/3}\ell^{2/3}$.

The proof has four stages.
\begin{enumerate}
  \item \emph{Structure} (\Cref{sec:central-recurrence}). The vanishing
    moments force $Z$ to agree, on a long interval around $n/2$, with a
    rational linear recurrence $y$ of order $O(n^{2/3}\ell^{1/3})$.
  \item \emph{Arithmetic} (\Cref{sec:recurrence-arithmetic}). Integrality of $Z$ on the
    agreement interval controls the characteristic roots of $y$.
  \item \emph{Short agreement} (\Cref{sec:prediction-and-filtering,sec:circles-and-far-sector,sec:roots-near-minus-one,sec:endpoint-test}).
    We build a polynomial of degree below $r$ whose moment against $Z$ is
    dominated by the first index where $Z$ and $y$ differ. This requires
    that few roots of $y$ lie near $-1$, which \Cref{sec:roots-near-minus-one} proves.
  \item \emph{Long agreement} (\Cref{sec:proof-of-main-theorem}). Then $y$ is periodic,
    and a polynomial of bounded degree finishes the proof.
\end{enumerate}

\subsection{Tests and their transforms}
We call any polynomial $f$ of degree at most $n$ a \emph{test}, and
\[
  \sum_{k=0}^n(-1)^k\binom nk f(n-2k)\,Z_k
\]
its \emph{pairing} with $Z$. Evaluating $f$ at $n-2k$ rather than at $k$
centers the grid at zero. By \Cref{lem:degree-and-moments}, the deficit is
at least $r$ exactly when every test of degree below $r$ pairs to zero with
$Z$. To reach a contradiction we must produce one test of degree below $r$
whose pairing with $Z$ is nonzero. Since we know nothing about $Z$ in
advance, the proof uses the vanishing pairings twice: first to learn the
structure of $Z$, then to build a test that contradicts them.

Tests interact with exponential sequences in a simple way. Attach to a test
$f$ the weights $\gamma_k=(-1)^k\binom nkf(n-2k)$. Their generating
function satisfies
\[
  \sum_{k=0}^n\gamma_kz^k=(1-z)^n(\Bt f)\bigl(x(z)\bigr),
  \qquad x(z)=n\frac{1+z}{1-z},
\]
where $\Bt f$, the \emph{binomial transform} of $f$, is a polynomial of
the same degree as $f$; $(\Bt f)(x)$ is the mean of $f(n-2Y)$ when $Y$
is binomially distributed with mean $(n-x)/2$. Thus the pairing of $f$
with the sequence $\lambda^k$ is $(1-\lambda)^n(\Bt f)(x_\lambda)$, where
$x_\lambda=x(\lambda)$ is the \emph{coordinate} of $\lambda$. Two
features of this formula drive the proof. On the unit circle the weight is
$|1-\lambda|^n=(2\cos(\theta/2))^n$, where $\theta$ is the angle of
$\lambda$ from $-1$: it is largest, $2^n$, at $\lambda=-1$, and decreases
with $\theta$. And a test can ignore $\lambda^k$ completely: it suffices
that $\Bt f$ vanish at $x_\lambda$, to the multiplicity of $\lambda$ if the
sequence has polynomial factors. Each such zero costs one degree.

\subsection{A recurrence near the center}
Put $M\approx\sqrt{n/r}$ and $R=r/M$, so that $n\approx M^3R$; we shall
take $R=\Lambda\ell$ with a large constant $\Lambda$. Since every test of
degree below $r$ pairs to zero with $Z$, Taylor's theorem shows that the
exponential $f(x)=e^{zx/\sqrt n}$ pairs with $Z$ to within $2^ne^{-r}$ for
$|z|\le c\sqrt r$ (\Cref{lem:small-transform}). Complex $z$ turns the
alternating weights $(-1)^k\binom nk$ into
$\binom nkp^k(1-p)^{n-k}e^{i(\pi+\theta)k}$: a binomial distribution of
bias $|p-\frac12|\lesssim\sqrt{r/n}$, twisted by a frequency $\pi+\theta$
with $|\theta|\lesssim\sqrt{r/n}\approx1/M$. Averaging over $p$ makes the
weights equal to one on a central interval of length about
$\sqrt{nr}=M^2R$. The result is a polynomial $\Acal(z)$ whose coefficients
equal $Z_k$ on that interval, with $|\Acal|=O(n)$ on the unit circle and
$|\Acal|\le e^{-cr}$ on an arc of length about $1/M$ around $-1$.

Integer sequences do not tolerate such a spectral hole. By the integer
Hankel determinant method in Carlson's proof of the P\'olya--Carlson
theorem~\cite{Carlson21}, the Hankel determinants of $Z$ of size
$D\approx M^2\ell$ on that interval are smaller than one, hence zero. So
$Z$ follows a linear recurrence with rational coefficients of order
$d=O(M^2\ell)$ on a centered interval of length $u_0\approx M^2R$
(\Cref{sec:central-recurrence}). This step needs the depth $e^{-cr}$ of
the hole to beat the size $e^{O(D/M)}$ of the auxiliary polynomials on the
hole, that is, $R\ge C\ell$. This is where the logarithm enters our
estimates.

Let $(y_k)_{k\in\Z}$ be the two-sided sequence that satisfies the
recurrence and equals $Z_k$ on that interval. After reflecting
$k\mapsto n-k$ if needed, either $y=Z$ on $0,\ldots,n$, or there is a
\emph{first defect} $t<n/2$: $y=Z$ on $I=\{t+1,\ldots,n-t-1\}$ and
$y_t\ne Z_t$. Put $u=n-2t$, so that $I$ has $u-1$ elements. We call the
values $Z_k-y_k$, $0\le k\le n$, the \emph{defects}; they vanish on $I$.
Write each characteristic root as $\lambda=-e^{\rho+i\theta}$ with
$|\theta|\le\pi$; we call $|\theta|$ its \emph{angle} and $|\rho|$ its
\emph{drift}.

\subsection{Arithmetic}
The recurrence is rational, not integral, but the integer values of $Z$
on $I$ supply four facts (\Cref{sec:recurrence-arithmetic}). By pigeonhole, $y$
satisfies a short integer recurrence: there is $A(z)=\sum_iA_iz^i$ with
integer coefficients, of degree below $2d$, with $\sum_i|A_i|\le8d^2$ and
$A_0\ne0$, such that $\sum_iA_iy_{k+i}=0$ for all $k$. At the first defect,
$A_0(Z_t-y_t)=\sum_iA_iZ_{t+i}$ is then a nonzero integer. Hankel matrices
of $Z$ on $I$ are nonsingular integer matrices, so the part of $y$ that
comes from a set of roots cannot be uniformly tiny on $2d-1$ consecutive
indices unless the set is small. A Mahler-measure estimate bounds the
total drift of the roots by $O(M^2\ell^2/u)$. Jensen's formula for $A$
bounds the number of roots within angle $s$ of any point by
$A_\theta s+B_\theta$, where $A_\theta=O(M^2\ell)$ and $B_\theta=O(M\ell)$.

\subsection{The endpoint test}
Suppose $u\le n/128$; longer agreement is treated at the end. Since
$Z=y$ on $I$, the vanishing pairing of a test $f$ splits as
\[
  0=\sum_k\gamma_ky_k+\sum_{k\notin I}\gamma_k(Z_k-y_k).
\]
We want the defect at $t$ to dominate the second sum, and the first sum
to be smaller still. The weight of the index $t$ is $\binom nt$. Define
the \emph{defect angle} $\alpha_t$ by $(2\cos(\alpha_t/2))^n=\binom nt$;
it is close to $2u/n$. A root at angle below $\alpha_t$ carries more
weight than the defect, so the test must cancel it exactly by a zero of
$\Bt f$. A root at a larger angle carries less, and its contribution is
bounded by a contour integral of the rational generating function of
$y$ (\Cref{lem:far-sector}). Thus the test needs one exact zero for each
root of angle up to slightly more than $\alpha_t$, and these must number
at most the available degree $m\approx r/\sqrt\Lambda$.

Roots of large drift are not cancelled, and they would spoil the contour
estimate. A rational filter moves them radially outward
(\Cref{sec:prediction-and-filtering}). The filtered word $X$ still has vanishing pairings
with every test whose transform vanishes at the coordinates of the moved
poles. On $I$ it splits into a recurrence $v$ with roots of small drift and
outward modes that decay away from the right end of $I$.

The test itself (\Cref{sec:endpoint-test}) is built in two stages. First, an
explicit polynomial $\psi$ of degree $m$ gives weights $\gamma_k$ that
equal $(-1)^t\binom nt$ at $k=t$ and are tiny, relative to $\binom nt$, to
the left of $t$ (\Cref{lem:geometric-tails}). Second, the unknown $y$ must
be removed to the left of $t$, where it may be large. Convolving $\gamma$
with the integer annihilator $A$ does this: in the pairing of the convolved
weights with $Z$, the values at $k<t$ enter only through the bounded
integers $\sum_iA_iZ_{k+i}$, and the term at $t$ becomes
$(-1)^t\binom ntA_0(Z_t-y_t)$, a nonzero integer multiple of $\binom nt$.
But convolution with $A$ multiplies the generating function by $A(z)$, that
is, the transform by the rational function $A(\lambda(x))$, where
$\lambda(x)=(x-n)/(x+n)$ inverts $x(z)$. So the convolved weights do not
come from a test of degree below $r$. We therefore replace $A(\lambda(x))$
by a polynomial with the following properties:
\begin{itemize}
  \item it matches $A(\lambda(x))$ to high order at $x=u$, the coordinate of
    $-q_t$ with $q_t=t/(n-t)$; the circle $|z|=q_t$ controls the weights near
    $t$ by Cauchy's estimate;
  \item it vanishes exactly at the coordinates of the roots of angle below
    $\alpha_s\approx\alpha_t(1+1/(10w))$, where $w=u/u_0$, and at the moved
    poles of the filter;
  \item it vanishes to high order at $x=-u$, which silences the weights near
    $n-t$.
\end{itemize}
Three errors remain: the Taylor error near $t$; the pairing with the roots
that are not cancelled, bounded by the contour estimate; and the weights
near $n-t$. Each is $o(\binom nt)$ (\Cref{thm:endpoint-comparison}), which contradicts
the vanishing pairing.

\subsection{Counting the roots near \texorpdfstring{$-1$}{-1}}
It remains to show that few roots lie near $-1$
(\Cref{sec:roots-near-minus-one}). Suppose many roots have angle below some
$\alpha_2\le\alpha_t$. Choose a slightly smaller angle $\theta_\Gamma$
and a contour that separates the roots of angle below $\theta_\Gamma$,
the \emph{inner} roots, from the others. For each index $\bar n+j$ near the
center ($\bar n=\lfloor n/2\rfloor$, $|j|<d$), Hermite interpolation produces a
test of degree below $r$ whose generating function reproduces $\nu^{\bar n+j}$
at the inner roots and vanishes at the roots with angle between
$\theta_\Gamma$ and a ray angle $\alpha$ slightly below $\alpha_2$. Its
pairing with the remaining roots is controlled by the contour bound,
because their weights are at most $(2\cos(\alpha/2))^n$, smaller than the
weight $(2\cos(\theta_\Gamma/2))^n$ of the inner roots. Its pairing with the
defects is small for the same reason. So the part of $y$ coming from the
inner roots is exponentially small at the center, and the Hankel
determinant fact bounds the number of inner roots. A quantile argument
turns this into a bound $O(M\ell)$ on all roots of angle below $\alpha_t$
(\Cref{lem:roots-near-minus-one}). Since $m\approx\sqrt\Lambda\,M\ell$, the
endpoint test can afford to cancel them.

\subsection{Long agreement}
If $u>n/128$, the recurrence agrees with signs on a block much longer than
its order. A Mahler-measure estimate and Voutier's height bound then make
all roots roots of unity. The root count, applied at a fixed angle, leaves
no root within angle $1/384$ of $-1$, so every root has bounded order. The
word is then periodic on $I$ with a bounded period. If $y=Z$ on all of
$0,\ldots,n$, a Lagrange test isolates each nonconstant frequency of the
periodic $y$, so $Z$ would be constant. Otherwise the difference of $Z$
and its shift by one period vanishes on $I$ and has bounded integer values
at the ends, and an explicit test of bounded degree isolates its first
nonzero value (\Cref{sec:proof-of-main-theorem}).

\subsection{Parameters}
The scales are linked by
\[
  r=\Lambda\ell M,\qquad n\approx\Lambda\ell M^3,
  \qquad\text{hence}\qquad r\approx\Lambda^{2/3}n^{1/3}\ell^{2/3}.
\]
The logarithms of the a priori bounds, which we call heights, are
$O(M\ell)$, and the test degree is $m\approx\sqrt\Lambda M\ell=r/\sqrt\Lambda$.
In the endpoint comparison the size of the test costs
$O(m(\log\Lambda+\log w))$, while the contour gain is at least $cwr$.
The principal ratios are therefore
\[
  \begin{array}{lcc}
    \text{Comparison} & \text{Ratio} & \text{Uniform bound}\\[3pt]
    \text{Height / test degree} & M\ell/m & O(\Lambda^{-1/2})\\[3pt]
    \text{Test degree / deficit} & m/r & O(\Lambda^{-1/2})\\[3pt]
    \text{Test growth / contour gain}
    & \dfrac{m(\log\Lambda+\log w)}{wr}
    & O\!\left(\dfrac{\log\Lambda}{\sqrt\Lambda}\right)
  \end{array}
\]
For the last row, $m/r\approx\Lambda^{-1/2}$ gives a ratio of order
$(\log\Lambda+\log w)/(w\sqrt\Lambda)$; use $1/w\le1$ and
$(\log w)/w\le1$ for $w\ge1$. A large absolute $\Lambda$ thus controls these ratios
even when the agreement length grows with $n$.
\Cref{thm:endpoint-comparison} also bounds the Taylor error and the opposite
end. The exponent $1/3$ comes from the
first step, which needs $r=RM$ with $R\ge C\ell$ and $M\approx\sqrt{n/r}$,
that is, $r^{3/2}\ge C\ell\sqrt n$. The root count leads to the same scale:
$O(M\ell)$ roots near $-1$ must be cancelled with $r/\sqrt\Lambda$
degrees.

\section{Preliminaries}\label{sec:preliminaries}

\subsection{Conventions}
Throughout, $n$ is a positive integer.
All logarithms are natural, and $\ell=\log(n+3)$. Constants denoted by
$c,C$, with or without subscripts, are positive and absolute; their
values may change between occurrences unless they carry a subscript.
We write $\T=\{z\in\C:|z|=1\}$. For a polynomial $g(z)=\sum_jg_jz^j$ we
write $[z^j]g=g_j$, $\|g\|_1=\sum_j|g_j|$ and
$\|g\|_2=(\sum_j|g_j|^2)^{1/2}$; the same notation is used for power
series. We write $k^{\underline j}=k(k-1)\cdots(k-j+1)$, with
$k^{\underline0}=1$, $\bar n=\lfloor n/2\rfloor$, $(x)_+=\max(x,0)$ for real
$x$, and $\mathbf1_{a=b}$ for the number that is $1$ if $a=b$ and $0$
otherwise. The Chebyshev polynomial of degree $m$ of the first kind is
$T_m$, with $T_m(\cos\theta)=\cos(m\theta)$. All intervals of
indices are intervals of integers. A block $a,\ldots,a+B-1$ of
integers, with $B\ge1$, is \emph{centered} in $0,\ldots,n$ when
$2a+B-1=n$. For a two-sided sequence $y=(y_k)_{k\in\Z}$ the shift is
$(\shift y)_k=y_{k+1}$, and for a polynomial $g(z)=\sum_jg_jz^j$ we put
$(g(\shift)y)_k=\sum_jg_jy_{k+j}$.

The following symbols keep one meaning throughout.
\begin{center}
  \small
  \begin{tabular}{lll}
    Symbol & Meaning & Defined in\\\hline
    $\ell$, $\bar n$, $\T$ & $\log(n+3)$, $\lfloor n/2\rfloor$, unit circle & \Cref{sec:preliminaries}\\
    $Z$ & sign word $Z_k=2F(k)-1$ & \eqref{eq:vanishing-moments}\\
    $\langle f,a\rangle$ & pairing of a test with a sequence & \eqref{eq:test-pairing}\\
    $\Bt f$, $x_\lambda$ & binomial transform, coordinate & \eqref{eq:binomial-transform-definition}, \eqref{eq:moebius-maps}\\
    $\alpha_t$ & defect angle & \Cref{def:defect-angle}\\
    $r$, $M$, $R$, $\Lambda$ & deficit and scales & \eqref{eq:deficit-scale}, \eqref{eq:scale-parameters}\\
    $d$, $P$, $y$, $u_0$ & central recurrence & \Cref{thm:central-recurrence}\\
    $t$, $u$, $I$, $\mathfrak e$ & agreement interval and defect & \Cref{sec:agreement-interval}\\
    $w$, $S$, $\kappa$ & agreement parameters & \eqref{eq:agreement-parameters}\\
    $A$, $d_A$, $H_A$ & short integer annihilator & \eqref{eq:short-annihilator}\\
    $T$ & total drift & \eqref{eq:total-drift}\\
    $A_\theta$, $B_\theta$ & angular-count constants & \Cref{prop:standing-bounds}\\
    $\sigma$, $K$, $X$ & filter scale, removed roots, filtered word & \eqref{eq:filter-scale}, \Cref{def:rational-filter}\\
    $v$, $E$, $H_0$ & retained recurrence, filtered defects, height & \Cref{lem:filter-properties}, \Cref{prop:filtered-defect-heights}\\
    $m$, $\mathcal L$ & endpoint test degree and logarithm & \eqref{eq:endpoint-parameters}
  \end{tabular}
\end{center}

\subsection{Degree, moments and tests}
The following collects standard degree facts. The degree identification
appears in von zur Gathen--Roche~\cite[p.~346 and Theorem~2.8]{GR97};
the moment formulation is the finite-difference characterization in
\cite[Proposition 2.1 and Theorem 2.2]{GR97}, written on a fixed grid.
One direction of its formulation by a zero at one is
Erd\'elyi's~\cite[Lemma 4.2]{Erdelyi15}.

\begin{lemma}[Degree and moments]\label{lem:degree-and-moments}
  For a Boolean profile $F$ on $0,\ldots,n$,
  $\deg_\R(F(|x|))=\deg q_F$.
  For an integer $N\ge0$ and real data $G$ on $0,\ldots,N$, let $q_G$ be the
  polynomial of degree at most $N$ interpolating $G$, and define, for integers
  $j\ge0$,
  \[
    m_j(G)=\sum_{x=0}^N(-1)^x\binom Nx x^jG(x).
  \]
  For an integer $0\le r\le N+1$, the conditions $m_j(G)=0$ for $0\le j<r$
  are equivalent to $\deg q_G<N+1-r$, with $\deg0=-\infty$.
\end{lemma}
\begin{proof}
  The degree identification is the fact cited above. For the fixed-grid
  moment formulation, the alternating sum
  $\sum_{x=0}^N(-1)^x\binom NxP(x)$ of a polynomial $P$ is
  $(-1)^N\Delta_{\rm fwd}^N P(0)$, where
  $\Delta_{\rm fwd}g(x)=g(x+1)-g(x)$. It vanishes for
  degree less than $N$ and equals $(-1)^NN!$ times the leading coefficient
  for degree $N$. If $q_G\ne0$, apply this to $P(X)=X^jq_G(X)$: the first
  nonzero moment is at $j=N-\deg q_G$.
\end{proof}

A \emph{test} is a complex polynomial $f$ of degree at most $n$. Its
\emph{pairing} with a complex sequence $a=(a_k)_{0\le k\le n}$, or with
the restriction of a two-sided sequence to $0,\ldots,n$, is
\begin{equation}\label{eq:test-pairing}
  \langle f,a\rangle=\sum_{k=0}^n(-1)^k\binom nkf(n-2k)\,a_k .
\end{equation}
Since $k\mapsto f(n-2k)$ is a polynomial of the same degree as $f$,
\Cref{lem:degree-and-moments} says: if a profile $F$ has deficit at least
$r$, then $\langle f,F\rangle=0$ for every test $f$ with $\deg f<r$. Since
$F$ is nonconstant, $\deg q_F\ge1$, so $r\le n-1$; and
$\langle f,1\rangle=0$ whenever $\deg f<n$. Hence the \emph{sign word}
$Z_k=2F(k)-1$ satisfies
\begin{equation}\label{eq:vanishing-moments}
  \langle f,Z\rangle=0\qquad\text{for every test $f$ with $\deg f<r$.}
\end{equation}

\subsection{Recurrences}
A two-sided complex sequence $y$ \emph{satisfies} a polynomial $P$ if
$P(\shift)y=0$. If $y$ satisfies a nonzero polynomial, its \emph{minimal
polynomial} is the monic polynomial that $y$ satisfies and that divides every
polynomial $y$ satisfies; since these polynomials form an ideal, it is the
monic polynomial of least degree that $y$ satisfies. Since $\shift$ is
invertible on two-sided sequences, the minimal polynomial does not vanish at
zero. On a finite word
$x_0,\ldots,x_{N-1}$, a \emph{monic recurrence} of order $d$ is a relation
$x_{k+d}+p_{d-1}x_{k+d-1}+\cdots+p_0x_k=0$ for $0\le k<N-d$; its
characteristic polynomial is $z^d+p_{d-1}z^{d-1}+\cdots+p_0$, and it is
\emph{regular} if $p_0\ne0$. If $P$ is
monic of degree $d$ and $P(0)\ne0$, every $d$ consecutive values
determine a unique two-sided solution. The solutions are the sequences
\begin{equation}\label{eq:recurrence-solutions}
  y_k=\sum_\lambda p_\lambda(k)\lambda^k\qquad(k\in\Z),
\end{equation}
where $\lambda$ runs over the distinct roots of $P$ and $p_\lambda$ is a
polynomial of degree below the multiplicity $m_\lambda$ of $\lambda$; the
minimal polynomial is $P$ exactly when every $p_\lambda$ has degree
$m_\lambda-1$. For a set $\mathcal Z$ of distinct roots, the
\emph{component} of $y$ on $\mathcal Z$ is
$y_{\mathcal Z}=\sum_{\lambda\in\mathcal Z}p_\lambda(k)\lambda^k$; equivalently, it is the
unique sequence $y_{\mathcal Z}$ that satisfies
$\prod_{\lambda\in\mathcal Z}(z-\lambda)^{m_\lambda}$ and for which $y-y_{\mathcal Z}$
satisfies $\prod_{\lambda\notin\mathcal Z}(z-\lambda)^{m_\lambda}$. A component of a
solution of $P$ again satisfies $P$, and also satisfies every polynomial
that $y$ satisfies.

\begin{lemma}[Hankel matrices]\label{lem:nonsingular-hankel}
  Let $y$ be a two-sided sequence with minimal polynomial $P$ of degree
  $d\ge1$. Every Hankel matrix $(y_{k+i+j})_{0\le i,j<d}$, $k\in\Z$, is
  nonsingular. Let $\mathcal Z$ be a set of roots of $P$ of total
  multiplicity $k_0$, and let $y'$ be the component of $y$ on the other
  roots. Then every matrix $(y'_{k+i+j})_{0\le i,j<d}$ has rank at most
  $d-k_0$.
\end{lemma}
\begin{proof}
  Suppose $\sum_{j<d}g_jy_{k+i+j}=0$ for $0\le i<d$, with
  $g(z)=\sum_{j<d}g_jz^j\ne0$. The sequence $g(\shift)y$ satisfies $P$
  and vanishes at $d$ consecutive indices, so it is zero. Then $y$
  satisfies $g$, of degree below $d$, contradicting minimality. For the
  second claim, $y'$ satisfies
  $P/\prod_{\lambda\in\mathcal Z}(z-\lambda)^{m_\lambda}$, of degree
  $d-k_0$ and nonzero at zero. Each column $(y'_{k+j+i})_{0\le i<d}$ is
  therefore determined linearly by its first $d-k_0$ entries, so the
  columns lie in a space of dimension at most $d-k_0$.
\end{proof}

\subsection{The binomial transform}
The \emph{binomial transform} of a test $f$ is the polynomial
\begin{equation}\label{eq:binomial-transform-definition}
  (\Bt f)(x)=\sum_{k=0}^n\binom nk
  \Bigl(\frac{n-x}{2n}\Bigr)^k\Bigl(\frac{n+x}{2n}\Bigr)^{n-k}f(n-2k).
\end{equation}
For real $|x|\le n$ it is the mean of $f(n-2Y)$, where $Y$ is binomially
distributed with parameters $n$ and $(n-x)/(2n)$; the mean of $n-2Y$ is
$x$. Define the inverse Möbius maps
\begin{equation}\label{eq:moebius-maps}
  x(z)=n\,\frac{1+z}{1-z},\qquad \lambda(x)=\frac{x-n}{x+n},
\end{equation}
and call $x_\lambda=x(\lambda)$ the \emph{coordinate} of a complex number
$\lambda\ne1$. The \emph{generating function} of a test is
\begin{equation}\label{eq:test-generating-function}
  \Sg_f(z)=\sum_{k=0}^n(-1)^k\binom nkf(n-2k)\,z^k .
\end{equation}

\begin{lemma}[Transforms]\label{lem:binomial-transform}
  Let $f$ be a test.
  \begin{enumerate}
    \item[(a)] $\deg\Bt f=\deg f$, and $\Bt$ is a bijection of the space of
      polynomials of degree at most $n$ onto itself.
    \item[(b)] $\Sg_f(z)=(1-z)^n(\Bt f)(x(z))$ for every $z\ne1$.
    \item[(c)] If $a_k=\sum_\lambda p_\lambda(k)\lambda^k$ with polynomials
      $p_\lambda$, then, with $\vartheta=\nu\,\partial_\nu$,
      \begin{equation}\label{eq:spectral-pairing}
        \langle f,a\rangle=\sum_\lambda
        \bigl[p_\lambda(\vartheta)\Sg_f\bigr]_{\nu=\lambda}.
      \end{equation}
      The term of $\lambda\ne1$ is zero when $\Bt f$ vanishes at
      $x_\lambda$ to order greater than $\deg p_\lambda$. The term of
      $\lambda=1$ is zero when $\deg f<n-\deg p_1$.
    \item[(d)] For complex numbers $\lambda\notin\{0,1\}$ and $x\ne-n$,
      \begin{equation}\label{eq:moebius-factor}
        1-\frac{\lambda(x)}\lambda=\frac{2n}{x+n}\cdot
        \frac{x-x_\lambda}{n-x_\lambda},\qquad 1-\lambda(x)=\frac{2n}{x+n}.
      \end{equation}
      Consequently, if $\Df(z)=\prod_{i=1}^K(1-z/\mu_i)$ with
      $\mu_i\notin\{0,1\}$, and
      $\Df^\sharp(x)=\prod_{i=1}^K(x-x_{\mu_i})/(n-x_{\mu_i})$, then
      $\Df(z)=(1-z)^K\Df^\sharp(x(z))$ for $z\ne1$.
  \end{enumerate}
\end{lemma}
\begin{proof}
  (a) Write $g(k)=f(n-2k)=\sum_ib_ik^{\underline i}$, and let $Y$ be
  binomial with parameters $n$ and $p=(n-x)/(2n)$. Since
  $\E Y^{\underline i}=n^{\underline i}p^i$, we get
  $(\Bt f)(x)=\sum_ib_in^{\underline i}p^i$. If $f$ has degree $j$ and
  leading coefficient $a$, then $b_j=(-2)^ja$, and the coefficient of $x^j$
  is $a\,n^{\underline j}/n^j\ne0$ for $j\le n$. So $\Bt$ is triangular
  with nonzero diagonal.

  (b) For $z\ne1$ we have $(n-x(z))/(2n)=-z/(1-z)$ and
  $(n+x(z))/(2n)=1/(1-z)$. Substitute these in \eqref{eq:binomial-transform-definition}.

  (c) The identity follows from $\vartheta^j\nu^k=k^j\nu^k$. For
  $\lambda\ne1$ the map $\nu\mapsto x(\nu)$ has nonzero derivative at
  $\lambda$, so by (b) $\Sg_f$ vanishes at $\lambda$ to the same order
  as $\Bt f$ at $x_\lambda$, and $p_\lambda(\vartheta)$ is a
  differential operator of order $\deg p_\lambda$. Finally,
  $\Sg_f(\nu)$ is $(1-\nu)^n$ times a polynomial in $(1+\nu)/(1-\nu)$ of
  degree $\deg f$, hence divisible by $(1-\nu)^{n-\deg f}$.

  (d) Both identities follow by clearing denominators:
  $(x+n)(x_\lambda-n)-(x-n)(x_\lambda+n)=2n(x_\lambda-x)$. For the
  consequence, put $x=x(z)$, so that $\lambda(x)=z$ and $1-z=2n/(x+n)$.
\end{proof}

In particular, a test pairs to zero with a solution of a recurrence
whenever its transform vanishes at the coordinate of every root
$\lambda\ne1$, to the multiplicity of the root, and, if $1$ is a root, of
multiplicity $m_1$, its degree is below $n-m_1+1$; indeed
$\deg p_\lambda\le m_\lambda-1$ for every root.

\subsection{Angles, drift and binomial weights}
Write a nonzero complex number as $\lambda=-e^\zeta$ with
$\zeta=\rho+i\theta$, $\rho\in\R$ and $|\theta|\le\pi$. We call $|\theta|$
the \emph{angle} of $\lambda$ (its angular distance from $-1$) and
$|\rho|=|\log|\lambda||$ its \emph{drift}. The coordinate of $\lambda$ is
\begin{equation}\label{eq:coordinate-tanh}
  x_\lambda=n\frac{1-e^\zeta}{1+e^\zeta}=-n\tanh(\zeta/2).
\end{equation}

\begin{lemma}[Coordinates near $-1$]\label{lem:coordinates-near-minus-one}
  Let $\zeta=\rho+i\theta$, with $\rho\in\R$ and $|\theta|\le\pi$, and
  put $\lambda=-e^\zeta$. Then the following hold.
  \begin{enumerate}
    \item[(a)] If $|\zeta_1|,|\zeta_2|\le1/5$, then
      $0.48n|\zeta_1-\zeta_2|\le|n\tanh(\zeta_1/2)-n\tanh(\zeta_2/2)|
      \le0.52n|\zeta_1-\zeta_2|$.
    \item[(b)] $|\tanh w|\le|w|(1+|w|^2)$ for $|w|\le1/8$, and
      $|\tanh(\zeta/2)|^2=(\cosh\rho-\cos\theta)/(\cosh\rho+\cos\theta)$
      whenever $\lambda\ne1$.
    \item[(c)] Put $s(\zeta)^2=2(\cosh\rho+\cos\theta)$. Then
      $|1-\lambda|^2=e^\rho s(\zeta)^2$, and, if $\lambda\ne1$, for every integer $j$,
      \begin{equation}\label{eq:binomial-weight}
        |\lambda|^{\bar n+j}|1-\lambda|^{-n}=e^{\rho(\bar n+j-n/2)}s(\zeta)^{-n}.
      \end{equation}
  \end{enumerate}
\end{lemma}
\begin{proof}
  (a) On $|w|\le1/10$ the derivative of $\tanh$ is
  $1-\tanh^2w$, and $|\tanh w|\le0.101$ by (b). So the
  derivative of $\zeta\mapsto n\tanh(\zeta/2)$ lies in the disk of center
  $n/2$ and radius $0.0052n$. Integrate along the segment.
  (b) Comparing power series, for $|w|\le1$ we have
  $|\sinh w-w|\le\frac29|w|^3$ and $|\cosh w-1|\le\frac34|w|^2$. So for
  $|w|\le1/8$,
  $|\tanh w|\le(|w|+\frac29|w|^3)/(1-\frac34|w|^2)\le|w|(1+|w|^2)$.
  The second identity is elementary.
  (c) $|1+e^\zeta|^2=1+2e^\rho\cos\theta+e^{2\rho}=e^\rho s(\zeta)^2$, and
  $|\lambda|^{\bar n+j}=e^{\rho(\bar n+j)}$.
\end{proof}

The quantity $s(\zeta)^n$ is the size of the binomial weights at
$\lambda$: by \eqref{eq:binomial-weight} with $\rho=0$,
$|1-\lambda|^n=(2\cos(\theta/2))^n$, which equals $2^n$ at $\lambda=-1$
and decreases with the angle.

\begin{definition}[Defect angle]\label{def:defect-angle}
  For an integer $0\le t<n/2$, let $\alpha_t\in(0,\pi)$ be defined by
  $(2\cos(\alpha_t/2))^n=\binom nt$.
\end{definition}

\begin{lemma}[Angles and weights]\label{lem:angles-and-weights}
  Let $\zeta=\rho+i\theta$, with $\rho\in\R$ and $|\theta|\le\pi$, put
  $\lambda=-e^\zeta$, and set $s(\zeta)=\sqrt{2(\cosh\rho+\cos\theta)}$.
  Then the following hold.
  \begin{enumerate}
    \item[(a)] If $0\le\theta'<\alpha<\pi$, then
      $n\log\bigl(\cos(\theta'/2)/\cos(\alpha/2)\bigr)\ge\frac n8(\alpha^2-\theta'^2)$.
    \item[(b)] If $|\rho|\le\tau\le1$, $0\le\alpha\le\pi/2$ and
      $\alpha\le|\theta|\le\pi$, then
      $s(\rho+i\theta)^n\le(2\cos(\alpha/2))^ne^{n\tau^2/2}$.
    \item[(c)] If $|\theta|\le\theta'<\pi$, then
      $s(\rho+i\theta)^{-n}\le(2\cos(\theta'/2))^{-n}$.
    \item[(d)] If $n\ge2$, $0\le t<n/2$ and $u=n-2t\le n/128$, then
      \[
        1.99\frac un\le\alpha_t\le2.017\frac un+\sqrt{\frac{8\log(n+1)}n}.
      \]
  \end{enumerate}
\end{lemma}
\begin{proof}
  (a) The left side is $n\int_{\theta'/2}^{\alpha/2}\tan x\,dx$, and
  $\tan x\ge x$.
  (b) We have $s^2\le2(\cosh\tau+\cos\alpha)\le4\cos^2(\alpha/2)+2\tau^2$,
  using $\cosh\tau\le1+\tau^2$ for $\tau\le1$, and
  $\cos^2(\alpha/2)\ge1/2$. Hence
  $s^n\le(2\cos(\alpha/2))^n(1+\tau^2)^{n/2}$.
  (c) Use $\cosh\rho\ge1$ and $\cos\theta\ge\cos\theta'$.

  (d) Hoeffding's inequality~\cite[Theorem 2]{Hoeffding63}, in the
  binomial form
  \begin{equation}\label{eq:hoeffding-binomial-bound}
    \binom nk\le2^ne^{-(n-2k)^2/(2n)}\qquad(0\le k\le n),
  \end{equation}
  gives $\cos(\alpha_t/2)\le e^{-v}$ with $v=u^2/(2n^2)$. With
  $\cos x\ge1-x^2/2$ this gives $\alpha_t^2/8\ge1-e^{-v}\ge v-v^2/2$,
  hence $\alpha_t\ge(2u/n)(1-u^2/(4n^2))^{1/2}\ge1.99u/n$.

  For the upper bound, put $s=\bar n-t$ and $b=n-2\bar n$, so that
  $u=2s+b$. The ratio $\binom n{\bar n}/\binom nt$ is the product over
  $0\le j<s$ of $(\bar n+b+j+1)/(\bar n-j)=1+(2j+1+b)/(\bar n-j)$, and each
  factor is at most $\exp((2j+1+b)/(t+1))$. Hence
  \[
    \log\frac{\binom n{\bar n}}{\binom nt}\le\frac{s(s+b)}{t+1}
    \le\frac{u^2}{2(n-u)}\le\frac{u^2}{2n}\Bigl(1+\frac{2u}n\Bigr).
  \]
  Since $\binom n{\bar n}\ge2^n/(n+1)$, we get
  $\cos(\alpha_t/2)\ge(n+1)^{-1/n}\exp(-u^2(1+2u/n)/(2n^2))$. Now
  $\log\cos x\le-x^2/2$ gives
  $\alpha_t^2\le(4u^2/n^2)(1+2u/n)+8\log(n+1)/n$, and
  $(1+1/64)^{1/2}<1.0085$.
\end{proof}

\subsection{Classical tools}
We use the following facts from complex analysis.
\begin{itemize}
  \item \emph{Cauchy's estimate.} If $G$ is holomorphic on a neighborhood
    of $|z|\le\rho$, then $|[z^k]G|\le\rho^{-k}\max_{|z|=\rho}|G|$; for a
    Laurent polynomial the same holds on any circle $|z|=\rho>0$.
  \item \emph{Bernstein--Walsh lemma for a disk}~\cite[Lemma 3.7]{Saff10}.
    A polynomial $p$ of degree at most $D$ with $|p|\le\Gamma$ on
    $|z-z_0|\le\rho$ satisfies $|p(z)|\le\Gamma(|z-z_0|/\rho)^D$ for
    $|z-z_0|\ge\rho$. (Apply the maximum principle to
    $p(z)\rho^D/(z-z_0)^D$ on the exterior of the disk.)
  \item \emph{Hermite's interpolation formula}; compare~\cite[(4.2) and the
    remark after it]{Saff10}, which treats simple zeros. If $\Phi$ is
    holomorphic on and inside a contour
    $\Gamma$, and $\Omega$ is a monic polynomial of degree $k$ with all
    zeros inside $\Gamma$, then
    \begin{equation}\label{eq:hermite-formula}
      \varpi(x)=\frac1{2\pi i}\oint_\Gamma\Phi(\xi)
      \frac{\Omega(\xi)-\Omega(x)}{\Omega(\xi)(\xi-x)}\,d\xi
    \end{equation}
    is the polynomial of degree below $k$ such that $\Phi-\varpi$ vanishes at
    each zero of $\Omega$ to its multiplicity. Indeed,
    $(\Omega(\xi)-\Omega(x))/(\xi-x)$ is a polynomial in $x$ of degree below
    $k$, and for $x$ inside $\Gamma$ Cauchy's formula gives
    $\Phi(x)-\varpi(x)=\Omega(x)\cdot\frac1{2\pi i}\oint_\Gamma
    \frac{\Phi(\xi)\,d\xi}{\Omega(\xi)(\xi-x)}$, where the integral is
    holomorphic in $x$.
  \item \emph{Jensen's formula.} If $p$ is holomorphic on a neighborhood
    of $|z|\le1$, then $\log|p(0)|\le\int_0^{2\pi}\log|p(e^{i\theta})|\,
    d\theta/(2\pi)$.
  \item \emph{Counting zeros.} If $p$ is holomorphic on a neighborhood of
    $|z-z_0|\le\rho$ and has $k$ zeros $a_1,\ldots,a_k$ in
    $|z-z_0|\le\rho/2$, listed with multiplicity, then
    $2^k|p(z_0)|\le\max_{|z-z_0|=\rho}|p|$. Indeed, the function
    $p(z)\prod_i\bigl(\rho^2-\overline{(a_i-z_0)}(z-z_0)\bigr)/
    \bigl(\rho(z-a_i)\bigr)$ is holomorphic on the disk and has the modulus
    of $p$ on its boundary, and its modulus at $z_0$ is
    $|p(z_0)|\prod_i\rho/|a_i-z_0|\ge2^k|p(z_0)|$. Apply the maximum
    principle.
  \item \emph{Carath\'eodory's inequality.} If $\Phi$ is holomorphic on
    $|z|<1$ and $\operatorname{Re}\Phi\ge0$ there, then
    $|[z^j]\Phi|\le2\operatorname{Re}\Phi(0)$ for $j\ge1$. Indeed, for
    $0<\varrho<1$ and $j\ge1$,
    $[z^j]\Phi\cdot\varrho^j=\pi^{-1}\int_0^{2\pi}\operatorname{Re}
    \Phi(\varrho e^{i\theta})e^{-ij\theta}\,d\theta$; bound the integrand by
    $\operatorname{Re}\Phi$ and let $\varrho\to1$.
\end{itemize}

\section{The central recurrence}\label{sec:central-recurrence}

\subsection{Standing assumptions}\label{sec:standing-assumptions}
From now until the end of \Cref{sec:proof-of-main-theorem} we fix a real number
$\Lambda_*\ge1$, an integer $n\ge1$, and a nonconstant profile $F$ on
$0,\ldots,n$ with deficit at least
\begin{equation}\label{eq:deficit-scale}
  r=\bigl\lceil\Lambda_*^{2/3}n^{1/3}\ell^{2/3}\bigr\rceil .
\end{equation}
Let $Z_k=2F(k)-1$ be its sign word, so that \eqref{eq:vanishing-moments}
holds, and put
\begin{equation}\label{eq:scale-parameters}
  M=\bigl\lfloor\sqrt{n/r}\bigr\rfloor,\qquad R=\frac rM,\qquad
  \Lambda=\frac R\ell .
\end{equation}
Then $r=\Lambda\ell M$. If $n$ is large in terms of $\Lambda_*$, then
$M\ge2$, and
\begin{equation}\label{eq:scale-comparison}
  M^3R\le n\le4M^3R,\qquad \Lambda_*\le\Lambda\le6\Lambda_* .
\end{equation}
Indeed $M^2r\le n<(M+1)^2r\le4M^2r$. Next
$R\ge r^{3/2}n^{-1/2}\ge\Lambda_*\ell$, and $M\ge\frac12\sqrt{n/r}$ together
with $r\le2\Lambda_*^{2/3}n^{1/3}\ell^{2/3}$ gives
$R\le2r^{3/2}n^{-1/2}\le2^{5/2}\Lambda_*\ell$.

\emph{Order of the constants.} Every requirement below has one of two
forms: $\Lambda\ge C$ for an absolute constant $C$, or $n\ge n_0$ for a
threshold $n_0$ depending on $\Lambda_*$ only. We say that a statement
holds \emph{for large $\Lambda$} if it holds under finitely many
requirements of these forms. The absolute constants of all lemmas are
fixed first; then $\Lambda_*$ is chosen large, which makes $\Lambda$ large
by \eqref{eq:scale-comparison}; then $n$ is chosen large in terms of
$\Lambda_*$. No other parameter is chosen.

\subsection{The theorem}
The first step is a quantitative form of the integer Hankel determinant
method in Carlson's proof of the P\'olya--Carlson
theorem~\cite[pp.~3--4]{Carlson21}.

\begin{theorem}[Central recurrence]\label{thm:central-recurrence}
  There are absolute constants $c_1,C_d$ such that the following holds
  for large $\Lambda$. There is a two-sided rational sequence $y$ whose
  minimal polynomial $P\in\Q[z]$ has degree $d$ with
  \[
    1\le d\le C_dM^2\ell,
  \]
  such that $y_k=Z_k$ for all $k$ in a centered block of length at
  least $u_0:=c_1M^2R$.
\end{theorem}

\subsection{A small transform and removal of the binomial weights}

\begin{lemma}[Small transform]\label{lem:small-transform}
  The entire function
  \[
    \mathcal G(z)=2^{-n}\sum_{k=0}^n(-1)^k\binom nk Z_k
    e^{(2k-n)z/\sqrt n}
  \]
  satisfies $|\mathcal G(z)|\le e^{-r}$ for
  $|z|\le e^{-3/2}\sqrt r$.
\end{lemma}
\begin{proof}
  The coefficient of $z^j$ is a nonzero scalar multiple of the
  alternating binomial sum of $Z_k(2k-n)^j$. This vanishes for $j<r$
  by \Cref{lem:degree-and-moments}. Also
  \[
    |\mathcal G(z)|\le
    \cosh(\operatorname{Re}z/\sqrt n)^n
    \le e^{(\operatorname{Re}z)^2/2}.
  \]
  Apply the maximum modulus principle to $\mathcal G(z)/z^r$ on the
  disk of radius $\sqrt r$: for $|z|\le e^{-3/2}\sqrt r$,
  $|\mathcal G(z)|\le e^{r/2}(|z|/\sqrt r)^r\le e^{-r}$.
\end{proof}

Set $c_0=e^{-3/2}$ and
\[
  \varepsilon_0=c_0\sqrt{r/n},\qquad
  \tau_0=\tanh(\varepsilon_0/2),\qquad p_\pm=(1\pm\tau_0)/2.
\]
For real $s_1,s_2$ let $p_{s_1}=(1+e^{-2s_1/\sqrt n})^{-1}$. The exact identity
\[
  \mathcal G(s_1+is_2)=\cosh(s_1/\sqrt n)^n e^{-is_2\sqrt n}
  \E_{Y\sim\operatorname{Bin}(n,p_{s_1})}
  [(-1)^Y Z_Y e^{2is_2Y/\sqrt n}]
\]
and \Cref{lem:small-transform} imply
\begin{equation}\label{eq:biased-binomial-smallness}
  \left|\sum_{k=0}^n\binom nk p^k(1-p)^{n-k}
  Z_ke^{i(\pi+\theta)k}\right|\le e^{-r}
  \quad(p\in[p_-,p_+],\ |\theta|\le\varepsilon_0).
\end{equation}
Indeed, with $p=p_{s_1}$ and $\theta=2s_2/\sqrt n$, the numbers $|s_1|,|s_2|$ are at
most $c_0\sqrt r/2$,
and the factor $\cosh(s_1/\sqrt n)^n$ is at least one.

Define
\[
  b_k=(n+1)\int_{p_-}^{p_+}\binom nk p^k(1-p)^{n-k}\,dp.
\]
The beta integral gives $0\le b_k\le1$ and
$\sum_kb_k=(n+1)\tau_0$. Integrating the derivative of the binomial
tail, or equivalently using the beta-binomial identity, gives
\begin{equation}\label{eq:flattening-weight-tails}
  1-b_k=\Pr\{\operatorname{Bin}(n+1,p_+)\le k\}
  +\Pr\{\operatorname{Bin}(n+1,p_-)\ge k+1\}.
\end{equation}
For the centered block
\[
  J_0=\{k:|k-n/2|\le n\tau_0/4\},
\]
both deviations in \eqref{eq:flattening-weight-tails} are at least $n\tau_0/4$.
Hoeffding's inequality~\cite[Theorem 2]{Hoeffding63}, in the Bernoulli form
$\Pr\{Y-\E Y\ge s\}\le e^{-2s^2/N}$, for a sum $Y$ of $N$ independent
Bernoulli variables and $s\ge0$, therefore gives
\begin{equation}\label{eq:flattening-weight-error}
  1-b_k\le2\exp\left(-\frac{n^2\tau_0^2}{8(n+1)}\right)
  \le2e^{-\kappa_0 r}\quad(k\in J_0),\qquad
  \kappa_0=c_0^2/256.
\end{equation}
Here $\tau_0\ge\varepsilon_0/4$ and $n^2/(n+1)\ge n/2$.

Now set
\begin{equation}\label{eq:central-polynomial-definition}
  \Acal(z)=\sum_{k=0}^n b_kZ_kz^k
  +\sum_{k\in J_0}(1-b_k)Z_kz^k.
\end{equation}
Its coefficients on $J_0$ are exactly $Z_k$; elsewhere they need not be
integral. Since $0\le b_k\le1$, and by integration of
\eqref{eq:biased-binomial-smallness} over $p$ together with
\eqref{eq:flattening-weight-error},
\begin{equation}\label{eq:central-polynomial-bounds}
  |\Acal(z)|\le4(n+1)\quad(z\in\T),\qquad
  |\Acal(-e^{i\theta})|\le3(n+1)e^{-\kappa_0r}\quad(|\theta|\le\varepsilon_0).
\end{equation}
The block length satisfies
\begin{equation}\label{eq:central-window-length}
  |J_0|\ge n\tau_0/2-1\ge(c_0/8)\sqrt{nr}-1.
\end{equation}

\subsection{Vanishing integer Hankel determinants}

The next lemma is a quantitative finite-block version of the integer-Hankel
determinant method in Carlson's proof~\cite[pp.~3--4]{Carlson21}.
We retain the small-arc width and the determinant size explicitly.
Bertrandias's rationality theorem is the form of this method that uses the
capacity of a set; see Dimitrov~\cite[Theorem 4 and the outline of its proof]{Dimitrov19}.

\begin{lemma}[Small arc and integer determinants]\label{lem:small-arc-determinants}
  Let $W\ge1$, $W_0>0$ and $0<\varepsilon\le1$. Let $\Xi(z)=\sum_{k\ge0}\xi_kz^k$ be a
  complex polynomial satisfying $|\Xi|\le W$ on $\T$ and
  $|\Xi|\le W_0$ on the arc of angular radius $\varepsilon$ about $-1$. Suppose that $D\ge4$ is an integer and
  \begin{equation}\label{eq:small-arc-conditions}
    W_0\le W e^{-3\varepsilon D},\qquad
    \varepsilon^2D\ge32\log(8W).
  \end{equation}
  Every determinant $\det(\xi_{b+i+j})_{0\le i,j<D}$ with integer $b\ge0$
  and integer entries is zero.
\end{lemma}
\begin{proof}
  Andr\'eief's integration identity, in the form recorded by
  Forrester~\cite[equation (1.7) and Section 2.2]{Forrester19}, expresses the
  Hankel determinant as an integral of two Vandermonde determinants.
  With $d\mu=d\theta/(2\pi)$, taking absolute
  values yields, for variables $z_1,\ldots,z_D\in\T$,
  \begin{align*}
    \left|\det(\xi_{b+i+j})\right|
    &\le\frac1{D!}\int_{\T^D}\prod_{k=1}^D|\Xi(z_k)|
    \prod_{i<j}|z_i-z_j|^2\,d\mu^D\\
    &=\det\left(\int_\T z^i\overline{z^j}|\Xi(z)|\,d\mu\right)_{0\le i,j<D}.
  \end{align*}
  The shift $b$ contributes only factors of modulus one. Replacing the
  monomials in this Gram matrix by any monic polynomials $Q_j$ of
  degrees $j$ preserves its determinant. Hadamard's inequality bounds it by
  \begin{equation}\label{eq:gram-determinant-bound}
    \prod_{j=0}^{D-1}\int_\T|Q_j(z)|^2|\Xi(z)|\,d\mu.
  \end{equation}

  Take $Q_0=1$, $Q_1=z$, and, for $m\ge1$, take
  \begin{align*}
    Q_{2m}(z)&=2^{1-m}(1+\cos\varepsilon)^m z^m
    T_m\left(\frac{z+z^{-1}-1+\cos\varepsilon}{1+\cos\varepsilon}\right),\\
    Q_{2m+1}(z)&=zQ_{2m}(z).
  \end{align*}
  The displayed expressions
  are monic polynomials of the stated degrees. Off the small arc the
  argument of $T_m$ lies in $[-1,1]$, and hence
  \[
    |Q_j(z)|\le2e^{-\varepsilon^2(j-1)_+/8}.
  \]
  This follows from $\log\cos(\varepsilon/2)\le-\varepsilon^2/8$.
  On the entire circle the argument is at least
  $-1-2\tan^2(\varepsilon/2)$. The identities
  $T_m(\cosh t)=\cosh(mt)$ and
  $\arcosh(1+2u^2)=2\arsinh u$, together with
  $\tan(\varepsilon/2)\le\varepsilon$, give
  $|Q_j(z)|\le2e^{j\varepsilon}$ everywhere on $\T$.
  Thus the $j$th factor of \eqref{eq:gram-determinant-bound} is at most
  \[
    4W e^{-\varepsilon^2(j-1)_+/4}+4W_0e^{2j\varepsilon}
    \le8W e^{-\varepsilon^2(j-1)_+/4}.
  \]
  The last step uses the first inequality of
  \eqref{eq:small-arc-conditions}, since
  $2j\varepsilon+\varepsilon^2j/4\le3\varepsilon D$ for $j<D$ and
  $\varepsilon\le1$.
  Thus the logarithm of the bound is at most
  $D\log(8W)-\varepsilon^2(D-1)(D-2)/8<0$, since
  $(D-1)(D-2)>D^2/4$ for $D\ge4$ and \eqref{eq:small-arc-conditions} holds.
  The determinant is an integer of modulus below one, so it is zero.
\end{proof}
\subsection{Extraction of a regular recurrence}

Finite Hankel rank and finite recurrences are closely related; see
Elkies~\cite[Proposition 1 and its corollary]{Elkies02} for the
homogeneous annihilator formulation. Here we start only with
\emph{consecutive} minors, and need a regular recurrence on a
\emph{centered} block. We give the trimming argument for these requirements.

\begin{lemma}[Symmetric trimming]\label{lem:symmetric-trimming}
  Let $D\ge2$ and $N\ge2D+1$ be integers, and let $x_0,\ldots,x_{N-1}$ be a
  word of signs $x_i\in\{-1,1\}$ whose consecutive $D$-by-$D$ Hankel determinants
  $\det(x_{k+i+j})_{0\le i,j<D}$ all vanish. After at most $D-1$ equal
  deletions at the two ends, the remaining word satisfies a regular
  monic rational recurrence of order $1\le d<D$.
  Every consecutive $d$-by-$d$ Hankel determinant
  on the remaining word is nonzero, and the recurrence is minimal.
\end{lemma}
\begin{proof}
  On a word $x_0,\ldots,x_{L-1}$ write
  $\mathcal H_d(k)=\det(x_{k+i+j})_{0\le i,j<d}$ for $0\le k\le L-2d+1$,
  with $\mathcal H_0(k)=1$. Call $\mathcal H_d(k)$ \emph{interior} if
  $1\le k\le L-2d$. For $0\le k\le L-2d-1$, the Desnanot--Jacobi identity
  (Dodgson condensation), applied to the first and last rows and columns of
  the $(d+1)$-by-$(d+1)$ Hankel matrix at $k$, gives
  \begin{equation}\label{eq:hankel-adjacent-minors}
    \mathcal H_{d+1}(k)\mathcal H_{d-1}(k+2)
    =\mathcal H_d(k)\mathcal H_d(k+2)-\mathcal H_d(k+1)^2 .
  \end{equation}
  The procedure maintains the following invariant: \emph{all order-$(d+1)$
  Hankel determinants on the current word vanish}. It holds for the given
  word with $d=D-1$. If it holds for some $d\ge1$, then
  \eqref{eq:hankel-adjacent-minors} gives
  $\mathcal H_d(k+1)^2=\mathcal H_d(k)\mathcal H_d(k+2)$, so a nonzero
  interior term $\mathcal H_d(k+1)$ forces both of its neighbors to be
  nonzero. Iterating, every term $\mathcal H_d(k)$ is nonzero, and we stop.
  If instead every interior term vanishes, delete one symbol at each end.
  The order-$d$ determinants of the shortened word are exactly the interior
  terms of the old one, so they all vanish, and the invariant holds with
  $d-1$ in place of $d$. After $j$ deletions the word has length $N-2j$
  and $d=D-1-j$, so $L-2d=N-2D+2\ge3$: interior terms and the identity
  \eqref{eq:hankel-adjacent-minors} are available at every stage. If we
  reach $d=0$, after $D-1$ deletions, the invariant says that every symbol
  vanishes, which is impossible for a nonempty word of signs.

  Suppose we stop at some $d\ge1$, so that every order-$d$ determinant is
  nonzero and every order-$(d+1)$ determinant vanishes. Fix $k$ with
  $0\le k\le L-2d-1$. The $(d+1)$-by-$(d+1)$ Hankel matrix at $k$ is
  singular, but its first $d$ columns are independent because their first
  $d$ rows form the nonsingular matrix at $k$ of order $d$. Hence its kernel
  is spanned by a vector $(p_0(k),\ldots,p_{d-1}(k),1)$, and
  $x_{h+d}+\sum_{j<d}p_j(k)x_{h+j}=0$ for $k\le h\le k+d$. The equations
  with $k+1\le h\le k+d$ determine $p(k+1)$ uniquely, because the order-$d$
  matrix at $k+1$ is nonsingular, so $p(k+1)=p(k)$. Thus one monic recurrence
  $x_{h+d}+\sum_{j<d}p_jx_{h+j}=0$ holds on the whole word. Its companion
  matrix $\mathsf C$, of determinant $(-1)^dp_0$, carries each order-$d$
  Hankel matrix to the next, so
  $\mathcal H_d(k+1)=(-1)^dp_0\mathcal H_d(k)$ and $p_0\ne0$. Thus the
  recurrence is regular. A recurrence of order $e<d$ would give a linear
  relation among the first $e+1$ columns of an order-$d$ Hankel matrix,
  making it singular. This proves minimality.
\end{proof}

\subsection{Proof of the central recurrence}
Take $W=4(n+1)$, $W_0=3(n+1)e^{-\kappa_0r}$ and $\varepsilon=\varepsilon_0$ in
\Cref{lem:small-arc-determinants}, applied to $\Acal$, with
\begin{equation}\label{eq:hankel-size-choice}
  D=4+\left\lceil\frac{32}{\varepsilon_0^2}\log(8W)\right\rceil.
\end{equation}
The second inequality of \eqref{eq:small-arc-conditions} holds by the choice
of $D$. The first follows from $\kappa_0r\ge3\varepsilon_0D$.
Since $\varepsilon_0\ge c_0/(2M)$ and $\log(8W)\le5\ell$, we have
$\varepsilon_0D\le5+320M\ell/c_0$ and $D\le5+640M^2\ell/c_0^2$. As
$\kappa_0r=\kappa_0\Lambda M\ell$, the condition $\kappa_0r\ge3\varepsilon_0D$
holds for large $\Lambda$. By \Cref{lem:small-arc-determinants} and
\eqref{eq:central-polynomial-bounds}, every $D$-by-$D$ Hankel determinant
$\det(Z_{b+i+j})_{0\le i,j<D}$ with $b,\ldots,b+2D-2\in J_0$ vanishes.

Apply \Cref{lem:symmetric-trimming} to the signs on $J_0$. Since
$\sqrt{nr}\ge Mr=M^2R$, \eqref{eq:central-window-length} gives
$|J_0|\ge(c_0/8)M^2R-1$, which exceeds $2D+1$ for large $\Lambda$. After at
most $D-1$ equal deletions at the two ends we
obtain a centered block of length at least $(c_0/8)M^2R-2D\ge(c_0/32)M^2R$,
again for large $\Lambda$, on which the signs satisfy a regular minimal
monic rational recurrence of order $d$ with $1\le d<D\le C_dM^2\ell$. Let
$P$ be its characteristic polynomial and $y$ the two-sided solution that
agrees with $Z$ on the block. Any polynomial satisfied by $y$ is satisfied
on the block, so minimality on the block makes $P$ the minimal polynomial
of $y$. Since the recurrence is rational and agrees with integers, $y$ is
rational. This proves \Cref{thm:central-recurrence} with
$c_1=c_0/32$. \qed

\subsection{The agreement interval}\label{sec:agreement-interval}
Let $y$, $P$ and $d$ be as in \Cref{thm:central-recurrence}. If $y_k=Z_k$
for all $0\le k\le n$, we speak of \emph{full agreement} and put $t=0$.
Otherwise let $t$ be the largest integer with
$0\le t<n/2$ such that $y_t\ne Z_t$ or $y_{n-t}\ne Z_{n-t}$. Replacing $F$
by the profile $k\mapsto F(n-k)$, which has the same deficit, and $y$ by
$k\mapsto y_{n-k}$, whose minimal polynomial is the monic reverse
$z^dP(1/z)/P(0)$, we may assume that $y_t\ne Z_t$. We call $t$ the
\emph{first defect}, and put $\mathfrak e=Z_t-y_t\ne0$. In both cases let
\begin{equation}\label{eq:first-defect}
  u=n-2t,\qquad I=\{t+1,\ldots,n-t-1\}.
\end{equation}
Thus $y=Z$ on the centered interval $I$ of length $u-1$. In the
first-defect case $I$ contains the block of \Cref{thm:central-recurrence};
in the full-agreement case $u-1=n-1$. Either way $u-1\ge u_0$.

\emph{Working blocks.} Long agreement (\Cref{sec:proof-of-main-theorem})
needs the results of
\Cref{sec:recurrence-arithmetic,sec:prediction-and-filtering,sec:circles-and-far-sector,sec:roots-near-minus-one}
on a shorter centered interval, so we state them for a general one. In
those sections $t$ is an integer with $0\le t<n/2$, $u=n-2t$ and
$I=\{t+1,\ldots,n-t-1\}$, subject to
\begin{itemize}
  \item[\textup{(W)}] $y_k=Z_k$ for all $k\in I$, and $u-1\ge u_0$.
\end{itemize}
A statement that also uses a defect at $t$ says so, and assumes
\begin{itemize}
  \item[\textup{(D)}] $y_t\ne Z_t$, and $\mathfrak e=Z_t-y_t$.
\end{itemize}
The agreement interval satisfies \textup{(W)}, and also \textup{(D)} in the
first-defect case; the short-agreement case applies \Cref{sec:endpoint-test}
to it. For $t$ and $u$ as above put
\begin{equation}\label{eq:agreement-parameters}
  w=\frac u{u_0}\ge1,\qquad S=\frac{u^2}n,\qquad \kappa=\frac un .
\end{equation}
By \eqref{eq:scale-comparison},
\begin{equation}\label{eq:agreement-parameter-ranges}
  \frac{c_1^2}4w^2r\le S\le c_1^2w^2r,\qquad
  \frac{c_1w}{4M}\le\kappa\le\frac{c_1w}M .
\end{equation}
If \textup{(W)} holds and $u\le n/128$, then, for large $\Lambda$,
\begin{equation}\label{eq:defect-angle-range}
  1.99\kappa\le\alpha_t\le2.02\kappa .
\end{equation}
Indeed, by \Cref{lem:angles-and-weights}(d) it suffices that
$\sqrt{8\log(n+1)/n}\le0.003\kappa$; and $\log(n+1)/n\le\ell/n\le1/(\Lambda M^3)$
by \eqref{eq:scale-comparison}, while $\kappa\ge c_1/(4M)$.
Write $Z$ for the (possibly reflected) sign word throughout; the deficit
condition \eqref{eq:vanishing-moments} is unchanged by the reflection.

\section{Arithmetic of the recurrence}\label{sec:recurrence-arithmetic}
The sequence $y$ is rational, and need not be integral outside $I$. The
integer values of $Z$ on $I$ nevertheless supply four facts.

\subsection{A short integer annihilator}

\begin{lemma}[Short annihilator]\label{lem:short-annihilator}
  Let $y$ be a two-sided rational sequence whose minimal polynomial
  $P\in\Q[z]$ has degree $d\ge1$, and suppose that $y$ takes values in
  $\{-1,1\}$ at $3d-1$ consecutive integers. There is a polynomial
  $A\in\Z[z]$ such that
  \[
    P\mid A,\qquad\deg A<2d,\qquad A(0)\ne0,\qquad\|A\|_1\le8d^2 .
  \]
\end{lemma}
\begin{proof}
  Translate the indices so that the signs are $y_0,\ldots,y_{3d-2}$, and
  map each $e\in\{0,\ldots,4d\}^{2d}$ to the vector
  $\bigl(\sum_{j<2d}e_jy_{i+j}\bigr)_{0\le i<d}$. Its coordinates are
  integers in $[-8d^2,8d^2]$, and
  $(4d+1)^{2d}=((4d+1)^2)^d>(16d^2+1)^d$. So two inputs have the same image,
  and their difference gives a nonzero integer polynomial $\widehat A$ of
  degree below $2d$, with coefficients of modulus at most $4d$, such that
  $(\widehat A(\shift)y)_i=0$ for $0\le i<d$. The sequence
  $\widehat A(\shift)y$ satisfies $P$ and vanishes at $d$ consecutive
  indices, so it vanishes everywhere. Minimality of $P$ gives
  $P\mid\widehat A$. Remove the largest power of $z$ dividing $\widehat A$; since
  $P(0)\ne0$, divisibility survives. The result $A$ has a nonzero integer
  constant coefficient and $\|A\|_1\le2d\cdot4d=8d^2$.
\end{proof}

The interval $I$ has length at least $u_0\ge3d-1$ for large $\Lambda$, so
\Cref{lem:short-annihilator} applies to the signs on $I$. We fix such an
$A(z)=\sum_{i=0}^{d_A}A_iz^i$ of degree $d_A$, so that $A_{d_A}$ is a nonzero
integer, and write $H_A=\|A\|_1$. Thus
\begin{equation}\label{eq:short-annihilator}
  A(\shift)y=0,\qquad d_A<2d,\qquad A_0\in\Z\setminus\{0\},\qquad
  H_A\le8d^2 .
\end{equation}
The defect becomes a nonzero integer after multiplication by $A_0$.

\begin{corollary}[Integral defect]\label{cor:integral-defect}
  For large $\Lambda$, under \textup{(W)} and \textup{(D)}, with $A$ fixed as in
  \eqref{eq:short-annihilator}, $A_0\mathfrak e=(A(\shift)Z)_t$ is a nonzero integer.
\end{corollary}
\begin{proof}
  The indices $t+i$, $1\le i\le d_A$, lie in $I$, where $Z=y$. Hence
  $(A(\shift)Z)_t-(A(\shift)y)_t=A_0(Z_t-y_t)=A_0\mathfrak e$, and
  $(A(\shift)y)_t=0$. The left side is an integer combination of signs,
  and $A_0\mathfrak e\ne0$.
\end{proof}

\subsection{Hankel determinants of a component}

\begin{corollary}[Small components have few roots]\label{cor:small-component-root-count}
  Assume \textup{(W)}. Let $\mathcal Z$ be a set of roots of $P$ of total multiplicity $k\ge1$,
  and let $y_\rmin$ be the component of $y$ on $\mathcal Z$. If
  $|y_\rmin|\le\epsilon$ at $2d-1$ consecutive integers of $I$, where
  $0<\epsilon\le1$, then
  $1\le(4\sqrt d)^d\epsilon^k$, and hence $k\log(1/\epsilon)\le d\log(4\sqrt d)$.
  For large $\Lambda$,
  \begin{equation}\label{eq:hankel-count}
    d\log(4\sqrt d)\le C_HM^2\ell^2
  \end{equation}
  for an absolute constant $C_H$.
\end{corollary}
\begin{proof}
  Let the window start at $b$, and put
  $\mathsf H=(Z_{b+i+j})_{0\le i,j<d}=(y_{b+i+j})$. It is an integer
  matrix, nonsingular by \Cref{lem:nonsingular-hankel}, so $|\det\mathsf H|\ge1$.
  Write $\mathsf H=\mathsf A+\mathsf E$, where $\mathsf A$ is the Hankel
  matrix of $y-y_\rmin$ and $\mathsf E$ that of $y_\rmin$. By
  \Cref{lem:nonsingular-hankel}, $\operatorname{rank}\mathsf A\le d-k$; its entries
  have modulus at most $|Z|+|y_\rmin|\le2$. Expand $\det(\mathsf A+\mathsf E)$
  multilinearly in its columns. A nonzero term uses at least $k$ columns
  from $\mathsf E$, since $d-k+1$ columns of $\mathsf A$ are dependent.
  Hadamard's inequality bounds each term by $(2\sqrt d)^d\epsilon^k$;
  there are at most $2^d$ terms. Hence
  $1\le|\det\mathsf H|\le(4\sqrt d)^d\epsilon^k$.
  The last bound uses $d\le C_dM^2\ell$ and $\log d\le C\ell$.
\end{proof}

\subsection{Bounded blocks and radial drift}
The following lemma uses only the signs on the agreement block.

\begin{lemma}[Arithmetic of a bounded block]\label{lem:bounded-block-arithmetic}
  Suppose a regular monic rational recurrence of order $d\ge1$ holds on
  signs $x_0,\ldots,x_{N-1}\in\{-1,1\}$.
  Let $\mathsf B_0=(x_{i+j})_{0\le i,j<d}$ be nonsingular, and set
  $\sigma_0=N-2d+1\ge1$.
  If $2^{\sigma_0}>d^{d/2}$, the recurrence coefficients are integers.
  For the Mahler measure $\mathcal M(P)=\prod_i\max(1,|\lambda_i|)$
  of its monic characteristic polynomial, with roots $\lambda_i$ counted with
  multiplicity, one always has
  \begin{equation}\label{eq:mahler-upper-bound}
    \log\mathcal M(P)\le\frac{d\log d}{\sigma_0}.
  \end{equation}
\end{lemma}
\begin{proof}
  Since $\mathsf B_0$ is a nonsingular integer sign matrix,
  $1\le|\det\mathsf B_0|\le d^{d/2}$ by Hadamard's inequality.
  The rational companion matrix $\mathsf C$ satisfies
  $\mathsf B_h=\mathsf B_0\mathsf C^h$, where $\mathsf B_h=(x_{h+i+j})_{0\le i,j<d}$ and $0\le h\le\sigma_0$.
  Every entry of every exterior power $\bigwedge^j(\mathsf C^{\sigma_0})$ has denominator
  dividing $\det\mathsf B_0$. Indeed, use $\mathsf C^{\sigma_0}=\mathsf B_0^{-1}\mathsf B_{\sigma_0}$, Cauchy--Binet,
  and Jacobi's complementary-minor formula for the minors of $\mathsf B_0^{-1}$.

  If a characteristic coefficient is nonintegral, choose a prime $p$
  at which its $p$-adic absolute value is at least $p$. That coefficient
  is an elementary symmetric function of the roots, so some product
  of $j$ roots has $p$-adic absolute value at least $p$.
  It is an eigenvalue of $\bigwedge^j\mathsf C$, and consequently
  $\bigwedge^j(\mathsf C^{\sigma_0})$ has an eigenvalue of absolute value at least $p^{\sigma_0}$.
  But all entries of that matrix have absolute value at most
  $p^{v_p(\det\mathsf B_0)}$. The maximum-entry matrix norm bounds its
  eigenvalues, even over an extension containing the roots. Hence
  $\sigma_0\le v_p(\det\mathsf B_0)$, contradicting
  $2^{\sigma_0}>d^{d/2}\ge|\det\mathsf B_0|$.

  For \eqref{eq:mahler-upper-bound}, let $j$ be the number of roots outside
  the closed unit disk, with multiplicity. The eigenvalues of
  $\bigwedge^j(\mathsf C^{\sigma_0})$ are the products of $j$ of the
  numbers $\lambda_i^{\sigma_0}$, so its spectral radius is
  $\mathcal M(P)^{\sigma_0}$; for $j=0$ this exterior power is the number
  one. Euclidean operator norms give
  \[
    \mathcal M(P)^{\sigma_0}
    \le\|\bigwedge^j\mathsf B_0^{-1}\|\,\|\bigwedge^j\mathsf B_{\sigma_0}\|
    =\frac{\|\bigwedge^{d-j}\mathsf B_0\|\,\|\bigwedge^j\mathsf B_{\sigma_0}\|}{|\det\mathsf B_0|}
    \le\frac{\|\mathsf B_0\|^{d-j}\,\|\mathsf B_{\sigma_0}\|^j}{|\det\mathsf B_0|}.
  \]
  Here the norm of $\bigwedge^i\mathsf B$ is the product of the $i$ largest
  singular values of $\mathsf B$, which is at most $\|\mathsf B\|^i$, and
  the singular values of $\mathsf B_0^{-1}$ are the reciprocals of those of
  $\mathsf B_0$. Finally $\|\mathsf B_h\|\le d$, since every entry is a sign,
  and $|\det\mathsf B_0|\ge1$. Taking logarithms proves
  \eqref{eq:mahler-upper-bound}.
\end{proof}

We use Voutier's explicit form~\cite[Corollary~2]{Voutier96} of
Dobrowolski's lower bound~\cite{Dobrowolski79}: a nonzero algebraic number
$\alpha$ of degree $d_\alpha\ge2$ that is not a root of unity satisfies
$\log\mathcal M(\alpha)>2/(\log 3d_\alpha)^3$, where $\mathcal M(\alpha)$ is
the Mahler measure of its minimal polynomial over $\Z$. Consequently a
monic integer polynomial $P$ of degree at most $d$, with nonzero constant
coefficient and at least one noncyclotomic irreducible factor, satisfies
\begin{equation}\label{eq:voutier-height-gap}
  \log\mathcal M(P)>\frac{2}{(\log 6d)^3}.
\end{equation}
Indeed, the Mahler measure is multiplicative and is at least one for every
monic integer factor, so it suffices to bound one monic noncyclotomic
irreducible factor $P_1$ of $P$. If $\deg P_1\ge2$, then $P_1$ is the
minimal polynomial of each of its roots, and these are nonzero and are not
roots of unity, since an irreducible monic integer polynomial with a root
of unity as a root is cyclotomic. Voutier's bound and $\deg P_1\le d$ give
$\log\mathcal M(P_1)>2/(\log 3d)^3$. If $\deg P_1=1$, then $P_1=z-a$ with
$a\in\Z\setminus\{0,\pm1\}$, so $\log\mathcal M(P_1)\ge\log2>2/(\log 6)^3$.

Write the roots of $P$, with multiplicity, as
$\lambda_i=-e^{\rho_i+i\theta_i}$, and let
\begin{equation}\label{eq:total-drift}
  T=\sum_{i=1}^d|\rho_i|=\sum_{i=1}^d\bigl|\log|\lambda_i|\bigr|
\end{equation}
be their \emph{total drift}.

\begin{corollary}[Total drift]\label{cor:total-drift}
  Under \textup{(W)}, $T(u-2d)\le2d\log d$.
\end{corollary}
\begin{proof}
  The signs on $I$ form a block of $N=u-1\ge2d$ integers of modulus one on
  which $y$ satisfies the regular monic rational recurrence with
  polynomial $P$. Its Hankel matrix at the start of $I$ is a nonsingular
  integer matrix by \Cref{lem:nonsingular-hankel}, so $|\det\mathsf B_0|\ge1$,
  and \eqref{eq:mahler-upper-bound} gives
  $(N-2d+1)\log\mathcal M(P)\le d\log d$, and $N-2d+1=u-2d$. The reversed block
  satisfies the recurrence with the monic reverse
  $P^{\rm rev}(z)=z^dP(1/z)/P(0)$, whose roots are the $1/\lambda_i$, so
  the same bound holds for $\mathcal M(P^{\rm rev})$. Finally
  $\log\mathcal M(P)+\log\mathcal M(P^{\rm rev})=T$.
\end{proof}

\begin{corollary}[Long agreement is cyclotomic]\label{cor:long-agreement-cyclotomic}
  There is an absolute constant $C_V$ such that, if $y$ agrees with signs
  on a block of
  \[
    N\ge C_Vd\log(2d)(\log 6d)^3
  \]
  consecutive integers, then $P$ is a product of cyclotomic polynomials. In
  particular $P\in\Z[z]$ and every root of $P$ is a root of unity.
\end{corollary}
\begin{proof}
  The Hankel matrix at the start of the block is nonsingular by
  \Cref{lem:nonsingular-hankel}. If $C_V$ is large, then
  $2^{N-2d+1}>d^{d/2}$, so
  \Cref{lem:bounded-block-arithmetic} makes $P$ integral, and
  $P(0)\ne0$. It also gives $\log\mathcal M(P)\le d\log d/(N-2d+1)$,
  which is below the right side of \eqref{eq:voutier-height-gap} if $C_V$ is
  large. So every monic irreducible factor of $P$ is cyclotomic.
\end{proof}

\subsection{Angular counts}
The angle of a root is the angular distance of its projection
$\lambda/|\lambda|$ from $-1$. Jensen's formula for a projected
annihilator bounds how many projections can crowd near any point.

\begin{lemma}[Angular count]\label{lem:angular-count}
  There is an absolute constant $C$ with the following property.
  Let $m\ge1$ be an integer and $h\ge1$, and let $A_\circ$ be a polynomial of degree
  below $m$ with $|A_\circ(0)|\ge1$ and $|A_\circ|\le e^h$ on $\T$. Then,
  for every $x\in\T$ and $s\ge0$, the arc of angular radius $s$ about $x$
  contains at most $C(ms+\sqrt{mh})$ zeros of $A_\circ$, counted with
  multiplicity.
\end{lemma}
\begin{proof}
  If $s\ge1$ or $h\ge m$, the degree bound suffices. Otherwise put
  $s'=\max(s,\sqrt{h/m})<1$, and let $k$ be the number of zeros, with
  multiplicity, on the arc $\mathcal J$ of angular radius $s'$ about $x$.
  We show that
  \begin{equation}\label{eq:arc-zero-count}
    k\log2\le h+4ms'+\frac{\pi h}{s'} .
  \end{equation}
  The arc has the fraction $s'/\pi$ of the length of $\T$. If
  $\log|A_\circ|<-\pi h/s'$ held on all of $\mathcal J$, then, since
  $\log|A_\circ|\le h$ on $\T$, the mean of $\log|A_\circ|$ over $\T$ would
  be less than $-h+h(1-s'/\pi)<0$, contradicting Jensen's formula and
  $|A_\circ(0)|\ge1$. So some $z_0\in\mathcal J$ has
  $|A_\circ(z_0)|\ge e^{-\pi h/s'}$. Every point of $\mathcal J$ lies in the
  disk $|z-z_0|\le2s'$. On the circle $|z-z_0|=4s'$ we have $|z|\le1+4s'$,
  so the Bernstein--Walsh lemma for the unit disk gives
  $|A_\circ(z)|\le e^h(1+4s')^m\le e^{h+4ms'}$. The zero count of
  \Cref{sec:preliminaries} with $\rho=4s'$ now gives
  \eqref{eq:arc-zero-count}.

  The original arc is contained in $\mathcal J$, and
  $ms'\le ms+\sqrt{mh}$, $h/s'\le\sqrt{mh}$ and $h\le\sqrt{mh}$.
  Substitute these bounds into \eqref{eq:arc-zero-count}.
\end{proof}

\subsection{The standing bounds}
Recall $A$ from \eqref{eq:short-annihilator}. Write $\lambda_i=|\lambda_i|\hat\lambda_i$
with $|\hat\lambda_i|=1$. Replacing in $A$ the factors $1-z/\lambda_i$, for the
roots of $P$ with multiplicity, by $1-z/\hat\lambda_i$ gives a polynomial
$A_\circ$ with the same degree and constant coefficient. On $\T$, the
ratio $|1-z/\hat\lambda_i|/|1-z/\lambda_i|$ is at most
$2/(1+e^{-\rho_i})\le e^{|\rho_i|}$, so $|A_\circ|\le8d^2e^T$ on $\T$.
Applying \Cref{lem:angular-count} with $m=2d$ and
$h=\max(1,\log(8d^2))+T$ gives part (c) of the following proposition.

\begin{proposition}[Standing bounds]\label{prop:standing-bounds}
  Assume \textup{(W)}, and fix $A$ as in \eqref{eq:short-annihilator}.
  There are absolute constants $C_T,C_A,C_B,C_h$ such that, for large
  $\Lambda$, the following hold.
  \begin{enumerate}
    \item[(a)] $d\le C_dM^2\ell$, and $d_A<2d\le J:=2C_dM^2\ell\le u_0/4<u/4$.
      The number $h_A=\max(1,\log H_A)$ satisfies $h_A\le C_h\ell$.
    \item[(b)] $T\le C_TM^2\ell^2/u\le\ell$.
    \item[(c)] For every point of $\T$ and every $s\ge0$, at most
      $A_\theta s+B_\theta$ roots of $P$, counted with multiplicity, have
      projections within angular distance $s$ of that point, where
      \[
        A_\theta=C_AM^2\ell,\qquad B_\theta=C_BM\ell ;
      \]
      the same holds for the reciprocals of the roots. Hence, for
      $a_1\le a_2$, at most $A_\theta(a_2-a_1)+2B_\theta$ roots have angle in
      $[a_1,a_2]$. The same bounds hold a fortiori for every submultiset of
      the roots.
  \end{enumerate}
\end{proposition}
\begin{proof}
  (a) The bound on $d$ is \Cref{thm:central-recurrence}; $J\le u_0/4$
  because $u_0=c_1\Lambda M^2\ell$; and $H_A\le8d^2$.
  (b) By \Cref{cor:total-drift}, since $u-2d\ge u/2$,
  $T\le4d\log d/u\le C_TM^2\ell^2/u$, and
  $C_TM^2\ell^2/u\le C_T\ell/(c_1\Lambda)\le\ell$.
  (c) Apply \Cref{lem:angular-count} as above, using $m=2d\le2C_dM^2\ell$
  and $h\le\log(8d^2)+1+\ell\le C\ell$, so $\sqrt{mh}\le CM\ell$. The
  angles in $[a_1,a_2]$ form two arcs of radius $(a_2-a_1)/2$. The
  projections of the reciprocals are the complex conjugates of those of
  the roots.
\end{proof}

For $\theta>0$ let $N_{\rm all}(\theta)$ be the number of roots of $P$,
with multiplicity, of angle below $\theta$.

\section{Prediction, filtering and boundary modes}\label{sec:prediction-and-filtering}
This section controls the recurrence outside the interval where it is
observed. \Cref{lem:recurrence-prediction} predicts a recurrence from a short
observation, with growth governed by the height of its annihilator.
Roots of large drift would make this growth too fast, and the rational
filter of \Cref{sec:rational-filter} moves them outward.
\Cref{prop:filtered-defect-heights} collects the resulting bounds.

\subsection{Prediction from an annihilator}

\begin{lemma}[Prediction]\label{lem:recurrence-prediction}
  Let $v$ be a two-sided sequence whose minimal polynomial has all its
  roots in $e^{-\eta}\le|\lambda|\le e^\eta$, where $\eta\ge0$. Suppose the
  sequence satisfies an integer polynomial $\widehat A$ of degree $a\ge1$
  with $\widehat A(0)\ne0$. Put $h=\max(1,\log\|\widehat A\|_1)$.
  If $|v_k|\le V$ at $a$ consecutive integers, then at every integer at
  distance $b\ge1$ from them,
  \begin{equation}\label{eq:prediction-bound}
    |v_k|\le V\sqrt{ab}\,\exp\bigl(h+\eta b+2\sqrt{2hb}\bigr).
  \end{equation}
\end{lemma}
\begin{proof}
  The nonzero end coefficients of $\widehat A$ are integers, so both have
  modulus at least one. Divide $\widehat A$ by $\widehat A(0)$; this keeps
  $|\widehat A|\le e^h$ on $\T$. Factor
  $\widehat A(z)=\prod_\alpha(1-z/\alpha)$ over its roots, and replace each factor
  with $|\alpha|<e^{-\eta}$ by $1-\overline\alpha z$. On $\T$ this
  multiplies the modulus of the factor by $|\alpha|<1$. Such $\alpha$ are
  not roots of the minimal polynomial of $v$. The result $B$ has degree
  $a$, $B(0)=1$, $|B|\le e^h$ on $\T$, no zeros in $|z|<e^{-\eta}$, and it
  is divisible by the minimal polynomial of $v$.

  Choose the branch of $\log B(e^{-\eta}z)$ that vanishes at $0$.
  The function $\Phi(z)=h-\log B(e^{-\eta}z)$ is holomorphic on $|z|<1$ with
  $\operatorname{Re}\Phi\ge0$, by the maximum principle. Carath\'eodory's
  inequality gives $|[z^j]\log B|\le2he^{\eta j}$ for $j\ge1$. Hence
  $1/B=\exp(-\log B)$ is majorized coefficientwise in modulus by
  $\exp(2he^\eta z/(1-e^\eta z))$. The coefficients of
  $\exp(2hz/(1-z))$ are positive, so Cauchy's estimate at radius
  $\varrho=\sqrt j/(\sqrt j+\sqrt{2h})$ bounds the $j$th one by
  $\varrho^{-j}\exp(2h\varrho/(1-\varrho))\le e^{2\sqrt{2hj}}$, because
  $-j\log\varrho\le\sqrt{2hj}$ and $2h\varrho/(1-\varrho)=\sqrt{2hj}$. Thus
  $|[z^j](1/B)|\le e^{\eta j+2\sqrt{2hj}}$.

  Let $w_b$ be the Taylor polynomial of $1/B$ of degree below $b$. Then
  $\Theta_b=Bw_b=1+\sum_{j=b}^{a+b-1}q_jz^j$ is divisible by the minimal
  polynomial of $v$, and
  $\|\Theta_b\|_2\le\max_\T|B|\cdot\|w_b\|_2\le e^h\sqrt b\,e^{\eta b+2\sqrt{2hb}}$.
  Translate the observed integers to $0,\ldots,a-1$. The identity
  $(\Theta_b(\shift)v)_{-b}=0$ reads $v_{-b}=-\sum_{j=b}^{a+b-1}q_jv_{j-b}$, and
  Cauchy--Schwarz gives \eqref{eq:prediction-bound} on the left. For the right
  side apply the same argument to $k\mapsto v_{-k}$ and to
  the reverse of the original integer polynomial $\widehat A$, which has
  the same degree and coefficient norm and a nonzero constant coefficient.
\end{proof}

\subsection{The rational filter}\label{sec:rational-filter}
Assume $u\le n/128$, and put
\begin{equation}\label{eq:filter-scale}
  \sigma=128\kappa=\frac{128u}n\le1 .
\end{equation}

\begin{definition}[Filter at threshold $\eta$]\label{def:rational-filter}
  Let $0<\eta\le\sigma/8$. The \emph{removed} roots are the roots
  $\lambda_i=-e^{\rho_i+i\theta_i}$, $1\le i\le K$, of $P$ of drift
  $|\rho_i|>\eta$, listed with multiplicity; the other roots are
  \emph{retained}. For each $i$ choose
  \begin{equation}\label{eq:moved-poles}
    \max(2|\rho_i|,\sigma)\le\tau_i\le2\max(2|\rho_i|,\sigma),
    \qquad \mu_i=-e^{\tau_i+i\theta_i},
  \end{equation}
  such that the $\mu_i$ are pairwise distinct. Put
  \[
    \Cf(z)=\prod_{i=1}^K\Bigl(1-\frac z{\lambda_i}\Bigr),\qquad
    \Df(z)=\prod_{i=1}^K\Bigl(1-\frac z{\mu_i}\Bigr),\qquad
    \frac{\Cf(z)}{\Df(z)}=\sum_{j\ge0}\phi_jz^j\quad(|z|<e^\sigma),
  \]
  and define the \emph{filtered word}
  $X_k=\sum_{j=0}^{n-k}\phi_jZ_{k+j}$ for $0\le k\le n$. Let $\Cf^\sharp$
  and $\Df^\sharp$ be the polynomials of \Cref{lem:binomial-transform}(d) for $\Cf$
  and $\Df$. If $K=0$, the filter is \emph{trivial}: $\Cf=\Df=1$ and
  $X=Z$.
\end{definition}

The removed roots are neither $0$ nor $1$, and $|\mu_i|=e^{\tau_i}>1$, so
$\Cf^\sharp$ and $\Df^\sharp$ are defined. The coordinates of the moved
poles are far from the origin: by \Cref{lem:coordinates-near-minus-one}(b),
$|\tanh((\tau_i+i\theta_i)/2)|\ge\tanh(\tau_i/2)$, so
\begin{equation}\label{eq:moved-coordinates}
  |x_{\mu_i}|\ge n\tanh(\sigma/2)\ge n\sigma/4=32u .
\end{equation}

\begin{lemma}[The filter]\label{lem:filter-properties}
  In the setting of \Cref{def:rational-filter}, the following hold.
  \begin{enumerate}
    \item[(a)] $K\le T/\eta$, $|\Cf/\Df|\le e^{5T+2\sigma K}$ on $\T$, and
      $|X_k|\le\sqrt{n+1}\,e^{5T+2\sigma K}$.
    \item[(b)] Every test $f$ with $\deg f<r$ whose transform is divisible
      by $\Df^\sharp$ satisfies $\langle f,X\rangle=0$.
    \item[(c)] Let $y_{\rm ret}$ be the component of $y$ on the retained
      roots, and put $v_k=\sum_{j\ge0}\phi_jy_{{\rm ret},k+j}$. Then $v$ has
      minimal polynomial $\prod(z-\lambda)^{m_\lambda}$ over the retained
      roots, $A(\shift)v=0$, and $\Df(\shift)v=\Cf(\shift)y$. We call $v$
      the \emph{retained recurrence}.
    \item[(d)] There is a two-sided sequence $\omega$ with
      $\Df(\shift)\omega=0$ and $X=v+\omega$ on $I$, provided $K<u-1$.
    \item[(e)] Under \textup{(D)}, if $K<u-1$, then
      $X_t-v_t=\mathfrak e+\omega_t$.
  \end{enumerate}
\end{lemma}
\begin{proof}
  (a) Each removed root has $|\rho_i|>\eta$, so $K\eta\le T$. By
  \eqref{eq:moved-poles}, $\tau_i\le4|\rho_i|+2\sigma$. On $\T$, write
  $z/\lambda_i=\varsigma e^{-\rho_i}$ and $z/\mu_i=\varsigma e^{-\tau_i}$ with
  $|\varsigma|=1$. As a function of $\operatorname{Re}\varsigma$, the
  quotient $|1-\varsigma e^{-\rho_i}|^2/|1-\varsigma e^{-\tau_i}|^2$ is a
  ratio of linear functions, so it is largest at $\varsigma=\pm1$. At
  $\varsigma=-1$ the square root of the quotient is
  $(1+e^{-\rho_i})/(1+e^{-\tau_i})\le e^{|\rho_i|+\tau_i}$; at $\varsigma=1$
  it is
  $|1-e^{-\rho_i}|/(1-e^{-\tau_i})\le|\rho_i|e^{|\rho_i|}/(\tau_ie^{-\tau_i})
  \le e^{|\rho_i|+\tau_i}$, since $\tau_i\ge2|\rho_i|$. Multiply over $i$,
  using $\sum_i|\rho_i|\le T$ and $\sum_i\tau_i\le4T+2\sigma K$.
  The function $\Cf/\Df$ is holomorphic on $|z|<e^\sigma$, so Parseval
  gives $\sum_j|\phi_j|^2\le e^{2(5T+2\sigma K)}$, and Cauchy--Schwarz bounds
  $X_k$.

  (b) Write $\Bt f=\Df^\sharp\Gamma$ with $\deg\Gamma<r-K$. By
  \Cref{lem:binomial-transform}(b) and (d),
  \[
    \Sg_f(z)\frac{\Cf(z)}{\Df(z)}
    =(1-z)^{n-K}\Gamma(x(z))\Cf(z)
    =(1-z)^n(\Cf^\sharp\Gamma)(x(z))=\Sg_{f'}(z),
  \]
  where $f'$ is the test with $\Bt f'=\Cf^\sharp\Gamma$, of degree below
  $r$. Comparing coefficients,
  $\langle f,X\rangle=\sum_{m=0}^nZ_m[z^m](\Sg_f\Cf/\Df)
  =\langle f',Z\rangle=0$.

  (c) The retained roots have modulus at most $e^\eta<e^\sigma$, so the
  series converges absolutely. On $y_{{\rm ret},k}=k^p\lambda^k$ the
  operator gives $[\vartheta^p((\Cf/\Df)(\nu)\nu^k)]_{\nu=\lambda}$, which is
  $(\Cf/\Df)(\lambda)k^p\lambda^k$ plus terms $k^{p'}\lambda^k$ with $p'<p$,
  and $(\Cf/\Df)(\lambda)\ne0$ for retained $\lambda$. So the operator is
  invertible on the solutions of the retained factor of $P$, and $v$ has
  the same minimal polynomial as $y_{\rm ret}$, which divides $P$ and hence
  $A$. Next, $\Df(\shift)v=\Cf(\shift)y_{\rm ret}$, and $\Cf(\shift)$
  annihilates the component of $y$ on the removed roots.

  (d) For $0\le k\le n-K$, collecting the coefficient of each $Z_{k+m}$
  gives $(\Df(\shift)X)_k=(\Cf(\shift)Z)_k$. With (c),
  $(\Df(\shift)(X-v))_k=(\Cf(\shift)(Z-y))_k$, which vanishes when
  $k,\ldots,k+K\in I$. So $X-v$ satisfies, on $I$, the recurrence of
  order $K$ with polynomial $\Df$, and $\Df(0)=1$. It extends to a unique
  two-sided solution $\omega$.

  (e) Since $\Cf(0)=1$ and $Z=y$ at $t+1,\ldots,t+K$, the identity in
  (d) at $k=t$ gives $(X-v)_t+\sum_{i=1}^K[z^i]\Df\cdot\omega_{t+i}=\mathfrak e$.
  Since $\Df(0)=1$ and $\Df(\shift)\omega=0$, the sum equals $-\omega_t$.
\end{proof}

\subsection{Model spaces and an annihilator close to one}
The outward modes $\omega$ are separated from $v$ by a polynomial that
annihilates $\omega$ and is close to $1$. We use the
Takenaka--Malmquist--Walsh basis of a finite-dimensional model space
\cite[Proposition 3.15 and Section 5.4]{GR15}.

\begin{lemma}[Model-space tail]\label{lem:model-space-tail}
  Let $K\ge1$ and $1/4\le q<1$, let $\Df_1(z)=\prod_{j=1}^K(1-\overline{a_j}z)$
  with distinct $a_j$ of modulus at most $q$, and let $f=\varrho_1/\Df_1$
  with $\deg\varrho_1<K$.
  Then
  \[
    \sum_{j\ge0}|[z^j]f|^2q^{-j}\le16^K\|f\|_2^2,\qquad
    \sum_{j\ge L}|[z^j]f|^2\le16^Kq^L\|f\|_2^2\quad(L\ge0).
  \]
\end{lemma}
\begin{proof}
  The functions
  $\mathfrak b_j(z)=\sqrt{1-|a_j|^2}\,(1-\overline{a_j}z)^{-1}
  \prod_{i<j}(z-a_i)/(1-\overline{a_i}z)$, $1\le j\le K$, are an
  orthonormal basis of the space of such $f$ in $\|\cdot\|_2$, by the
  cited model-space representation and basis formula. Put $\varrho=q^{-1/2}$. For $|a|\le q$,
  $\|\sqrt{1-|a|^2}(1-\overline a\varrho z)^{-1}\|_2^2
  =(1-|a|^2)/(1-|a|^2/q)\le1+q\le2$, the middle expression being
  increasing in $|a|$. On $|z|=\varrho$ the Blaschke factor
  $(z-a)/(1-\overline az)$ has modulus at most
  $(\varrho-|a|)/(1-\varrho|a|)\le(\varrho-q)/(1-\varrho q)
  =q^{-1/2}+1+q^{1/2}<4$. Hence
  $\|\mathfrak b_j(\varrho\,\cdot)\|_2^2\le2\cdot16^{j-1}$, and expanding $f$ in the
  basis with Cauchy--Schwarz gives
  $\|f(\varrho\,\cdot)\|_2^2\le16^K\|f\|_2^2$. This is the first bound; the
  second follows by dropping the first $L$ terms.
\end{proof}

\begin{lemma}[Completion]\label{lem:divisible-completion}
  Let $\Df_1$ be as in \Cref{lem:model-space-tail}, with every $a_j\ne0$, so
  that $\deg\Df_1=K$. Let $N\ge K$ be an integer such that
  $\epsilon_N:=\sqrt{2N}\,4^Kq^{N/4}\le1/8$.
  There is a polynomial $\Theta$ with
  \[
    \Df_1\mid\Theta,\qquad\deg\Theta\le N,\qquad\Theta(0)=1,\qquad
    \sum_{j=1}^N|[z^j]\Theta|\,q^{-j/4}\le\epsilon_N.
  \]
  Its reciprocal satisfies
  $\sum_{j\ge0}|[z^j]\Theta^{-1}|q^{-j/4}\le8/7$.
\end{lemma}
\begin{proof}
  Put $\mathsf a_N=16^Kq^N\le\epsilon_N^2/(2N)\le1/4$.
  In the space $\mathcal P_N$ of polynomials of degree at most $N$, with
  the coefficient inner product, let $\mathsf V$ be the subspace of multiples of
  $\Df_1$ and $\mathsf W=\mathsf V^\perp$, of dimension $K$. Let
  $\iota g(z)=z^N\overline{g(1/\overline z)}$ and let $\Pi_N$ truncate a
  power series after degree $N$. We claim that $\mathsf W$ consists of
  the $\iota\Pi_Nf$ with $f$ as in \Cref{lem:model-space-tail}. Put
  $\Df_1^*(z)=\prod_j(z-a_j)$. On $\T$,
  $\overline{\Df_1^*(z)}=\overline z^K\Df_1(z)$, so for every polynomial $p$
  the inner product of $f=\varrho_1/\Df_1$ with $\Df_1^*p$ is that of
  $\varrho_1$ with $z^Kp$, which is zero. When $\deg(\Df_1^*p)\le N$,
  truncation does not change this inner product, and
  $\iota(\Df_1^*p)=\Df_1\cdot z^{N-K}\overline{p(1/\overline z)}$ runs over
  $\mathsf V$ as $p$ runs over $\mathcal P_{N-K}$. Since $\iota$ is an
  antiunitary involution, $\iota\Pi_Nf\perp\mathsf V$. The map
  $f\mapsto\Pi_Nf$ is injective, because $f=O(z^{N+1})$ would force
  $\varrho_1=\Df_1f=0$, so the dimensions agree.

  For $g=\iota\Pi_Nf\in\mathsf W$ we have $|g(0)|=|[z^N]f|$ and
  $\|g\|_2=\|\Pi_Nf\|_2$. By \Cref{lem:model-space-tail},
  $|[z^N]f|^2\le \mathsf a_N\|f\|_2^2$ and
  $\|f\|_2^2\le\|\Pi_Nf\|_2^2+q\mathsf a_N\|f\|_2^2$, so
  $|g(0)|^2\le\frac43\mathsf a_N\|g\|_2^2$. Let $g_*$ be the orthogonal projection of
  $1$ onto $\mathsf W$ and $t_*=\|g_*\|_2^2=g_*(0)$. Then
  $t_*^2\le\frac43\mathsf a_Nt_*$, so $t_*\le\frac43\mathsf a_N\le\frac13$. The polynomial
  $\Theta=(1-g_*)/(1-t_*)$ lies in $\mathsf V$, has $\Theta(0)=1$, and
  $\|\Theta-1\|_2^2=t_*/(1-t_*)\le2\mathsf a_N$. Cauchy--Schwarz and
  $q^{-j/4}\le q^{-N/4}$ give the bound $\epsilon_N$. The last claim is the
  geometric series in the weighted $\ell^1$ algebra.
\end{proof}

\subsection{Separating the outward modes}

\begin{proposition}[Separation]\label{prop:boundary-mode-separation}
  Let $B\ge1$, and let $Y=v+\omega$ on $1,\ldots,B$, where $v$ and $\omega$
  are two-sided sequences, $|Y_k|\le H$ there with $H\ge1$, the minimal
  polynomial of $v$ has roots of modulus in $[e^{-\eta},e^\eta]$, and
  $\Df(\shift)\omega=0$ for a polynomial $\Df$ of degree $K\ge1$, with
  $\Df(0)=1$ and distinct roots, all of modulus at least $e^\sigma$, where
  $0<\sigma\le1$. Suppose $v$ has an integer annihilator $\widehat A$
  of degree $a\ge1$ with $\widehat A(0)\ne0$, and put
  $h=\max(1,\log\|\widehat A\|_1)$. Assume that, with
  $N=\lceil(4K\log4+2\log(2B)+4\log(12H)+12)/\sigma\rceil$,
  \begin{equation}\label{eq:separation-conditions}
    \eta\le\frac\sigma{32},\qquad N+a\le B,\qquad
    16^Ke^{-\sigma B}\le\frac12 .
  \end{equation}
  Then, with
  $V=2He^h\sqrt{a(N+2)}\exp(\sigma(N+1)/8+64h/\sigma)$,
  \[
    |v_k|\le V\quad(0\le k\le B),\qquad
    |\omega_k|\le2V\sqrt{2B}\,4^Ke^{-\sigma(B-k)/2}\quad(k\le B).
  \]
\end{proposition}
\begin{proof}
  Write $\Df(z)=\prod_{j=1}^K(1-\overline{\mathfrak a_j}z)$, with distinct
  $\mathfrak a_j$ of modulus at most $q:=e^{-\sigma}$; note $q\ge1/4$. With $\epsilon=1/(12H)\le1/8$,
  the choice of $N$ gives $\sigma N\ge4K\log4+2\log(2N)+4\log(1/\epsilon)$,
  which is equivalent to $\epsilon_N\le\epsilon$; it also gives $N\ge K$.
  Let $\Theta$ be the polynomial of
  \Cref{lem:divisible-completion}. Then $\Theta(\shift)\omega=0$, so
  $\tilde v:=\Theta(\shift)v=\Theta(\shift)Y$ on $1,\ldots,U$, where $U=B-N\ge a$, and
  $|\tilde v_k|\le(1+\epsilon)H$ there, since $\|\Theta\|_1\le1+\epsilon$. The
  sequence $\tilde v$ is a
  solution of the recurrence of $v$, so \Cref{lem:recurrence-prediction} bounds it at
  distance $b\ge1$ from $1,\ldots,U$ by
  $(1+\epsilon)He^h\sqrt{a(1+b)}\exp(\eta b+2\sqrt{2hb})$; this bound also
  holds on $1,\ldots,U$.

  Write $\Theta^{-1}=\sum_j\varkappa_jz^j$, so that
  $\sum_j|\varkappa_j|e^{\sigma j/4}\le8/7$ by \Cref{lem:divisible-completion}. Since
  the roots of $v$ have modulus at most $e^\eta<e^{\sigma/4}$, we have
  $v_k=\sum_{j\ge0}\varkappa_j\tilde v_{k+j}$: on $v_k=k^p\lambda^k$ this is
  $[\vartheta^p(\Theta^{-1}(\nu)\Theta(\nu)\nu^k)]_{\nu=\lambda}$. For
  $0\le k\le B$ the index $k+j$ has distance at most $N+j+1$ from
  $1,\ldots,U$. Using $\sqrt{N+j+2}\le\sqrt{N+2}\,e^{\sigma j/16}$, which
  holds since $\sigma(N+2)\ge8$,
  $2\sqrt{2h(N+j+1)}\le\sigma(N+1+j)/16+64h/\sigma$, and
  $\eta\le\sigma/32$, the factors depending on $j$ total at most
  $e^{5\sigma j/32}$. Summing against $|\varkappa_j|$ gives
  $|v_k|\le\frac87(1+\epsilon)He^h\sqrt{a(N+2)}
  e^{3\sigma(N+1)/32+64h/\sigma}\le V$.

  Hence $|\omega_k|\le H+V\le2V$ for $1\le k\le B$. The reversed sequence
  $x_j=\omega_{B-j}$ satisfies a recurrence of order $K$ whose roots
  $\overline{\mathfrak a_j}$ have modulus at most $q$; its generating function is a
  function $f$ as in \Cref{lem:model-space-tail}. By the second bound there,
  $\sum_{j\ge B}|x_j|^2\le\frac12\|f\|_2^2$, so
  $\|f\|_2^2\le2B(2V)^2$; the first bound gives
  $|x_j|\le4^Kq^{j/2}\|f\|_2$. This is the bound for $\omega_{B-j}$, $j\ge0$.
\end{proof}

\subsection{Heights on a short agreement interval}
Put $c_\eta=\min(1,c_1)/400$, and call a threshold $\eta$
\emph{admissible} if
\begin{equation}\label{eq:admissible-threshold}
  \frac{c_\eta}M\le\eta\le\frac\kappa8 .
\end{equation}
Since $\kappa/8=\sigma/1024$, an admissible threshold satisfies the
condition $\eta\le\sigma/8$ of \Cref{def:rational-filter}. For a filter at an
admissible threshold, let
\[
  E_k=X_k-v_k\qquad(0\le k\le n),
\]
so that $E=\omega$ on $I$. We call $E$ the \emph{filtered defects}. By a
\emph{height} we mean the logarithm of an a priori bound; the next
proposition shows that all heights are $O(M\ell)$. With
$R_v$ we denote the rational generating function of the retained
recurrence at the center,
\begin{equation}\label{eq:retained-generating-function}
  R_v(z)=\sum_{j\ge0}v_{\bar n+j}z^j=-\sum_{j\ge1}v_{\bar n-j}z^{-j}.
\end{equation}
The first series converges inside all poles and the second outside them;
both equal the same proper rational function, with poles at the
reciprocals of the retained roots. (For $v_k=k^p\lambda^k$ both are
$(\lambda\partial_\lambda)^p[\lambda^{\bar n}/(1-\lambda z)]$.)

\begin{proposition}[Heights]\label{prop:filtered-defect-heights}
  There are absolute constants $C_K,C_E$ such that the following holds for
  large $\Lambda$. Let $u\le n/128$, let $\eta$ be admissible, consider
  the filter at threshold $\eta$, and put $H_0=C_EM\ell/w$.
  \begin{enumerate}
    \item[(a)] $K\le T/\eta\le C_KM\ell/(\Lambda w)$, in particular $K<u-1$,
      and $|X_k|\le e^\ell$ for $0\le k\le n$.
    \item[(b)] $|v_k|\le e^{H_0}$ for $t\le k\le n-t-1$.
    \item[(c)] For $1\le s<u$ and $0\le s'\le t$,
      \begin{equation}\label{eq:defect-envelopes}
        |E_{n-t-s}|\le e^{H_0-64\kappa s},\qquad
        |E_{t-s'}|,\ |E_{n-t+s'}|\le e^{H_0+\kappa s'/2}.
      \end{equation}
    \item[(d)] With $q_t=t/(n-t)$, each of the three sums
      \begin{equation}\label{eq:defect-weighted-sums}
        \sum_{k\le t}q_t^{\,t-k}|E_k|,\qquad
        \sum_{k\ge n-t}q_t^{\,k-(n-t)}|E_k|,\qquad
        \sum_{t<k<n-t}q_t^{\,k-(n-t)}|E_k|
      \end{equation}
      is at most $e^{H_0}/\kappa$; here $q_t\le e^{-2\kappa}$ and
      $q_t^{-1}\le e^{2.01\kappa}$.
    \item[(e)] Under \textup{(D)}, $E_t=\mathfrak e+\omega_t$ with
      $|\omega_t|\le e^{H_0-64S}$.
    \item[(f)] If $4\eta\le\tau\le1/16$, then $|R_v(z)|\le e^{C_EM\ell}$ on
      $|z|=e^{2\tau}$.
  \end{enumerate}
\end{proposition}
\begin{proof}
  Throughout we use \Cref{prop:standing-bounds}, \eqref{eq:agreement-parameter-ranges}, and
  $w=u/u_0\le n/u_0\le4M/c_1$, so that $\ell\le CM\ell/w$ and
  $\log(1/\kappa)\le C\ell$.

  (a) By \Cref{lem:filter-properties}(a) and $\eta\ge c_\eta/M$,
  $K\le T/\eta\le C_TM^3\ell^2/(c_\eta u)$, and $u=wc_1\Lambda M^2\ell$. Also
  $\sigma K\le128\kappa TM/c_\eta\le128C_TM^3\ell^2/(c_\eta n)
  \le128C_T\ell/(c_\eta\Lambda)$ and $T\le C_T\ell/(c_1\Lambda)$, so
  $5T+2\sigma K\le\ell/2$ for large $\Lambda$, while
  $\sqrt{n+1}\le e^{\ell/2}$.

  (b), (e) If $K=0$, then $X=Z$, $v=y$ and $\omega=0$. Then (e) is
  \Cref{lem:filter-properties}(e), and (b) holds with $V=e^{2h_A+3}\sqrt{d_A}$, because
  $|y|=1$ on $I$ and, by \Cref{lem:recurrence-prediction} at distance one,
  $|y_t|\le V$; then $\log V\le C\ell$. Otherwise apply \Cref{prop:boundary-mode-separation} to the block
  $I$, numbered $1,\ldots,B$ with $B=u-1$ and index $j$ standing for
  $t+j$, with $Y=X$, $H=e^\ell$, $a=d_A$, $h=h_A$, and the polynomial $A$
  of \eqref{eq:short-annihilator}, which satisfies the hypotheses of
  \Cref{lem:recurrence-prediction} by \Cref{lem:filter-properties}(c). The conditions
  \eqref{eq:separation-conditions} hold for large $\Lambda$: $\eta\le\kappa/8$;
  $\sigma N\le C(K+\ell)$ while $\sigma B\ge64S$ and
  $S\ge\frac{c_1^2}4\Lambda\ell M$; and $a<2d\le u/4$. The
  index $j=0$ is $t$, and $B-j=s-1$ at the index $n-t-s$. So (b) holds with
  $\log V\le C(\ell+K+h_An/u)\le CM\ell/w$, the first bound in (c) holds,
  and (e) follows from \Cref{lem:filter-properties}(e) with
  $e^{-\sigma B/2}=e^{-64S+64\kappa}$.

  (c) It remains to bound $E=X-v$ outside $I$. By (b), $|v|\le V$ on the
  $u\ge d_A$ integers $t,\ldots,n-t-1$. \Cref{lem:recurrence-prediction}, with
  $\eta\le\kappa/8$, $2\sqrt{2h_Ab}\le\kappa b/8+16h_A/\kappa$ and
  $\sqrt b\le\sqrt{2/\kappa}\,e^{\kappa b/4}$, bounds $v$ at distance $b$ by
  $V\sqrt{2d_A/\kappa}\,e^{h_A+16h_A/\kappa}e^{\kappa b/2}$. The index
  $t-s'$ has distance $s'$, and $n-t+s'$ has distance $s'+1$. With
  $|X|\le e^\ell$ this proves the second bound in (c).

  (d) Since $q_t=(1-\kappa)/(1+\kappa)$ and $\kappa\le1/128$, we have
  $q_t\le e^{-2\kappa}$ and $q_t^{-1}\le e^{2.01\kappa}$. By (c) the three
  sums are geometric series of ratio at most $e^{-3\kappa/2}$, $e^{-3\kappa/2}$
  and $e^{-61\kappa}$, and $1/(1-e^{-3\kappa/2})\le1/\kappa$.

  (f) By (b) and \Cref{lem:recurrence-prediction}, for $j\ge1$,
  $|v_{\bar n-j}|\le V\sqrt{d_Aj}\,e^{h_A+\eta j+2\sqrt{2h_Aj}}$, whether or not
  $\bar n-j\ge t$. On $|z|=e^{2\tau}$ use the second series in
  \eqref{eq:retained-generating-function}, $\eta\le\tau/4$,
  $2\sqrt{2h_Aj}\le\tau j/2+4h_A/\tau$ and
  $\sum_{j\ge1}je^{-\tau j}=(4\sinh^2(\tau/2))^{-1}\le\tau^{-2}$:
  \[
    |R_v(z)|\le V\sqrt{d_A}\,\tau^{-2}e^{h_A+4h_A/\tau}.
  \]
  Since $\tau\ge4\eta\ge4c_\eta/M$, we have $4h_A/\tau\le CM\ell$ and
  $\log(1/\tau)\le C\ell$.
\end{proof}

\section{Transforms on circles and the far sector}\label{sec:circles-and-far-sector}
Tests are estimated only through their transforms, apart from one
explicit polynomial in \Cref{sec:endpoint-test}. This section provides the two
tools for this: coefficient bounds from a single circle, and a contour
bound for the part of a pairing that comes from roots outside a sector.

\subsection{One circle and one monotone weight}
By \Cref{lem:binomial-transform}(b), the coefficients of a test are the
coefficients of $(1-z)^n(\Bt f)(x(z))$, which Cauchy's estimate reads off
any circle $|z|=\varrho\ne1$. On such a circle the parallelogram law
$|1-z|^2+|1+z|^2=2(1+\varrho^2)$ gives
\begin{equation}\label{eq:parallelogram-identity}
  |1-z|^2\Bigl(1+\frac{|x(z)|^2}{n^2}\Bigr)=2(1+\varrho^2).
\end{equation}
Let $x_{-\varrho}=x(-\varrho)=n(1-\varrho)/(1+\varrho)$ be the coordinate of
the point $-\varrho$ of the circle. Since
$x(z)-x_{-\varrho}=2n(z+\varrho)/((1-z)(1+\varrho))$ and
$\varrho|1-z|^2+|z+\varrho|^2=\varrho(1+\varrho)^2$,
\begin{equation}\label{eq:circle-gaussian}
  \frac{|1-z|^2}{(1+\varrho)^2}
  =\frac{n^2-x_{-\varrho}^2}{n^2-x_{-\varrho}^2+|x(z)-x_{-\varrho}|^2}.
\end{equation}
In particular $|1-z|\le1+\varrho$ on the circle, and by
\eqref{eq:parallelogram-identity} the point $x_{-\varrho}$ has the smallest modulus
of $x(z)$ on the circle.

\begin{lemma}[One monotone weight]\label{lem:monotone-weight}
  Let $D\ge0$ be an integer, and let $G$ be a polynomial of degree at most $D$ with $|G(x)|\le\Gamma$ for all
  complex $|x|\le x_0$, where $0<x_0\le n$ and $D\le x_0^2/(2n)$. Then for every
  $z\ne1$,
  \begin{equation}\label{eq:monotone-weight-bound}
    \bigl|(1-z)^nG(x(z))\bigr|\le\Gamma\bigl(2(1+|z|^2)\bigr)^{n/2}
    \mathcal W(|x(z)|),\qquad
    \mathcal W(X)=\Bigl(1+\frac{X^2}{n^2}\Bigr)^{-n/2}
    \max\Bigl(1,\frac X{x_0}\Bigr)^D,
  \end{equation}
  and $\mathcal W$ is nonincreasing on $X\ge0$. In particular, on a
  circle $|z|=\varrho$ with $\varrho>0$ and $\varrho\ne1$, the left side is at most
  $\Gamma(1+\varrho)^n\max(1,|x_{-\varrho}|/x_0)^D$.
\end{lemma}
\begin{proof}
  The Bernstein--Walsh lemma gives $|G(x)|\le\Gamma\max(1,|x|/x_0)^D$, and
  \eqref{eq:parallelogram-identity} evaluates $|1-z|^n$. On $X\ge x_0$ the
  logarithmic derivative of $\mathcal W$ is
  $D/X-nX/(n^2+X^2)\le0$, because $nX^2/(n^2+X^2)\ge x_0^2/(2n)\ge D$
  there. On the circle, $|x(z)|\ge|x_{-\varrho}|$ and
  $(2(1+\varrho^2))^{1/2}(1+x_{-\varrho}^2/n^2)^{-1/2}=1+\varrho$.
\end{proof}

For $1\le t<n/2$, put $q_t=t/(n-t)$. The terms of
$(1+q_t)^n=\sum_k\binom nkq_t^k$ increase up to $k=t$ and decrease after
it, so
\begin{equation}\label{eq:saddle-point-bounds}
  (1+q_t)^n\le(n+1)\binom ntq_t^{\,t},\qquad
  (1+q_t^{-1})^n\le(n+1)\binom ntq_t^{-(n-t)},
\end{equation}
and $x_{-q_t}=u$, $x_{-1/q_t}=-u$, where $u=n-2t$.

\subsection{The centered contour identity}
Let $v$ be a two-sided sequence whose minimal polynomial has all roots of
drift at most $\eta$, and let $R_v$ be as in \eqref{eq:retained-generating-function}.
For a test $f$, put $\widetilde\Sg_f(z)=z^{\bar n}\Sg_f(1/z)$, a Laurent
polynomial.

\begin{lemma}[Contour identity]\label{lem:contour-identity}
  Let $\tau>\eta$ and $\Phi(z)=R_v(z)\widetilde\Sg_f(z)/z$, where $R_v$ is given
  by the first series of \eqref{eq:retained-generating-function} on
  $|z|=e^{-\tau}$ and by the second on $|z|=e^\tau$. Then
  \[
    \langle f,v\rangle
    =\frac1{2\pi i}\oint_{|z|=e^{-\tau}}\Phi(z)\,dz
    -\frac1{2\pi i}\oint_{|z|=e^{\tau}}\Phi(z)\,dz,
  \]
  with both circles positively oriented.
\end{lemma}
\begin{proof}
  Write $\gamma_k=[z^k]\Sg_f$, so that
  $\widetilde\Sg_f(z)=\sum_k\gamma_kz^{\bar n-k}$. On $|z|=e^{-\tau}$ use the
  first series in \eqref{eq:retained-generating-function}: the integral is the
  constant term of the product, $\sum_{j\ge0}v_{\bar n+j}\gamma_{\bar n+j}$. On
  $|z|=e^\tau$ use the second: the integral is
  $-\sum_{j\ge1}v_{\bar n-j}\gamma_{\bar n-j}$. Their difference is
  $\langle f,v\rangle$.
\end{proof}

\subsection{Averaged rays}
The contour of \Cref{lem:far-sector} below has two radial sides, and
the rational function $R_v$ may be large near its poles. We choose the
angle of the sides by averaging the logarithmic potential of the poles.

\begin{lemma}[Averaged rays]\label{lem:averaged-rays}
  There is an absolute constant $C_G$ with the following property.
  Let $0<\tau\le1/16$, and let $b_1,\ldots,b_N$ be nonzero complex numbers,
  listed with repetition, with $|\log|b_j||\le\tau/4$. Suppose that for
  every $x\in\T$ and $s>0$ at most $As+B$ of the $b_j/|b_j|$ lie within
  angular distance $s$ of $x$. Put $R_0=e^{2\tau}$ and
  \[
    G(z)=\sum_{j=1}^Ng(z,b_j),\qquad
    g(z,b)=\log\Bigl|\frac{R_0^2-\overline bz}{R_0(z-b)}\Bigr| .
  \]
  Let $\mathcal J=[a_0,a_0+4\tau]\subset(0,\pi)$.
  There is $\alpha\in\mathcal J$ such that, for $|\rho|\le\tau$,
  $G(-e^{\rho\pm i\alpha})\le C_G(A\tau+B)$ and $-e^{\rho\pm i\alpha}$ differs from
  every $b_j$.
  The same bound holds on $|z|=e^{\pm\tau}$.
\end{lemma}
\begin{proof}
  Expanding $|R_0^2-\overline bz|^2-R_0^2|z-b|^2$ gives the exact identity
  \[
    e^{2g(z,b)}=1+\frac{(R_0^2-|b|^2)(R_0^2-|z|^2)}{R_0^2|z-b|^2}.
  \]
  On $e^{-\tau}\le|z|\le e^\tau$ the fraction in this identity is
  nonnegative and its numerator is $O(\tau^2)$. If $\theta$ is the
  angular distance between $z/|z|$ and $b/|b|$, then
  \[
    |z-b|^2=(|z|-|b|)^2+2|zb|(1-\cos\theta)\ge c\theta^2.
  \]
  Consequently, uniformly for $|\rho|\le\tau$,
  \[
    g(-e^{\rho\pm i\alpha},b)\le f_\tau(\theta)
    :=\tfrac12\log\bigl(1+C\tau^2/\theta^2\bigr).
  \]
  Here $\int_{\R}f_\tau(|x|)\,dx=O(\tau)$, by scaling, and
  $f_\tau(\theta)\le C\tau^2/\theta^2$ away from zero.

  Average this majorant over $\alpha\in\mathcal J$, first for the rays
  $-e^{i\alpha}$. The projected poles within angular distance $8\tau$
  of the midpoint of this arc number at most $8A\tau+B$. Each contributes
  $O(1)$ on average, by the integral bound. A projected pole at distance
  $q>8\tau$ from the midpoint is at distance at least $3q/4$ from every
  ray. The bands $2^{k+3}\tau<q\le2^{k+4}\tau$ therefore contribute at most
  \[
    C\sum_{k\ge0}4^{-k}(2^kA\tau+B)=O(A\tau+B).
  \]
  The same argument applies to $-e^{-i\alpha}$. Choose $\alpha$ for which
  the sum of both majorants is at most its average. Since these majorants
  are uniform in $\rho$, this bounds both radial segments. As $g(z,b)=+\infty$
  at $z=b$, the two rays avoid every $b_j$.

  On $|z|=e^{\pm\tau}$ the radial distance from every pole is at least
  $c\tau$. Thus $g(z,b)\le C\tau^2/(\tau^2+\theta^2)$. Count the poles
  with $\theta\le\tau$ directly and sum the bands
  $2^k\tau<\theta\le2^{k+1}\tau$ as above to obtain the circle bound.
\end{proof}

\subsection{The far sector}\label{sec:far-sector}
In the rest of this section, $u\le n/128$, $v$ is the retained recurrence
of a filter at an admissible threshold $\eta$, and
$v_k=\sum_\lambda p_\lambda(k)\lambda^k$ over the retained roots. For a
polynomial $G$ and $x_0>0$ write
$\|G\|_{x_0}=\max\{|G(x)|:x\in\C,\ |x|\le x_0\}$.

\begin{lemma}[Far sector]\label{lem:far-sector}
  There is an absolute constant $C_F$ such that the following holds for
  large $\Lambda$. Let $4\eta\le\tau$, $4\tau\le\alpha_1$ and
  $\alpha_1+4\tau\le1/16$. There is $\alpha\in[\alpha_1,\alpha_1+4\tau]$,
  not depending on $f$, such that no retained root has angle exactly
  $\alpha$ and every test $f$ of degree $D\le n\alpha_1^2/2$ satisfies, with
  $x_0=\frac43n\alpha$ and $v_{\rm near}$ the component of $v$ on the retained
  roots of angle below $\alpha$,
  \begin{equation}\label{eq:far-sector-expansion}
    \langle f,v\rangle=\langle f,v_{\rm near}\rangle+\mathcal E,
    \qquad
    |\mathcal E|\le(2\cos(\alpha/2))^n\,\|\Bt f\|_{x_0}\,
    e^{n\tau^2/2+C_FM^2\ell\tau}.
  \end{equation}
\end{lemma}
By \eqref{eq:spectral-pairing},
$\langle f,v_{\rm near}\rangle=\sum_{\lambda:\ {\rm angle}<\alpha}
[p_\lambda(\vartheta)\Sg_f]_{\nu=\lambda}$.
\begin{proof}
  The poles of $R_v$ are the reciprocals $b_j$ of the retained roots, with
  multiplicity; $|\log|b_j||\le\eta\le\tau/4$, and
  \Cref{prop:standing-bounds}(c) gives the angular count with $A=A_\theta$,
  $B=B_\theta$. Let $\alpha$ be given by \Cref{lem:averaged-rays} for the
  interval $[\alpha_1,\alpha_1+4\tau]$; by that lemma, no pole lies on the two
  rays. Let
  $\Sigma_{\rm far}=\{-e^{\rho'+i\theta'}:|\rho'|\le\tau,\ \alpha\le|\theta'|\le\pi\}$
  be the far annular sector. The point $1/\lambda$ has the same angle as
  $\lambda$. Let $v_{\rm far}$ be the component of $v$ at the retained roots
  of angle above $\alpha$, with generating function $R_{\rm far}$; by
  \eqref{eq:spectral-pairing}, $\mathcal E=\langle f,v_{\rm far}\rangle$. The poles
  of $R_{\rm far}$ lie in the interior of $\Sigma_{\rm far}$, and $R_v-R_{\rm far}$
  has no pole in $\Sigma_{\rm far}$. In the coordinate $\zeta=\log(-z)$ the
  annulus $e^{-\tau}\le|z|\le e^\tau$ and $\Sigma_{\rm far}$ are rectangles,
  whose horizontal sides $\operatorname{Im}\zeta=\pm\pi$ are identified by
  periodicity. So \Cref{lem:contour-identity} for $v_{\rm far}$ and Cauchy's
  theorem, applied to $R_{\rm far}\widetilde\Sg_f/z$ off $\Sigma_{\rm far}$ and to
  $(R_v-R_{\rm far})\widetilde\Sg_f/z$ on $\Sigma_{\rm far}$, give
  $\mathcal E=-(2\pi i)^{-1}\oint_{\partial\Sigma_{\rm far}}\Phi(z)\,dz$ with
  $\partial\Sigma_{\rm far}$ positively oriented.

  Multiplying $R_v$ by the Blaschke product
  $\prod_jR_0(z-b_j)/(R_0^2-\overline{b_j}z)$, which has modulus one on
  $|z|=R_0$, and applying the maximum principle with
  \Cref{prop:filtered-defect-heights}(f), gives $|R_v(z)|\le e^{C_EM\ell+G(z)}$ for
  $|z|<R_0$. By \Cref{lem:averaged-rays},
  $|R_v|\le e^{C_EM\ell+C_G(A_\theta\tau+B_\theta)}$ on $\partial\Sigma_{\rm far}$.
  Since $\tau\ge4\eta\ge4c_\eta/M$, we have
  $M\ell\le M^2\ell\tau/(4c_\eta)$, so this is at most
  $\frac14e^{C_FM^2\ell\tau}$ for a suitable $C_F$. Also
  $\oint_{\partial\Sigma_{\rm far}}|dz|/|z|\le4\pi+4\tau$.

  Let $z=-e^{\rho'+i\theta'}\in\partial\Sigma_{\rm far}$ and $\nu=1/z$. Then
  $|\nu|=e^{-\rho'}$ and $\widetilde\Sg_f(z)=\nu^{-\bar n}(1-\nu)^n(\Bt f)(x(\nu))$,
  where $|x(\nu)|=n|\tanh((\rho'+i\theta')/2)|$. Apply
  \Cref{lem:monotone-weight} to $G=\Bt f$ with the radius $x_0$; its
  degree condition holds because $D\le n\alpha_1^2/2\le x_0^2/(2n)$. Since
  $|\nu|^{-\bar n}(2(1+|\nu|^2))^{n/2}=e^{\rho'(\bar n-n/2)}(4\cosh\rho')^{n/2}
  \le e^{\tau/2}(4\cosh\rho')^{n/2}$,
  \[
    |\widetilde\Sg_f(z)|\le e^{\tau/2}\|\Bt f\|_{x_0}(4\cosh\rho')^{n/2}
    \mathcal W(|x(\nu)|).
  \]
  For fixed $\rho'$, $|x(\nu)|$ increases with $|\theta'|$ by
  \Cref{lem:coordinates-near-minus-one}(b), and $\mathcal W$ is nonincreasing, so the right
  side is largest at $|\theta'|=\alpha$. There
  $|x(\nu)|\le0.5155n\alpha(1+(0.5155\alpha)^2)\le0.52n\alpha<x_0$, since
  $|\rho'+i\alpha|\le\alpha\sqrt{1+1/16}$, so
  $\mathcal W(|x(\nu)|)=(1+|x(\nu)|^2/n^2)^{-n/2}$, and
  $1+|\tanh((\rho'+i\alpha)/2)|^2=2\cosh\rho'/(\cosh\rho'+\cos\alpha)$ turns the right
  side into $e^{\tau/2}\|\Bt f\|_{x_0}s(\rho'+i\alpha)^n$. By
  \Cref{lem:angles-and-weights}(b), $|\widetilde\Sg_f|\le e^{\tau/2}\|\Bt f\|_{x_0}
  (2\cos(\alpha/2))^ne^{n\tau^2/2}$ on $\partial\Sigma_{\rm far}$. Multiply the
  three bounds and use $(4\pi+4\tau)e^{\tau/2}<8\pi$.
\end{proof}

\section{Counting the roots near \texorpdfstring{$-1$}{-1}}\label{sec:roots-near-minus-one}
Throughout this section $u\le n/128$.

\subsection{The root count}
\begin{figure}[htbp]
  \centering
  \begin{tikzpicture}[font=\small]
    \path[fill=gray!12,even odd rule]
    (0,0) circle (1.9) (0,0) circle (1.5);
    \path[fill=orange!20]
    (120:1.9) arc[start angle=120,end angle=240,radius=1.9]
    -- (240:1.5) arc[start angle=240,end angle=120,radius=1.5] -- cycle;
    \path[fill=linkblue!20]
    (155:1.9) arc[start angle=155,end angle=205,radius=1.9]
    -- (205:1.5) arc[start angle=205,end angle=155,radius=1.5] -- cycle;
    \draw[gray!45] (0,0) circle (1.5) (0,0) circle (1.9);
    \draw[gray,densely dotted] (0,0) circle (1.7);
    \draw[gray!55] (-2.15,0) -- (2.05,0);
    \foreach \a in {120,240}
    \draw[orange!65!black,dashed] (0,0) -- (\a:2.08);
    \foreach \a in {155,205}
    \draw[linkblue] (0,0) -- (\a:2.08);
    \node[anchor=east] at (-2.02,0.98) {$|\theta|=\theta_\Gamma$};
    \node[anchor=south] at (-1.05,1.94) {$|\theta|=\alpha$};
    \fill (0,0) circle (1pt);
    \node[anchor=north west] at (0.04,-0.04) {$0$};
    \node[anchor=north] at (-1.7,-0.04) {$-1$};
    \foreach \a/\rad in {166/1.76,179/1.83,199/1.61}
    \fill[linkblue] (\a:\rad) circle (1.8pt);
    \foreach \a/\rad in {136/1.62,147/1.79,222/1.65,234/1.81}
    \fill[orange!65!black] (\a:\rad) circle (1.8pt);
    \foreach \a/\rad in {42/1.75,87/1.59,280/1.79,322/1.63}
    \fill[gray!75!black] (\a:\rad) circle (1.8pt);
    \fill[linkblue!20] (2.5,1.25) rectangle (2.7,1.45);
    \node[anchor=west,align=left,text width=5.5cm] at (2.9,1.35)
    {\textbf{Inner:} $|\theta|<\theta_\Gamma$\\
    Recover the component by Hermite interpolation.};
    \fill[orange!20] (2.5,-0.1) rectangle (2.7,0.1);
    \node[anchor=west,align=left,text width=5.5cm] at (2.9,0)
    {\textbf{Outer zone:} $\theta_\Gamma<|\theta|<\alpha$\\
    Cancel by exact zeros of $\Bt f$.};
    \fill[gray!12] (2.5,-1.45) rectangle (2.7,-1.25);
    \node[anchor=west,align=left,text width=5.5cm] at (2.9,-1.35)
    {\textbf{Far:} $|\theta|\ge\alpha$\\
    Estimate by contour integration.};
  \end{tikzpicture}
  \caption{The three sectors for retained roots $\lambda=-e^{\rho+i\theta}$.
    Angles are measured from the negative real axis; the dotted circle is
    $|\lambda|=1$. Angular widths and radial drift are exaggerated.
  The separating rays contain no retained roots.}
  \label{fig:root-sectors}
\end{figure}
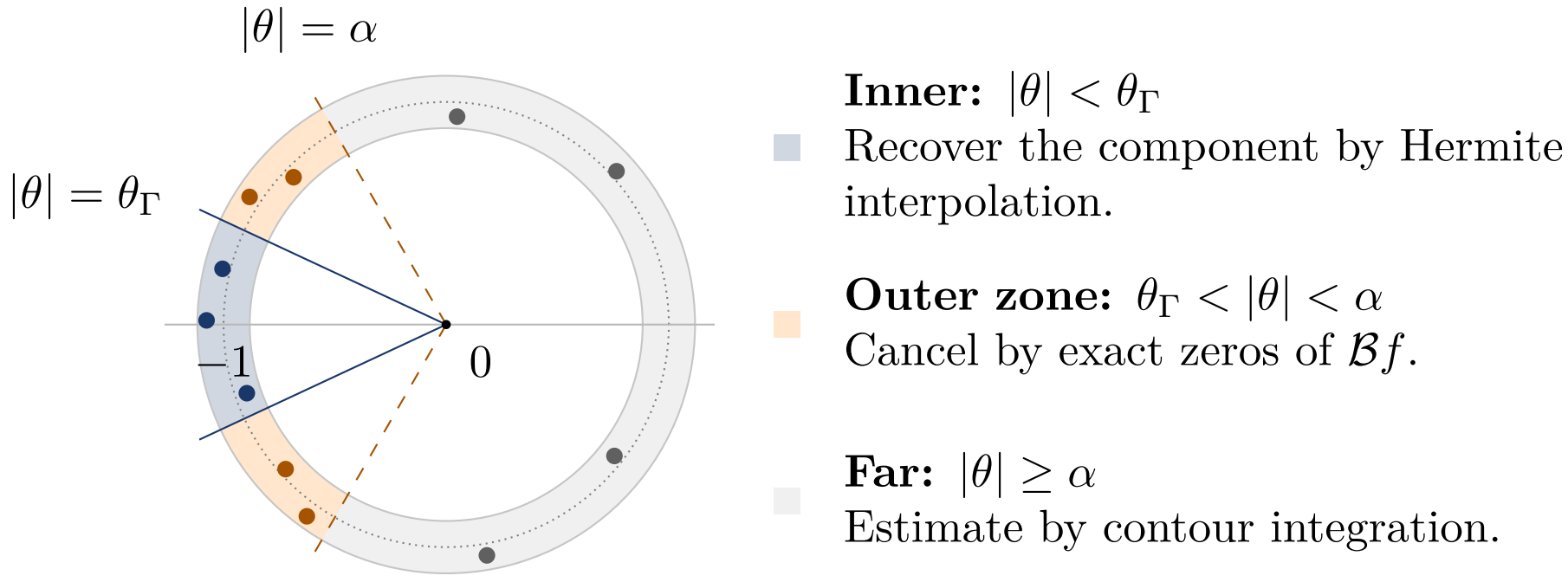

The proof splits the retained roots into three zones by angle:
\emph{inner} (below an angle $\theta_\Gamma$), the \emph{outer zone}
(between $\theta_\Gamma$ and a ray angle $\alpha$), and \emph{far} (at
least $\alpha$); see \Cref{fig:root-sectors}.
A test of degree below $r$ can reproduce the sequence
$\lambda^{\bar n+j}$ on the inner roots and vanish on the outer zone. Its
pairing with the far roots is small, because their binomial weight
$(2\cos(\alpha/2))^n$ is smaller than the inner weight
$(2\cos(\theta_\Gamma/2))^n$; the outer zone provides this gap. Its
pairing with the filtered word is zero, so its pairing with the
retained recurrence equals minus its pairing with the filtered defects
$E=X-v$. That is small: $E$ lives outside $I$, apart from outward modes
that are tiny away from the right end of $I$, and outside $I$ the weights
are at most $\binom nt\le(2\cos(\alpha/2))^n$. So the inner component of $y$ is
exponentially small at the center, and \Cref{cor:small-component-root-count}
bounds the number of inner roots. The interpolation costs the logarithm
in \eqref{eq:root-count-hypotheses}.

\begin{lemma}[Root count]\label{lem:root-count}
  There is an absolute constant $C_L$ such that the following holds for
  large $\Lambda$. Let $\tau\ge1/M$ and $16\tau\le\alpha_2\le\alpha_t$. Put
  $D_0=N_{\rm all}(\alpha_2)+8T/\tau$, and suppose
  \begin{equation}\label{eq:root-count-hypotheses}
    D_0<r\qquad\text{and}\qquad
    n\tau\alpha_2\ge C_LD_0\log\frac{\alpha_2}\tau .
  \end{equation}
  Then the number of roots of $P$ of angle at most $\alpha_2-16\tau$,
  with multiplicity, is at most
  \begin{equation}\label{eq:root-count-bound}
    \frac{4C_HM^2\ell^2}{n\tau\alpha_2}+\frac{8T}\tau .
  \end{equation}
\end{lemma}

By \eqref{eq:defect-angle-range}, $\alpha_2\le\alpha_t\le2.02\kappa<1/16$. The proof uses the
angles $\theta_1=\alpha_2-16\tau<\theta_2=\alpha_2-8\tau<\alpha_1=\alpha_2-4\tau$
and the gain
\begin{equation}\label{eq:root-count-gain}
  \mathfrak g=\frac n8(\alpha_1^2-\theta_2^2)-\frac{n\tau^2}2
  =n\tau\Bigl(\alpha_2-\frac{13}2\tau\Bigr)\ge\frac{n\tau\alpha_2}2 .
\end{equation}
Since $\log(\alpha_2/\tau)\ge\log16$ and $\alpha_2\ge16\tau$,
\eqref{eq:root-count-hypotheses} gives $2^{21}D_0\le n\alpha_2^2$ once
$C_L\ge2^{16}$.

\begin{proof}
  \emph{Step 0: a contour.} Let $\mathcal R_*$ be the roots of $P$ of
  angle below $\alpha_2$ and drift at most $2\alpha_2$, with multiplicity,
  and let $N_*=|\mathcal R_*|\le N_{\rm all}(\alpha_2)<r$. Write each root as
  $-e^{\zeta}$ with $|\operatorname{Im}\zeta|\le\pi$. For a real $b$ and an
  interval $\mathcal J$ of length $|\mathcal J|$,
  $\int_{\mathcal J}\log^+(|\mathcal J|/|s-b|)\,ds\le2|\mathcal J|$.
  Consequently, for $N_*$ real numbers $b_i$, some $s\in\mathcal J$
  satisfies $\sum_{i,\pm}\log^+(|\mathcal J|/|s\pm b_i|)\le4N_*$, and at
  that $s$ each product $\prod_i|s\pm b_i|$ is at least
  $(|\mathcal J|/64)^{N_*}$, since $e^4<64$.

  Apply this to the imaginary parts of the $\zeta$ of the roots in
  $\mathcal R_*$ on $\mathcal J=[\theta_1+\tau,\theta_2]$, of length $7\tau$, and
  to their real parts on $\mathcal J=[\tau/8,\tau/4]$, of length $\tau/8$.
  This chooses $\theta_\Gamma$ and $\eta_\Gamma$ such that, on the boundary
  $\Gamma_\zeta$ of the rectangle $|\operatorname{Re}\zeta|\le\eta_\Gamma$,
  $|\operatorname{Im}\zeta|\le\theta_\Gamma$,
  \begin{equation}\label{eq:rectangle-separation}
    \prod_{\lambda=-e^{\zeta_\lambda}\in\mathcal R_*}|\zeta-\zeta_\lambda|
    \ge h_*^{N_*},\qquad h_*=\frac\tau{512}.
  \end{equation}
  Let $\Gamma$ be the image of $\Gamma_\zeta$ under the coordinate map
  $\zeta\mapsto x_{-e^\zeta}=-n\tanh(\zeta/2)$. A number
  $\lambda=-e^\zeta\ne1$ has its coordinate inside $\Gamma$, respectively
  on $\Gamma$, exactly when $\zeta$ lies inside, respectively on,
  $\Gamma_\zeta$: the map $\zeta\mapsto-e^\zeta$ is injective on
  $-\pi<\operatorname{Im}\zeta\le\pi$, and $\lambda\mapsto x_\lambda$ is a
  M\"obius map.

  Filter at the threshold $\eta_\Gamma$. It is admissible, because
  $c_\eta/M\le\tau/8\le\eta_\Gamma\le\tau/4\le\alpha_2/64\le\kappa/8$ by
  \eqref{eq:defect-angle-range}. It removes $K\le T/\eta_\Gamma\le8T/\tau$ roots;
  call the others \emph{retained}. Let
  $\alpha\in[\alpha_1,\alpha_2]$ be the angle given by
  \Cref{lem:far-sector} with $\alpha_1$, $\tau$ and $\eta=\eta_\Gamma$;
  its hypotheses hold since $4\eta_\Gamma\le\tau$, $4\tau\le\alpha_1$ and
  $\alpha_2\le1/16$.

  Since $\eta_\Gamma\le\alpha/16$, every point of $\Gamma_\zeta$ has
  $|\zeta|\le1.01\alpha$, so \Cref{lem:coordinates-near-minus-one}(b) puts $\Gamma$ in
  $|x|\le0.52n\alpha$. No root of $P$ has its $\zeta$ on $\Gamma_\zeta$:
  the roots in $\mathcal R_*$ by \eqref{eq:rectangle-separation}, and every other root
  has angle at least $\alpha_2>\theta_\Gamma$ or drift above
  $2\alpha_2>\eta_\Gamma$. The retained roots of angle below $\theta_\Gamma$,
  the \emph{inner} roots, $k'$ of them with multiplicity, lie inside
  $\Gamma_\zeta$. The retained roots of angle between $\theta_\Gamma$ and
  $\alpha$, the \emph{outer zone} roots, the removed roots, and the moved
  poles $\mu_i$, of drift at least $\sigma>\eta_\Gamma$, lie outside. Let
  $m'$ be the number of retained roots of angle below $\alpha$. Then
  $m'+K\le D_0<r$, and $D_0\le2^{-20}n\alpha_1^2$ since
  $\alpha_1\ge3\alpha_2/4$. If $k'=0$, every root counted in
  \eqref{eq:root-count-bound} is removed, and the claim holds. Assume
  henceforth $k'>0$.

  \emph{Step 1: the test.} Fix an integer $j$ with $|j|<d$, and put
  $\Phi_j(\nu)=\nu^{\bar n+j}(\Df/\Cf)(\nu)$. We want a test whose generating
  function agrees with $\Phi_j$ at the inner roots, to full multiplicity.
  By \Cref{lem:binomial-transform}(b), the function of $x$ whose generating
  function is $\Phi_j$ is
  \[
    \Psi_j(x)=U_j(x)\frac{\Df(\lambda(x))}{\Cf(\lambda(x))},\qquad
    U_j(x)=\frac{(x-n)^{\bar n+j}(x+n)^{n-\bar n-j}}{(2n)^n}:
  \]
  indeed $(1-\nu)^nU_j(x(\nu))=\nu^{\bar n+j}$ and $\lambda(x(\nu))=\nu$. Here
  $0\le \bar n+j\le n$, so $U_j$ is a polynomial. By
  \Cref{lem:binomial-transform}(d), the factors $(1-z)^K$ cancel in the filter
  ratio:
  \begin{equation}\label{eq:filter-ratio}
    \frac{\Df(\lambda(x))}{\Cf(\lambda(x))}=\frac{\Df^\sharp(x)}{\Cf^\sharp(x)},
    \qquad\text{so}\qquad\Psi_j=\frac{U_j\Df^\sharp}{\Cf^\sharp}.
  \end{equation}
  Let $\Omega_\rmin$ and $\Omega_\rmout$ be the monic polynomials whose
  roots are the coordinates of the inner and of the outer zone roots, with
  multiplicity. The function
  $\Psi^\circ_j=\Psi_j/(\Df^\sharp\Omega_\rmout)=U_j/(\Cf^\sharp\Omega_\rmout)$ is
  holomorphic on and inside $\Gamma$, since the zeros of $\Cf^\sharp$ and
  $\Omega_\rmout$ are outside. Let $\varpi$ be its Hermite interpolant
  \eqref{eq:hermite-formula} at the zeros of $\Omega_\rmin$, of degree below $k'$,
  and put
  \[
    G_j=\Df^\sharp\Omega_\rmout\varpi,\qquad f_j=\Bt^{-1}G_j .
  \]
  Then $\deg G_j<K+m'\le D_0<r$ and $\Df^\sharp\mid G_j$. To full
  multiplicity, $G_j-\Psi_j$ vanishes at the inner coordinates, and $G_j$
  vanishes at the outer zone coordinates.

  \emph{Step 2: the exact identity.} By \Cref{lem:binomial-transform}(b),
  $\Sg_{f_j}(\nu)-\Phi_j(\nu)=(1-\nu)^n(G_j-\Psi_j)(x(\nu))$, which vanishes
  to full order at the inner roots; $\Sg_{f_j}$ vanishes to full order at
  the outer zone roots and at each $\mu_i$. For a sequence
  $a_k=\sum_\lambda p_\lambda(k)\lambda^k$ and a function $\Phi$
  holomorphic near its roots, write
  $[a,\Phi]=\sum_\lambda[p_\lambda(\vartheta)\Phi]_{\nu=\lambda}$,
  so that $[a,\nu^k]=a_k$ and, by \eqref{eq:spectral-pairing},
  $\langle f,a\rangle=[a,\Sg_f]$. Let $y_\rmin$ and $v_\rmin$ be the
  components of $y$ and $v$ on the inner roots. By the definition of $v$ in
  \Cref{lem:filter-properties}(c), $v_{\rmin,k}=\sum_{i\ge0}\phi_iy_{\rmin,k+i}$. Hence
  $[v_\rmin,\Phi]=[y_\rmin,(\Cf/\Df)\Phi]$: both
  sides are linear in $\Phi$ and depend only on finitely many derivatives of
  $\Phi$ at the inner roots, and they agree for $\Phi=\nu^k$, $k\ge0$. Hermite
  interpolation realizes every prescribed collection of these derivatives
  by a polynomial, so linearity extends the identity from the monomials to
  every $\Phi$ holomorphic near the inner roots. Therefore
  \[
    [v_\rmin,\Sg_{f_j}]=[v_\rmin,\Phi_j]
    =[y_\rmin,\nu^{\bar n+j}]=y_{\rmin,\bar n+j}.
  \]
  Put $\Pi=\langle f_j,v\rangle$ and $x_0=\frac43n\alpha$. In
  \Cref{lem:far-sector}, $v_{\rm near}$ is the component of $v$ on the inner
  and outer zone roots, so $\langle f_j,v_{\rm near}\rangle=[v_{\rm near},\Sg_{f_j}]$.
  The outer zone roots contribute nothing to it, by \Cref{lem:binomial-transform}(c), and
  the inner roots contribute $y_{\rmin,\bar n+j}$. Its degree condition
  holds since $D_0\le2^{-20}n\alpha_1^2$. Hence
  \begin{equation}\label{eq:inner-identity}
    |y_{\rmin,\bar n+j}-\Pi|
    \le(2\cos(\alpha/2))^n\,\|G_j\|_{x_0}\,e^{n\tau^2/2+C_FM^2\ell\tau}.
  \end{equation}
  On the other hand, $\langle f_j,X\rangle=0$ by \Cref{lem:filter-properties}(b), and
  $X=v+E$, so
  \begin{equation}\label{eq:inner-defect-pairing}
    \Pi=-\langle f_j,E\rangle .
  \end{equation}

  \emph{Step 3: one product estimate.} Write
  $\Omega=\Omega_\rmin\Omega_\rmout$. For $|x|=x_0$, Hermite's formula and
  \eqref{eq:filter-ratio} give
  \begin{equation}\label{eq:hermite-exterior}
    G_j(x)=-\frac{\Df^\sharp(x)\Omega(x)}{2\pi i}
    \oint_\Gamma\frac{U_j(\xi)}{\Cf^\sharp(\xi)\Omega(\xi)(\xi-x)}\,d\xi ;
  \end{equation}
  the term of \eqref{eq:hermite-formula} without $\Omega_\rmin(x)$ integrates to
  zero because $x$ is outside $\Gamma$. The circle $|x|=x_0$ has distance
  at least $0.8n\alpha$ from $\Gamma$, whose length is at most $2.4n\alpha$
  by \Cref{lem:coordinates-near-minus-one}(a). A point $\xi\in\Gamma$ is the coordinate of
  $-e^z$ for some $z\in\Gamma_\zeta$, and $|U_j(\xi)|=|e^z|^{\bar n+j}|1+e^z|^{-n}$.
  So \eqref{eq:binomial-weight} and \Cref{lem:angles-and-weights}(c) give
  $|U_j|\le(2\cos(\theta_\Gamma/2))^{-n}e^{(d+1)\eta_\Gamma}$ on $\Gamma$. It
  remains to bound the single ratio
  $\Df^\sharp(x)\Omega(x)/(\Cf^\sharp(\xi)\Omega(\xi))$.

  For a root $-e^{z_i}$ of $P$ put $\delta_i=|z-z_i|$. A retained root
  $\lambda=-e^{z_i}$ in $\Omega$ contributes the factor
  $(x-x_\lambda)/(\xi-x_\lambda)$. Its numerator is at most $1.86n\alpha_2$
  and, by \Cref{lem:coordinates-near-minus-one}(a), its denominator is at least
  $0.48n\delta_i$. For each removed root $\lambda_i=-e^{z_i}$, with moved pole
  $\mu_i=-e^{z_i'}$, uniformly for $|x|=x_0$ and $\xi\in\Gamma$,
  \begin{equation}\label{eq:removed-factor}
    \Bigl|\frac{x-x_{\mu_i}}{n-x_{\mu_i}}\cdot
    \frac{n-x_{\lambda_i}}{\xi-x_{\lambda_i}}\Bigr|
    \le C\Bigl(1+\frac\sigma{\delta_i}\Bigr).
  \end{equation}
  To see this, write both factors without the coordinates:
  \[
    \frac{x-x_{\mu_i}}{n-x_{\mu_i}}
    =\frac12\bigl(1-e^{-z_i'}\bigr)+\frac x{2n}\bigl(1+e^{-z_i'}\bigr),
    \qquad
    \frac{\xi-x_{\lambda_i}}{n-x_{\lambda_i}}=\frac{1-e^{z-z_i}}{1+e^z}.
  \]
  By \eqref{eq:moved-poles} and $|z|\le1.01\alpha_2\le\sigma/5$, which holds
  since $\alpha_2\le2.02\kappa=2.02\sigma/128$,
  $|z_i'|\le5|z_i|+2\sigma\le5\delta_i+3\sigma$. Since
  $\operatorname{Re}z_i'>0$ and $x_0/(2n)=\frac23\alpha$, the first factor
  has modulus at most
  $\frac12|1-e^{-z_i'}|+\frac23\alpha|1+e^{-z_i'}|\le C\min(1,\delta_i+\sigma)$.
  The second has modulus at least $c\min(1,\delta_i)$; for this use
  $|1-e^{a+ib}|^2=(1-e^a)^2+4e^a\sin^2(b/2)$ with
  $|b|=|\operatorname{Im}(z-z_i)|\le\pi+1/50$, and $|1+e^z|\le3$. This
  proves \eqref{eq:removed-factor}. Neither form involves the coordinates
  themselves, so the bound is uniform also when $\lambda_i$ or $\mu_i$ is
  near $1$, where the coordinates are large.

  Let $\mathcal R'\subseteq\mathcal R_*$ consist of the roots in $\Omega$
  and the $K_*$ removed roots in $\mathcal R_*$. On $\Gamma_\zeta$ every
  factor on the left of \eqref{eq:rectangle-separation} is at most $3.3\alpha_2$, so
  \eqref{eq:rectangle-separation} gives
  $\prod_{i\in\mathcal R'}\delta_i\ge h_*^{N_*}(3.3\alpha_2)^{-(N_*-|\mathcal R'|)}$.
  A retained factor costs $C\alpha_2/\delta_i$, and \eqref{eq:removed-factor}
  costs $C\sigma/\delta_i$ for a removed root in $\mathcal R_*$, since
  $\delta_i\le3.3\alpha_2\le\sigma$. Every removed root outside $\mathcal R_*$
  has $\delta_i\ge8\tau$ and costs at most $C\sigma/\tau$. Thus
  \begin{equation}\label{eq:product-estimate}
    \Bigl|\frac{\Df^\sharp(x)\Omega(x)}{\Cf^\sharp(\xi)\Omega(\xi)}\Bigr|
    \le\Bigl(\frac{C\alpha_2}{h_*}\Bigr)^{N_*}
    \Bigl(\frac{C\sigma}{\alpha_2}\Bigr)^{K_*}
    \Bigl(\frac{C\sigma}\tau\Bigr)^{K-K_*}
    \le\Bigl(\frac{C\alpha_2}{h_*}\Bigr)^{N_*}
    \Bigl(\frac{C\sigma}\tau\Bigr)^K .
  \end{equation}
  Combining \eqref{eq:hermite-exterior}--\eqref{eq:product-estimate} with
  the maximum principle gives
  \begin{equation}\label{eq:interpolant-bound}
    \log\|G_j\|_{x_0}\le-n\log\bigl(2\cos(\theta_\Gamma/2)\bigr)
    +(d+1)\eta_\Gamma+L_1,\qquad
    L_1=C\Bigl[N_*\log\frac{\alpha_2}\tau+K\log\frac\sigma\tau+1\Bigr].
  \end{equation}

  \emph{Step 4: the defect pairing.} By \Cref{lem:binomial-transform}(b), the
  number $\gamma_k=(-1)^k\binom nkf_j(n-2k)$ is the coefficient of $z^k$ in
  $(1-z)^nG_j(x(z))$. Apply \Cref{lem:monotone-weight} to $G_j$, with the
  radius $x_0$ and the degree $D_0\le x_0^2/(2n)$, on the circles
  $|z|=q_t$ and $|z|=1/q_t$, where $|x_{-q_t}|=|x_{-1/q_t}|=u$. Cauchy's
  estimate and \eqref{eq:saddle-point-bounds} give, for every $k$,
  \begin{equation}\label{eq:test-coefficient-bound}
    |\gamma_k|\le(n+1)\binom nt\max\Bigl(1,\frac u{x_0}\Bigr)^{D_0}
    \|G_j\|_{x_0}\min\bigl(q_t^{\,t-k},q_t^{\,k-(n-t)}\bigr).
  \end{equation}
  The first two factors after $n+1$ are harmless:
  \begin{equation}\label{eq:binomial-factor-bound}
    \binom nt\max\Bigl(1,\frac u{x_0}\Bigr)^{D_0}\le(2\cos(\alpha/2))^n .
  \end{equation}
  If $u\le x_0$, this is $\binom nt=(2\cos(\alpha_t/2))^n$ with
  $\alpha\le\alpha_t$. Otherwise \eqref{eq:hoeffding-binomial-bound} gives
  $\binom nt\le2^ne^{-u^2/(2n)}$; the function $X\mapsto e^{-X^2/(2n)}X^{D_0}$
  decreases on $X\ge x_0$, because $x_0^2\ge nD_0$; and
  $2^ne^{-x_0^2/(2n)}=2^ne^{-8n\alpha^2/9}\le(2\cos(\alpha/2))^n$. By
  \eqref{eq:inner-defect-pairing}, \eqref{eq:test-coefficient-bound}, \eqref{eq:binomial-factor-bound}
  and the three sums of \Cref{prop:filtered-defect-heights}(d),
  \begin{equation}\label{eq:inner-defect-bound}
    |\Pi|\le3(n+1)\kappa^{-1}(2\cos(\alpha/2))^n\|G_j\|_{x_0}e^{H_0}.
  \end{equation}

  \emph{Step 5: conclusion.} Every height is $O(M^2\ell\tau)$. Indeed
  $(d+1)\eta_\Gamma\le C_dM^2\ell\tau$; next
  $H_0+\log(3(n+1)/\kappa)\le CM\ell\le CM^2\ell\tau$, because
  $\tau\ge1/M$; and $K\log(\sigma/\tau)\le CM\ell/\Lambda$, because
  $K\le8MT\le CM\ell/(\Lambda w)$ by \Cref{prop:standing-bounds}(b) and
  $\sigma/\tau\le128\kappa M\le128c_1w$. Also $N_*\le D_0$,
  $\alpha\ge\alpha_1$ and $\theta_\Gamma\le\theta_2$. So
  \eqref{eq:inner-identity}, \eqref{eq:interpolant-bound},
  \eqref{eq:inner-defect-bound} and \Cref{lem:angles-and-weights}(a) with
  $\theta'=\theta_\Gamma$ give, for every $|j|<d$,
  \[
    \log|y_{\rmin,\bar n+j}|\le-\mathfrak g
    +C\Bigl[D_0\log\frac{\alpha_2}\tau+M^2\ell\tau\Bigr].
  \]
  Since $\alpha_2\ge16\tau\ge16/M$ and $n\ge M^3\Lambda\ell$, we have
  $M^2\ell\tau\le n\tau\alpha_2/(16\Lambda)$. By \eqref{eq:root-count-hypotheses}
  for a suitable $C_L$, large $\Lambda$, and \eqref{eq:root-count-gain}, the right
  side is at most $-\mathfrak g/2\le-n\tau\alpha_2/4$.

  The $2d-1$ integers $\bar n+j$ lie in $I$, because $u\ge u_0\ge4J\ge8d$.
  The inner roots, with full multiplicity, include every retained root of
  angle at most $\theta_1$: such a root has angle below $\theta_\Gamma$
  and drift at most $\eta_\Gamma$, so it lies in $\mathcal R_*$ and not on
  $\Gamma_\zeta$. By \Cref{cor:small-component-root-count} and
  \eqref{eq:hankel-count} with $\epsilon=e^{-n\tau\alpha_2/4}$,
  $k'\le4C_HM^2\ell^2/(n\tau\alpha_2)$. Every other root of angle at most
  $\theta_1$ is removed, and $K\le8T/\tau$.
\end{proof}

\subsection{Roots of angle below the defect angle}

\begin{lemma}[Roots near $-1$]\label{lem:roots-near-minus-one}
  There is an absolute constant $C_N$ such that, for large $\Lambda$, if
  $u\le n/128$ then $N_{\rm all}(\alpha_t)\le C_NM\ell$.
\end{lemma}
\begin{proof}
  Put $N_D=\lceil C_{\rm q}M\ell\rceil$ with $C_{\rm q}=\max(600(C_B+1),36C_A)$, and
  $\tau=N_D/(36A_\theta)$. Then $B_\theta\le N_D/600$ and $\tau\ge1/M$.
  Suppose $N_{\rm all}(\alpha_t)\ge N_D$, and list the angles of the roots,
  with multiplicity, as $\theta''_1\le\theta''_2\le\cdots$. Set
  $\alpha_2=\theta''_{N_D}$ and $\gamma_D=A_\theta\alpha_2/N_D$. By
  \Cref{prop:standing-bounds}(c), $\theta''_j\ge(j-2B_\theta)/A_\theta$, so
  $\gamma_D\ge0.99$ and $16\tau\le\alpha_2<\alpha_t$. Also
  \Cref{prop:standing-bounds}(c) bounds the number of angles between
  $\theta''_{\lceil N_D/2\rceil}$ and $\theta''_{N_D}$ by
  $A_\theta(\alpha_2-\theta''_{\lceil N_D/2\rceil})+2B_\theta$. Hence
  \[
    \alpha_2-\theta''_{\lceil N_D/2\rceil}\ge\frac{N_D/2-2B_\theta}{A_\theta}
    >16\tau .
  \]
  Thus at least $N_D/2$ roots have angle at most $\alpha_2-16\tau$, whereas
  $N_{\rm all}(\alpha_2)<N_D$. We check \eqref{eq:root-count-hypotheses}. By
  \Cref{prop:standing-bounds}(b), $8T/\tau\le8MT\le CM\ell/\Lambda\le N_D$, so
  $D_0<2N_D<r$. The gain is
  \[
    n\tau\alpha_2=\frac{n\gamma_DN_D^2}{36A_\theta^2}\ge c\,\gamma_D\Lambda M\ell,
  \]
  because $n\ge M^3\Lambda\ell$, while
  $D_0\log(\alpha_2/\tau)\le2N_D\log(36\gamma_D)\le C\gamma_DM\ell$. So
  \eqref{eq:root-count-hypotheses} holds for large $\Lambda$, and
  \Cref{lem:root-count} bounds the number of roots of angle at most
  $\alpha_2-16\tau$ by
  \[
    \frac{4C_HM^2\ell^2}{n\tau\alpha_2}+\frac{8T}\tau\le\frac{CM\ell}\Lambda
    <\frac{N_D}2,
  \]
  a contradiction. Thus $N_{\rm all}(\alpha_t)<N_D\le C_NM\ell$.
\end{proof}

\section{The endpoint test}\label{sec:endpoint-test}
Let $\gamma$ be the coefficient vector of a test that equals
one at the defect $t$ and is small to its left. Convolving it with
the integer annihilator $A$ gives a vector $\gamma^A$ whose pairing with
the defects on the left half is essentially $(-1)^t\binom ntA_0\mathfrak e$, a nonzero
integer multiple of $\binom nt$: the annihilator removes the unknown
recurrence and leaves bounded filtered data. But $\gamma^A$ is not the
vector of a test of degree below $r$. So $\gamma^A$ is replaced, near the
index $t$, by the vector $\gamma'$ of a genuine test $f_A$ whose transform
carries prescribed zeros. Since $f_A$ pairs to zero with the filtered
word, the main term must be cancelled by three error terms,
\eqref{eq:four-term-split}, and each of them is small.

\subsection{Parameters}
Throughout this section $t$ satisfies \textup{(W)} and \textup{(D)}, and
$u\le n/128$. Put
\begin{equation}\label{eq:endpoint-parameters}
  m=\Bigl\lfloor\frac r{\sqrt\Lambda}\Bigr\rfloor,\qquad
  \mathcal L=\log\frac Sm,\qquad \epsilon_w=\frac1{10w}\le\frac1{10}.
\end{equation}
By \eqref{eq:agreement-parameter-ranges}, with $c_S=c_1^2/4$, uniformly for $w\ge1$,
\begin{equation}\label{eq:endpoint-parameter-ranges}
  \frac mS\le\frac1{c_S\sqrt\Lambda\,w^2},\qquad
  \mathcal L=\frac12\log\Lambda+2\log w+O(1),\qquad
  \epsilon_wS\ge\frac{c_Swr}{10}.
\end{equation}
For large $\Lambda$, these give $\mathcal L\ge200$, $m\le e^{-200}S$ and
$23m<r$. On a logarithmic scale every height of \Cref{prop:filtered-defect-heights} is
$O(M\ell)$, whereas $m\ge\sqrt\Lambda M\ell/2$; this comparison absorbs
all errors below.

Put $\tau_s=\epsilon_w\alpha_t/4$, and filter at the threshold
$\eta_s=\tau_s/4$. It is admissible, since $\eta_s=\alpha_t/(160w)$ lies
between $c_\eta/M$ and $\kappa/8$ by \eqref{eq:defect-angle-range} and
\eqref{eq:agreement-parameter-ranges}. Write $K$ for the number of removed roots, so
that $K\le C_KM\ell/(\Lambda w)$, and let $X$, $v$, $\omega$ and $E$ be
as in \Cref{sec:prediction-and-filtering}. Let
$\alpha_s\in[\alpha_t(1+\epsilon_w),\alpha_t(1+2\epsilon_w)]$ be the angle
given by \Cref{lem:far-sector} with $\alpha_1=\alpha_t(1+\epsilon_w)$,
$\tau=\tau_s$ and $\eta=\eta_s$. Call the retained roots of angle below
$\alpha_s$ the \emph{near-sector} roots, and let $N_s$ be their number,
with multiplicity.

\subsection{The geometric-tail polynomial}
For $0<q<1$ and integers $m'\ge0$, define the polynomial
\[
  \psi_{m',q}(j)=\sum_{a=0}^{m'}(-1)^a\binom{m'}aq^{m'-a}(1-q)^a\binom{j-1}a .
\]
It has degree $m'$ in $j$ and $\psi_{m',q}(0)=1$.
These are normalized Meixner polynomials:
\[
  \psi_{m',q}(j)=q^{m'}M_{m'}(j-1;1,q),
\]
as follows by expanding the standard hypergeometric formula
\cite[\href{https://dlmf.nist.gov/18.20.E7}{(18.20.7)}]{NIST:DLMF}.
Their classical orthogonality supplies the tail estimate below; we also
need a bound on a complex disk.

\begin{lemma}[Geometric tails]\label{lem:geometric-tails}
  For $0<q<1$ and every integer $m'\ge0$,
  \[
    \sum_{j\ge1}q^j|\psi_{m',q}(j)|\le\frac{q^{(m'+2)/2}}{1-q}.
  \]
  Moreover, with $g=1-q$, for every $V_0\ge1$,
  \[
    \max_{|j|\le V_0}|\psi_{m',q}(j)|\le\Bigl[16\Bigl(1+\frac{gV_0}{m'}\Bigr)\Bigr]^{m'}
    \qquad(m'\ge1).
  \]
\end{lemma}
\begin{proof}
  The Meixner squared-norm formula
  \cite[\href{https://dlmf.nist.gov/18.19.T1}{Table 18.19.1}]{NIST:DLMF},
  with $\beta=1$, $c=q$ and $x=j-1$, gives
  \[
    \sum_{j\ge1}q^j\psi_{m',q}(j)^2
    =q^{2m'+1}\frac{q^{-m'}}{1-q}=\frac{q^{m'+1}}{1-q}.
  \]
  Cauchy--Schwarz
  with $\sum_{j\ge1}q^j=q/(1-q)$ gives the tail bound. For the local bound,
  $\psi_{m',q}(j)=[z^{m'}]\bigl((1+z)^{j-1}(qz-g)^{m'}\bigr)$; apply Cauchy's
  estimate on $|z|=x:=m'/(4(V_0+m'+1))$. Indeed,
  $|\log(1+z)|\le-\log(1-x)\le x/(1-x)\le m'/(4(V_0+1))$,
  so $|(1+z)^{j-1}|\le e^{m'/4}$ for $|j|\le V_0$; also
  $|qz-g|\le|z|+g$.
\end{proof}

Put $q_t=t/(n-t)$, $g=1-q_t$, so that $\kappa\le g\le2.01\kappa$, and
$\psi=\psi_{m,q_t}$. Since $\binom n{t-j}/\binom nt\le q_t^j$ and
$q_t\le e^{-g}$, \Cref{lem:geometric-tails} gives
\begin{equation}\label{eq:left-tail-bound}
  \sum_{j=1}^t\frac{\binom n{t-j}}{\binom nt}|\psi(j)|
  \le\frac{q_t^{(m+2)/2}}g\le g^{-1}e^{-gm/2}.
\end{equation}

The test is $f(x)=\psi((x-u)/2)$, for which $f(n-2k)=\psi(t-k)$; it has
degree $m$. Write $H_\psi=\Bt f$, and put
$V_f=\|f\|_{16u}$ and $V_H=\|H_\psi\|_{16u}$, maxima over the complex disk
$|x|\le16u$. The
binomial theorem gives, with $a_x=(n+x)/(2n)$,
\[
  H_\psi(x)=[z^m]\Bigl((q_tz-g)^m(1+z)^{t-1-n}(1+a_xz)^n\Bigr).
\]
Let $|x|\le16u$ and $|z|=\varrho\le1/4$. Here $|a_x|\le1$ and
$na_x-(n-t+1)=(x-u)/2-1$. Since $|1+w'|\le e^{\operatorname{Re}w'+|w'|^2/2}$
for every $w'$, and $|1+w'|\ge e^{\operatorname{Re}w'-4|w'|^2}$ for
$|w'|\le1/4$, the last two factors have modulus at most
$\exp\bigl((|x-u|/2+1)\varrho+5(n+1)\varrho^2\bigr)$. Take $V_0=8.5u$ and
$\varrho=m/(4(V_0+m+1))$. Since $|x-u|\le17u$, and $m(n+1)\le2u^2$ because
$m\le e^{-200}S$, this exponent is at most $3m/4$. Also
$|q_tz-g|\le\varrho+g$ and $(\varrho+g)/\varrho\le9(1+gV_0/m)$. So Cauchy's
estimate on $|z|=\varrho$ and $9e^{3/4}<24$ give
$|H_\psi(x)|\le(24(1+gV_0/m))^m$. The test obeys the stronger bound of
\Cref{lem:geometric-tails}, since $|(x-u)/2|\le V_0$. With $gu\le2.01S$,
$S\ge m$ and $24\cdot18.085<e^7$, this gives
\begin{equation}\label{eq:local-norm-bound}
  \log V_f,\ \log V_H\le m\log\Bigl(24\Bigl(1+\frac{8.5gu}m\Bigr)\Bigr)
  \le m\mathcal L+7m .
\end{equation}

\subsection{Approximating the annihilator with prescribed zeros}
Let $Q_c$ be the monic polynomial whose zeros are the coordinates of the
near-sector roots, with multiplicity. A near-sector root
$\lambda=-e^{\rho+i\theta}$ has $|\rho|\le\eta_s$ and
$|\theta|<\alpha_s\le1.2\alpha_t\le2.43u/n$ by \eqref{eq:defect-angle-range}, so by
\eqref{eq:coordinate-tanh} and \Cref{lem:coordinates-near-minus-one}(b) its coordinate
satisfies $|\operatorname{Re}x_\lambda|=n|\sinh\rho|/(\cosh\rho+\cos\theta)
\le0.52n\eta_s<u/8$ and $|x_\lambda|\le0.51n\alpha_s(1+(0.51\alpha_s)^2)\le1.25u$. The
zeros of $\Df^\sharp$ are the
$K$ coordinates $x_{\mu_i}$, of modulus at least $32u$ by
\eqref{eq:moved-coordinates}. Set $\beta=4m$, $B_Q=\beta+N_s+K$,
$N_T=16m$, and
\[
  Q(x)=\Bigl(\frac{x+u}{2u}\Bigr)^\beta\frac{Q_c(x)\Df^\sharp(x)}{Q_c(u)\Df^\sharp(u)},
  \qquad
  G_A(x)=Q(x)\sum_{j=0}^{N_T}\mathsf g_j(x-u)^j,
\]
where the $\mathsf g_j$ are the Taylor coefficients at $u$ of $A(\lambda(x))/Q(x)$.
Every zero of $Q$ is at distance at least $7u/8$ from $u$. Thus
$A(\lambda(x))/Q(x)$ is holomorphic on $|x-u|\le u/2$, where
$|\lambda(x)|\le1$ and $|Q|,|Q|^{-1}\le3^{B_Q}$. Consequently, by Cauchy's
estimate and the remainder of the Taylor series,
\[
  |\mathsf g_j|\le H_A3^{B_Q}(2/u)^j,\qquad
  |A(\lambda(x))-G_A(x)|\le2H_A9^{B_Q}(2\varsigma/u)^{N_T+1}
  \quad(|x-u|\le\varsigma\le u/4).
\]
Also $|Q(x)|\le20^{B_Q}$ on $|x|\le15.5u$, so
$|G_A(x)|\le H_A60^{B_Q}34^{N_T+1}$ there, and
\begin{equation}\label{eq:product-norm}
  |G_A(x)H_\psi(x)|\le M_0:=H_A60^{B_Q}34^{N_T+1}V_H\qquad(|x|\le15.5u).
\end{equation}
The product $G_AH_\psi$ vanishes at every near-sector coordinate and at
every zero of $\Df^\sharp$, to full multiplicity, and at $x=-u$ to order
$\beta$. Let $f_A=\Bt^{-1}(G_AH_\psi)$, of degree
$D_A\le m+B_Q+N_T$. Put
\begin{equation}\label{eq:endpoint-coefficient-vectors}
  \gamma_l=(-1)^l\binom nl\psi(t-l),\qquad
  \gamma^A_k=\sum_{i=0}^{d_A}A_i\gamma_{k-i},\qquad
  \gamma'_k=(-1)^k\binom nkf_A(n-2k),
\end{equation}
with $\gamma_l=0$ outside $0\le l\le n$. By \Cref{lem:binomial-transform}(b), their
generating functions are $(1-z)^nH_\psi(x(z))$, $A(z)(1-z)^nH_\psi(x(z))$
and $(1-z)^nG_A(x(z))H_\psi(x(z))$, and $\lambda(x(z))=z$. The polynomial
$f_A$ itself is never evaluated.

\subsection{The comparison}

\begin{theorem}[Endpoint comparison]\label{thm:endpoint-comparison}
  For large $\Lambda$, if \textup{(W)} and \textup{(D)} hold and $u\le n/128$,
  then $N_s>m$.
\end{theorem}
\begin{proof}
  Suppose that $N_s\le m$. By \Cref{prop:filtered-defect-heights} and
  $m\ge\sqrt\Lambda M\ell/2$, for large $\Lambda$,
  \begin{equation}\label{eq:endpoint-absorption}
    K\le m,\qquad (C_h+8)\ell+H_0\le m,\qquad gm\ge\kappa m\ge14\ell,
    \qquad\mathcal L\ge200 .
  \end{equation}
  Hence $B_Q\le6m$ and $D_A\le23m<r$. Since $\log H_A\le C_h\ell$ by
  \Cref{prop:standing-bounds}(a), \eqref{eq:local-norm-bound} and
  \eqref{eq:product-norm} give
  \[
    \log M_0\le C_h\ell+m\mathcal L+C_*m,\qquad C_*=105,
  \]
  because $60\le e^5$ and $34\le e^4$ give
  $B_Q\log60+(N_T+1)\log34\le30m+68m$. Moreover
  $m\mathcal L/S=\mathcal Le^{-\mathcal L}$ is negligible.

  The polynomial $\Df^\sharp$ divides $G_AH_\psi$, and $D_A<r$, so
  $\langle f_A,X\rangle=0$ by \Cref{lem:filter-properties}(b), that is,
  $\sum_k\gamma'_k(v_k+E_k)=0$. Split the sum over $E$ at $\bar n$ and insert
  $\gamma^A$ on the left half:
  \begin{equation}\label{eq:four-term-split}
    \sum_{k\le \bar n}\gamma^A_kE_k
    =\sum_{k\le \bar n}(\gamma^A_k-\gamma'_k)E_k
    -\sum_{k=0}^n\gamma'_kv_k-\sum_{k>\bar n}\gamma'_kE_k .
  \end{equation}
  Step 1 evaluates the left side; Steps 2--4 bound the three terms on the
  right.

  \emph{Step 1: the main term.} Since $\gamma_l=0$ for $l<0$, exchanging
  the order of summation gives exactly
  \begin{equation}\label{eq:main-term}
    \sum_{k\le \bar n}\gamma^A_kE_k=\sum_{l\le \bar n}\gamma_lY_l,\qquad
    Y_l=\sum_{0\le i\le\min(d_A,\bar n-l)}A_iE_{l+i}.
  \end{equation}
  By \Cref{prop:standing-bounds}(a), $d_A<J\le u/4$, so $t+d_A<\bar n$.
  \begin{itemize}
    \item $l<t$: then $Y_l=(A(\shift)E)_l=(A(\shift)X)_l$, since
      $A(\shift)v=0$. So $|Y_l|\le H_A\max|X|\le8d^2e^\ell\le e^{5\ell}$: the
      unknown recurrence has disappeared. By \eqref{eq:left-tail-bound} and
      $g^{-1}\le e^\ell$, these terms total at most
      $\binom nte^{6\ell-gm/2}\le\binom nte^{-\ell}$.
    \item $l=t$: the indices $t+i$, $1\le i\le d_A$, satisfy
      $n-t-(t+i)\ge u/2$, so \Cref{prop:filtered-defect-heights}(c) gives
      $|E_{t+i}|\le e^{H_0-32S}$. With \Cref{prop:filtered-defect-heights}(e),
      $Y_t=A_0\mathfrak e+O(H_Ae^{H_0-32S})$, and $\gamma_t=(-1)^t\binom nt$.
    \item $t<l\le \bar n$: every $E_{l+i}$ has $t<l+i\le \bar n$, so
      $|Y_l|\le H_Ae^{H_0-32S}$. Here $|t-l|\le u/2$, so $|\psi(t-l)|\le V_f$,
      and $\binom nl\le\binom ntq_t^{-(l-t)}\le\binom nte^{1.01S}$. By
      \eqref{eq:local-norm-bound} and \eqref{eq:endpoint-absorption}, these terms total
      at most $\binom nte^{-30.5S}$.
  \end{itemize}
  Hence
  \begin{equation}\label{eq:main-term-bound}
    \Bigl|\binom nt^{-1}(-1)^t\sum_{k\le \bar n}\gamma^A_kE_k-A_0\mathfrak e\Bigr|
    \le e^{-\ell}+e^{-30S}.
  \end{equation}

  \emph{Step 2: the comparison term.} The difference of the two vectors
  has the generating polynomial
  \[
    \sum_k(\gamma^A_k-\gamma'_k)z^k=\bigl(A(z)-G_A(x(z))\bigr)(1-z)^nH_\psi(x(z)).
  \]
  Bound it on the circle $|z|=q_t$, where $|A(z)|\le H_A$ and $x_{-q_t}=u$.
  Use the cutoff $\varsigma=\frac u2e^{-\mathcal L/4}\le\frac u8$, for which
  $\varsigma^2/n=\frac m4e^{\mathcal L/2}$.
  \begin{itemize}
    \item $|x(z)-u|\le\varsigma$: the Taylor bound gives
      $|A(z)-G_A(x(z))|\le2H_A9^{B_Q}(2\varsigma/u)^{N_T+1}$. With
      $N_T=16m$, $B_Q\le6m$, $|1-z|\le1+q_t$ and \eqref{eq:local-norm-bound}, the
      product with $|H_\psi|\le V_H$ has logarithm at most
      $n\log(1+q_t)+\log2+C_h\ell-3m\mathcal L+25m
      \le n\log(1+q_t)-\frac52m\mathcal L$, since $9\le e^3$.
    \item $|x(z)-u|>\varsigma$ and $|x(z)|\le15u$: by
      \eqref{eq:circle-gaussian} and $n^2-u^2\le n^2$,
      $|1-z|^n\le(1+q_t)^n(1+\varsigma^2/n^2)^{-n/2}
      \le(1+q_t)^ne^{-\varsigma^2/(4n)}$, while $|A-G_A||H_\psi|\le2M_0$.
    \item $|x(z)|>15u$: apply \Cref{lem:monotone-weight} with $x_0=15u$ to
      $H_\psi$ and to $G_AH_\psi$, whose degrees are at most
      $D_A\le23m\le x_0^2/(2n)$. By \eqref{eq:monotone-weight-bound} the product
      $|1-z|^n|A-G_A||H_\psi|$ is at most
      $2M_0(1+q_t)^n\bigl((1+\kappa^2)/(1+225\kappa^2)\bigr)^{n/2}$. As
      $\kappa\le1/128$, the ratio is at most $1/(1+200\kappa^2)$, and
      $(1+200\kappa^2)^{n/2}\ge e^{50n\kappa^2}$, so the product is at most
      $2M_0(1+q_t)^ne^{-50S}$.
  \end{itemize}
  Since $\mathcal L\ge200$, all three cases give at most
  $(1+q_t)^ne^{-\frac52m\mathcal L}$. Cauchy's estimate and
  \eqref{eq:saddle-point-bounds} yield
  \begin{equation}\label{eq:comparison-coefficients}
    |\gamma^A_k-\gamma'_k|\le(n+1)\binom nt\,q_t^{\,t-k}\,
    e^{-\frac52m\mathcal L}\qquad(k\ge0).
  \end{equation}
  Sum this against $|E_k|$ over $k\le \bar n$. The indices $k\le t$ are covered
  by the first sum in \eqref{eq:defect-weighted-sums}. For $t<k\le \bar n$,
  $q_t^{\,t-k}\le e^{1.01S}$ and $|E_k|\le e^{H_0-32S}$. By
  \eqref{eq:endpoint-absorption} and $\log((n+1)/\kappa)\le2\ell$,
  \begin{equation}\label{eq:comparison-term}
    \Bigl|\sum_{k\le \bar n}(\gamma^A_k-\gamma'_k)E_k\Bigr|
    \le\binom nte^{-2m\mathcal L}.
  \end{equation}

  \emph{Step 3: the retained term.} The transform $G_AH_\psi$ of $f_A$
  vanishes at the near-sector coordinates to full multiplicity, so
  $\langle f_A,v_{\rm near}\rangle=0$ in \eqref{eq:far-sector-expansion}, by
  \eqref{eq:spectral-pairing} and \Cref{lem:binomial-transform}(c).
  \Cref{lem:far-sector} applies with $\alpha_1=\alpha_t(1+\epsilon_w)$,
  $\tau=\tau_s$, $\eta=\eta_s$ and $D=D_A\le23m\le n\alpha_1^2/2$. Its disk
  is $|x|\le\frac43n\alpha_s\le15.5u$, where $|G_AH_\psi|\le M_0$. By
  \Cref{lem:angles-and-weights}(a) and (d),
  $(2\cos(\alpha_s/2))^ne^{n\tau_s^2/2}\le\binom nte^{-\frac n5\epsilon_w\alpha_t^2}
  \le\binom nte^{-0.79\epsilon_wS}$. So
  \[
    \Bigl|\sum_{k=0}^n\gamma'_kv_k\Bigr|
    \le\binom nt\exp\bigl(-0.79\epsilon_wS+C_h\ell+m\mathcal L+C_*m
    +C_FM^2\ell\tau_s\bigr).
  \]
  Here $\epsilon_wS\ge c_Swr/10$ by \eqref{eq:endpoint-parameter-ranges}, while
  $C_h\ell+C_FM^2\ell\tau_s\le CM\ell$, and, by \eqref{eq:endpoint-parameter-ranges} with $\log w\le w$,
  $m\mathcal L+C_*m\le C(1+\log\Lambda)wr/\sqrt\Lambda$. For large $\Lambda$ the
  exponent is at most $-\epsilon_wS/2\le-\ell$.

  \emph{Step 4: the opposite end.} Bound $(1-z)^nG_A(x(z))H_\psi(x(z))$ on
  the circle $|z|=1/q_t$, where $x_{-1/q_t}=-u$.
  \begin{itemize}
    \item $|x(z)+u|\le u$: the polynomial $G_AH_\psi/(x+u)^\beta$ is at most
      $M_0u^{-\beta}$ on $|x+u|=u$, hence inside. With
      \eqref{eq:circle-gaussian}, the product is at most
      $(1+q_t^{-1})^nM_0\sup_{X\ge0}(X/u)^\beta\exp(-X^2/(4n))
      \le(1+q_t^{-1})^nM_0(2\beta/S)^{\beta/2}$, because
      $X^{2k}e^{-cX^2}\le(k/c)^k$ for $c>0$ and $k=\beta/2$, as $k!\le k^k$.
    \item $|x(z)+u|>u$ and $|x(z)|\le15u$: \eqref{eq:circle-gaussian} gives
      the factor $(1+u^2/n^2)^{-n/2}\le e^{-S/4}$ in place of the supremum.
    \item $|x(z)|>15u$: \Cref{lem:monotone-weight} gives the factor
      $e^{-50S}$, as in Step 2.
  \end{itemize}
  Since $\beta=4m$, the first factor is
  $(8m/S)^{2m}=\exp(-2m\mathcal L+2m\log8)\le\exp(-2m\mathcal L+6m)$, and it is the
  largest of the three. Cauchy's estimate and \eqref{eq:saddle-point-bounds} yield
  \[
    |\gamma'_k|\le(n+1)\binom nt\,q_t^{\,k-(n-t)}e^{C_h\ell-m\mathcal L+(C_*+6)m}
    \qquad(k\ge0).
  \]
  Every $k>\bar n$ occurs in the second or third sum of \eqref{eq:defect-weighted-sums}.
  With \eqref{eq:endpoint-absorption}, the last term of \eqref{eq:four-term-split} is
  at most $\binom nte^{-m\mathcal L+(C_*+7)m}\le\binom nte^{-m\mathcal L/4}$.

  \emph{Conclusion.} By \Cref{cor:integral-defect},
  \eqref{eq:main-term-bound}, \eqref{eq:four-term-split},
  \eqref{eq:comparison-term} and Steps 3--4,
  \[
    1\le|A_0\mathfrak e|\le e^{-\ell}+e^{-30S}+e^{-2m\mathcal L}+e^{-\ell}
    +e^{-m\mathcal L/4},
  \]
  which is below $1/2$ for large $\Lambda$. This contradiction proves the theorem.
\end{proof}

\begin{remark}
  The convolved vector $\gamma^A$ enters only on the left half $k\le \bar n$,
  where the annihilator turns the defects into bounded filtered data
  (Step~1). Beyond $\bar n$ the proof uses only the genuine test vector
  $\gamma'$. Its generating function has a zero of order $\beta$ at the
  coordinate $x=-u$ of the point $-1/q_t$, and this makes the weights
  near $n-t$ small (Step~4).
\end{remark}

\section{Proof of the main theorem}\label{sec:proof-of-main-theorem}
We keep the standing assumptions of \Cref{sec:standing-assumptions} and the
agreement interval of \Cref{sec:agreement-interval}, and show that they are
contradictory for large $\Lambda$.

\subsection{Short agreement}
Suppose there is a first defect and $u\le n/128$. It satisfies \textup{(W)}
and \textup{(D)}, so we may use the filter, the
angle $\alpha_s$ and the near-sector roots of \Cref{sec:endpoint-test}.
\Cref{lem:roots-near-minus-one} bounds the number of
roots of angle below $\alpha_t$ by $C_NM\ell$. By
\Cref{prop:standing-bounds}(c), the roots with angle in $[\alpha_t,\alpha_s]$
number at most $2A_\theta\epsilon_w\alpha_t+2B_\theta$, which is $O(M\ell)$
because $A_\theta\alpha_t\le2.02C_AM^2\ell\kappa=O(wM\ell)$ by
\eqref{eq:defect-angle-range} and \eqref{eq:agreement-parameter-ranges}, and
$\epsilon_w=1/(10w)$. Thus, for large $\Lambda$,
\[
  N_s\le CM\ell\le\frac{\sqrt\Lambda}2M\ell\le m,
\]
so \Cref{thm:endpoint-comparison} excludes this case.

\subsection{Long agreement}
Suppose now that $u>n/128$; this includes full agreement.

A recurrence that agrees with signs on so long a block is
cyclotomic. A root of unity of large order lies near $-1$, which the root
count excludes. So the sign word has a bounded period on $I$, and the
difference of $Z$ and its shift by the period vanishes on $I$. An
explicit test of bounded degree isolates its first nonzero value.

\emph{The recurrence is cyclotomic.} The signs on $I$ form a block of
length $u-1>n/128-1$. Since $d\le C_dM^2\ell$, the threshold of
\Cref{cor:long-agreement-cyclotomic} is $O(M^2\ell^5)$, which is $o(n)$ by
\eqref{eq:scale-comparison}. So every root of $P$ is a root of unity,
$P\in\Z[z]$, and $T=0$. In particular every filter is trivial.

\emph{An empty arc.} Let $u'$ be the largest integer with $u'\le n/128$
and $u'\equiv n\pmod2$, put $t'=(n-u')/2$, and let
$I'=\{t'+1,\ldots,n-t'-1\}\subseteq I$. For large $\Lambda$ its length $u'-1$
exceeds $u_0$, so $(t',u',I')$ satisfies the working hypothesis
\textup{(W)} of \Cref{sec:agreement-interval}, and the results of
\Cref{sec:recurrence-arithmetic,sec:prediction-and-filtering,sec:circles-and-far-sector,sec:roots-near-minus-one}
hold with $(t',u',I')$ in place of $(t,u,I)$. By \Cref{lem:angles-and-weights}(d),
$\alpha_{t'}\ge1.99u'/n>1/128$ for large $\Lambda$.
\Cref{lem:roots-near-minus-one} gives
$N_{\rm all}(1/128)\le N_{\rm all}(\alpha_{t'})\le C_NM\ell<r$.
Apply \Cref{lem:root-count} with
$\alpha_2=1/128$ and $\tau=1/3072\ge1/M$, so that $\alpha_2-16\tau=1/384$.
Its hypotheses \eqref{eq:root-count-hypotheses} hold for large $\Lambda$:
$D_0=N_{\rm all}(1/128)\le C_NM\ell<r$, and $n\tau\alpha_2=n/393216$
exceeds $C_LD_0\log24$. By \eqref{eq:root-count-bound}, the number of roots of
angle at most $1/384$ is at most $4C_HM^2\ell^2\cdot393216/n<1$. Hence $P$
has no root within angle $1/384$ of $-1$.

\emph{Bounded period.} For every integer $q\ge2$ there is a primitive
$q$th root of unity at angle at most $4\pi/q$ from $-1$. Indeed, the
angle of $e^{2\pi ia/q}$ from $-1$ is $(\pi/q)|2a-q|$. If $q$ is odd, take
$a=(q+1)/2$. If $4\mid q$, take $a=q/2+1$, which is odd, and every common
odd prime factor of $a$ and $q$ divides $2a-q=2$. If $q=2q'$ with $q'$ odd,
take $a=q'+2$, which is odd and coprime to $q'$ (for $q=2$ the root is
$-1$ itself). By \Cref{cor:long-agreement-cyclotomic}, $P$ is a product of
cyclotomic polynomials $\Phi_q$, each irreducible over $\Q$. If
$\Phi_q\mid P$, then all primitive $q$th roots of unity are roots of $P$,
including one within angle $4\pi/q$ of $-1$; since $4\pi/4826<1/384$, this
forces $q<4826$. So every root of $P$ has order dividing the constant
$Q_0=\operatorname{lcm}(1,\ldots,4825)$. Write
$y_k=\sum_\zeta p_\zeta(k)\zeta^k$ with $\zeta^{Q_0}=1$ and
$\deg p_\zeta<d$. For each residue $s$ modulo $Q_0$, the sequence
$j\mapsto y_{s+Q_0j}$ is a polynomial in $j$ of degree below $d$. On $I$ it
takes values $\pm1$ at more than $u/Q_0-2>2d$ consecutive integers $j$. A
polynomial $g$ of degree below $d$ with $g^2=1$ at more than $2d$ points
satisfies $g^2\equiv1$, so it is constant. Hence $y$ is $Q_0$-periodic on
$\Z$ with values $\pm1$, and its minimal polynomial $P$ divides
$z^{Q_0}-1$; in particular $P$ has simple roots and $d\le Q_0$.

\emph{Full agreement is impossible.} Suppose $y=Z$ on $0,\ldots,n$, and
write $y_k=\sum_\zeta c_\zeta\zeta^k$. For a root $\zeta\ne1$ of $P$, let
$f_\zeta$ be the test whose transform is the Lagrange polynomial that
equals one at $x_\zeta$ and vanishes at the coordinates of the other roots
$\ne1$ of $P$. Its degree is below $Q_0<r$, so
$\langle f_\zeta,Z\rangle=0$. By \Cref{lem:binomial-transform}(b) and (c),
$\langle f_\zeta,y\rangle=c_\zeta(1-\zeta)^n$. Hence $c_\zeta=0$ for every
$\zeta\ne1$, so $y$ and $Z$ are constant on $0,\ldots,n$, contradicting
the assumption that $F$ is nonconstant.

\emph{The period difference.} So there is a first defect $t$. Put
$N=n-Q_0$, and
\[
  Z'_i=Z_{i+Q_0}-Z_i\in\{-2,0,2\}\qquad(0\le i\le N).
\]
For large $\Lambda$, $Q_0<u-1$. Periodicity of $y$ and agreement on $I$ give
$Z'_i=0$ for $t<i<N-t$, and $Z'_t=y_{t+Q_0}-Z_t=y_t-Z_t=-\mathfrak e\in\{-2,2\}$. Put
$u_*=N-2t=u-Q_0>n/129$. For a polynomial $f$ of degree below $r-Q_0$, let
\[
  \chi(k)=\frac{(-1)^{Q_0}k^{\underline{Q_0}}f(k-Q_0)-(n-k)^{\underline{Q_0}}f(k)}
  {n^{\underline{Q_0}}},\qquad\deg\chi<r .
\]
The identities $\binom N{k-Q_0}=\binom nkk^{\underline{Q_0}}/n^{\underline{Q_0}}$ and
$\binom Ni=\binom ni(n-i)^{\underline{Q_0}}/n^{\underline{Q_0}}$, and the change of
variables $k=i+Q_0$ in the first term, give
\[
  \sum_{i=0}^N(-1)^i\binom Nif(i)Z'_i
  =\sum_{k=0}^n(-1)^k\binom nk\chi(k)Z_k=0,
\]
by \Cref{lem:degree-and-moments}. Thus $Z'$ has vanishing moments, on the
grid $0,\ldots,N$, through degree $r-Q_0-1$.

\emph{An elementary endpoint test.} Put $q_0=129/130$. Both exterior
binomial tails, normalized by $\binom Nt$, are bounded by powers of $q_0$:
$\binom N{t-s}/\binom Nt\le(t/(N-t))^s\le q_0^s$, since
$t/(t+u_*)\le129/131$, and the same holds for $\binom N{N-t+s}$. Let
$\phi_L=\psi_{L,q_0}$ be the polynomial of \Cref{lem:geometric-tails}, so
that $\sum_{s\ge1}q_0^s|\phi_L(s)|\le q_0^{(L+2)/2}/(1-q_0)$. Choose a fixed
$L$ making this less than $1/16$, put $b'=L+1$, and use the polynomial
\[
  \varphi(i)=\phi_L(t-i)\Bigl(1+\frac{t-i}{u_*}\Bigr)^{b'},
\]
of degree $2L+1$, with $\varphi(t)=1$. On the left tail, multiplication by
$(1+s/u_*)^{b'}$ changes the normalized absolute sum
$\sum_{s\ge1}q_0^s|\phi_L(s)|$ by $O(1/u_*)$: expand this factor and sum
the fixed polynomial moments of $|\phi_L|$ against $q_0^s$. On the right
tail $i=N-t+s$ the factor is $(-s/u_*)^{b'}$, while
$|\phi_L(-u_*-s)|\le C(L)(u_*+1+s)^L$, with $C(L)$ depending only on $L$, so the normalized sum is
$O(u_*^{L-b'})=O(1/u_*)$; the term $s=0$ vanishes. Since $u_*\to\infty$ and
$|Z'_i|\le2$, both tails together contribute less than $\binom Nt/2$ to
$\sum_i(-1)^i\binom Ni\varphi(i)Z'_i$ for large $\Lambda$, whereas the term $i=t$ has
modulus $2\binom Nt$. So the moment is nonzero. This contradicts the
vanishing moments, because $2L+1<r-Q_0$ for large $\Lambda$. The argument also
covers $t=0$, where the left tail is empty.

\subsection{Conclusion}
Every agreement length leads to a contradiction. Hence there is an
absolute constant $\Lambda_*$ such that, for every sufficiently large $n$,
no nonconstant profile on $0,\ldots,n$ has deficit at least
$\lceil\Lambda_*^{2/3}n^{1/3}\ell^{2/3}\rceil$. Thus
$\Delta(n)\le\Lambda_*^{2/3}n^{1/3}\ell^{2/3}$ for large $n$. Here
$\ell=\log(n+3)\le2\log n$ for $n\ge3$. Since $\log n>0$ for $n\ge2$, enlarging the constant to cover the
finitely many remaining $n\ge2$ gives the bound for all $n\ge2$.
For $n=1$, every nonconstant profile has degree one, so $\Delta(1)=0$
and the bound holds as well. This proves \Cref{thm:one-third-degree-bound}. \qed

\emph{Effectivity.} The proof uses no compactness or limiting argument:
each absolute constant comes from finitely many explicit inequalities, and
Voutier's bound is explicit. So $\Lambda_*$ and the threshold on $n$ can
in principle be computed. We have not computed them, and they are
enormous. For example, the long-agreement case uses the period
$Q_0=\operatorname{lcm}(1,\ldots,4825)>10^{2095}$ and needs $Q_0<r$, so it
applies only to deficits above $10^{2095}$.

\section*{Disclosure of AI tools}
\addcontentsline{toc}{section}{Disclosure of AI tools}
AI tools were used extensively in this work, including
Claude Code (Anthropic) and Codex (OpenAI).
They searched for and tested proof strategies, drafted proofs and research
notes, wrote the search and validation code and the Lean formalization, checked
the literature, and drafted and revised this text. The authors directed the
work, reviewed the final text, and take responsibility for its content.

\clearpage
\bibliographystyle{alphaurl}
\bibliography{references}
\end{document}